\pdfoutput=1
\documentclass[12pt]{emulateapj}
\usepackage{lscape}
\usepackage{graphicx}
\usepackage{multirow}
\usepackage{mathtools}
\usepackage{epsf}
\usepackage{longtable}
\usepackage{amsmath}

\newcommand{\kms}{km s$^{-1}$ }

\newcommand{\Msu}{$M_{\odot}$ }

\newcommand{\degs}{$^{\circ}$ }
\newcommand{\degn}{$^{\circ}$}

\newcommand{\hi}{{\rm H\,}{{\sc i}}}
 
\newcommand{\his}{{\rm H\,}{{\sc i }}}
\newcommand{\hiis}{{\rm H\,}{{\sc ii}} }
\newcommand{\ci}{\ion{C}{1}}

\newcommand{\ois}{\ion{O}{1} }

\newcommand{\cii}{\ion{C}{2}}
\newcommand{\cis}{\ion{C}{1} }
\newcommand{\ciis}{\ion{C}{2} }

\newcommand{\sii}{\ion{S}{2}}  
\newcommand{\siis}{\ion{S}{2} }  
\newcommand{\znii}{\ion{Zn}{2}}
\newcommand{\zniis}{\ion{Zn}{2} }
\newcommand{\siii}{\ion{Si}{2}}  
\newcommand{\siiis}{\ion{Si}{2} }  
\newcommand{\mgii}{\ion{Mg}{2}}
\newcommand{\mgiis}{\ion{Mg}{2} }  
\newcommand{\niii}{\ion{Ni}{2}}  
\newcommand{\niiis}{\ion{Ni}{2} }  
\newcommand{\crii}{\ion{Cr}{2}}  
\newcommand{\criis}{\ion{Cr}{2} }  
  
\newcommand{\piis}{\ion{P}{2} }  
\newcommand{\feii}{\ion{Fe}{2}}  
\newcommand{\feiis}{\ion{Fe}{2} }  
  
\newcommand{\cuiis}{\ion{Cu}{2} }

\newcommand{\mAA}{\AA \,}

\begin{document}	\title{METAL: The Metal Evolution, Transport, and Abundance in the Large Magellanic Cloud Hubble program. II. Variations of interstellar Depletions and dust-to-gas ratio within the LMC }

\author{Julia Roman-Duval\altaffilmark{1}, Edward B. Jenkins\altaffilmark{2}, Kirill Tchernyshyov\altaffilmark{3}, Benjamin Williams\altaffilmark{3}, Christopher J.R. Clark\altaffilmark{1}, Karl D. Gordon\altaffilmark{1}, Margaret Meixner\altaffilmark{4}, Lea Hagen\altaffilmark{1}, Joshua Peek\altaffilmark{1}, Karin Sandstrom\altaffilmark{5}, Jessica Werk\altaffilmark{3}, Petia Yanchulova Merica-Jones\altaffilmark{5}}
\altaffiltext{1}{Space Telescope Science Institute, 3700 San Martin Drive, Baltimore, MD 21218; duval@stsci.edu}
\altaffiltext{2}{Princeton University Observatory, Peyton Hall, Princeton University, Princeton, NJ 08544-1001 USA}
\altaffiltext{3}{Department of Astronomy, Box 351580, University of Washington, Seattle, WA 98195, USA}
\altaffiltext{4}{SOFIA Science Mission Operations, NASA Ames Research Center, Building N232, M/S 232-12, P.O. Box 1, Moffett Field, CA 94035-0001}
\altaffiltext{5}{Center for Astrophysics and Space Sciences, Department of Physics, University of California, 9500 Gilman Drive, La Jolla, San Diego, CA 92093, USA}
\altaffiltext{6}{NASA Goddard Space Flight Center, Greenbelt, MD 20771, USA}

\begin{abstract}

A key component of the baryon cycle in galaxies is the depletion of metals from the gas to the dust phase in the neutral ISM. The METAL (Metal Evolution, Transport and Abundance in the Large Magellanic Cloud) program on the {\it Hubble Space Telescope} acquired UV spectra toward 32 sightlines in the half-solar metallicity LMC, from which we derive interstellar depletions (gas-phase fractions) of Mg, Si, Fe, Ni, S, Zn, Cr, and Cu. The depletions of different elements are tightly correlated, indicating a common origin. Hydrogen column density is the main driver for depletion variations. Correlations are weaker with volume density, probed by \cis fine structure lines, and distance to the LMC center. The latter correlation results from an East-West variation of the gas-phase metallicity. Gas in the East, compressed side of the LMC encompassing 30 Doradus and the Southeast \his over-density is enriched by up to $+$0.3 dex, while gas in the West side is metal-deficient by up to $-$0.5 dex. Within the parameter space probed by METAL, no correlation with molecular fraction or radiation field intensity are found. We confirm the factor 3-4 increase in dust-to-metal and dust-to-gas ratios between the diffuse ($\log$ N(H) $\sim$ 20 cm$^{-2}$) and molecular ($\log$ N(H) $\sim$ 22 cm$^{-2}$) ISM observed from far-infrared, 21 cm, and CO observations. The variations of dust-to-metal and dust-to-gas ratios with column density have important implications for the sub-grid physics of chemical evolution, gas and dust mass estimates throughout cosmic times, and for the chemical enrichment of the Universe measured via spectroscopy of damped Lyman-$\alpha$ systems.

\end{abstract}

\maketitle

\section{Introduction} \label{introduction}

\indent The transfer of metals between interstellar gas and dust constitutes an important component of the baryon cycle in galaxies, the incessant recycling of gas, dust, and metals between stars, the interstellar medium, and galaxy halos. Recent observations and modeling have shown that interstellar dust can grow and evolve in the interstellar medium (ISM). The evolution of the dust content of galaxies over cosmic times \citep{morgan2003, boyer2012, rowlands2012, rowlands2014, zhukovska2013} cannot be explained by balance of the dust production rates in evolved stars \citep{bladh2012, riebel2012, srinivasan2016} and supernova remnants \citep{matsuura2011} and the dust destruction rates in interstellar shocks \citep{jones1994, jones1996}. This so-called dust budget crisis can be resolved by dust growth in the ISM, via accretion of gas-phase metals onto pre-existing dust grains \citep{zhukovska2008, draine2009, mckinnon2016}, effectively modifying the relation between dust and gas mass. The dust-to-metal ratio (D/M) and the dust-to-gas ratio (D/H $=$ D/M $\times$ Z, where Z is the metallicity) are fundamental parameters resulting from the interstellar gas-dust cycle, and are expected to substantially vary with environment, in particular metallicity \citep{asano2013, feldmann2015}. \\
\indent Owing to the key role of dust in the radiative transfer, chemistry, and thermodynamics, galaxy evolution cannot be understood without accounting for dust, and thus for the interstellar gas-dust cycle. Because 30\%-50\% of stellar light is absorbed by dust in the optical-ultraviolet (UV) and re-emitted in the far-infrared (FIR), our understanding of dust is critical to correct for reddening effects, interpret observations of galaxies, and trace their stellar, metal, dust, and gas content over cosmic times. Additionally, a comprehensive understanding of how metals deplete from the gas to the dust phase via dust formation in the ISM, and from the dust to the gas phase via dust destruction by sputtering in Supernova (SN) shock waves, is critical to understand the chemical enrichment of the universe over cosmic times. Indeed, the chemical enrichment of the universe is traced by quasar absorption spectroscopy through damped Lyman-$\alpha$ systems \citep[DLAs, e.g., ][]{rafelski2012}, and the resulting DLA neutral gas abundances have to be corrected for depletion effects, particularly above 1\% solar metallicity.\\
\indent D/M and D/H can either be estimated in two ways. The first method consists in using emission-based tracers of gas (\his 21 cm and CO rotational transitions) and dust (FIR emission) as in, e.g., \citet{remyruyer2014, devis2019}. However, the degeneracies between the dust opacity, dust mass, and CO dark-gas \citep[e.g.,][]{RD2014, galliano2018} preclude an unambiguous characterization of the variations of D/M and D/H with metallicity using these emission-based ISM tracers. The second approach compares chemical abundances in neutral interstellar gas based on UV spectroscopy to stellar abundances of young stars recently formed out of ISM \citep[e.g., ][]{luck1998}, which can be used as a proxy for the total (gas + dust) abundances of the ISM \citep[i.e., interstellar depletions, as in][]{savage1996, jenkins2009, tchernyshyov2015}. In order to understand how metals deplete from the gas into the dust phase, and thus how D/M varies with environment, a detailed census of metals in neutral gas and dust from UV spectroscopy is therefore required.\\
\indent Interstellar depletions are the logarithm of the fraction of metals in the gas-phase. Thus, the depletion for element X is

\begin{equation}\label{depletion_equation}
\delta(X)  =  \log_{10}(X/H)_{\mathrm{gas}} -  \log_{10}(X/H)_{\mathrm{total}}
\end{equation}

\noindent where $(X/H)_{\mathrm{gas}}$ is the abundance of $X$ in the gas-phase and $(X/H)_{\mathrm{total}}$ is the total abundance of $X$ (gas+dust), assumed to be equal to the abundance of $X$ in the photospheres of young stars that have formed out of the ISM recently. Metallicities estimated from the emission lines of \hiis regions suffer from relatively large systematics due to 1) the need to estimate a temperature from line ratios and temperature fluctuations inside these regions, 2) the poorly understood discrepancy between recombination and collisionnally excited lines, and 3) the presence of dust in \hiis regions, removing some of the metals from the gas.\\
\indent In the Milky Way \citep{jenkins2009} and SMC \citep{jenkins2017}, the fraction of metals in the gas-phase decreases with increasing hydrogen volume density and column density, albeit at different rates for different elements. In this paper, we derive interstellar depletions in the Large Magellanic Cloud (LMC) using recent observations with the {\it Hubble Space Telescope} obtained as part of the METAL large Cycle 24 program \citep[GO-14675, see][]{RD2019}. The LMC metallicity \citep[1/2 solar, see][]{russell1992} lies approximately midway between that of the MW (solar) and the SMC \citep[1/5 solar, see][]{russell1992}, and provides the link between the large differences in dust properties seen between the MW and SMC, which is below the ''critical metallicity'' \citep{feldmann2015} where the dust-to-gas ratio departs from a linear scaling with metallicity \citep{remyruyer2014, RD2014, RD2017, chiang2018}, the PAH fraction is an order of magnitude lower than in the MW \citep{sandstrom2010}, and the UV extinction curves distinctively lacks a 2175 \mAA bump and are steeper in the FUV than in the MW \citep{gordon2003}. In addition, the LMC's gas disk is thinner \citep[120 pc,][]{elmegreen2001} and less inclined than the SMC, alleviating confusion in velocity and distance structure along the line-of-sight.\\
\indent In this paper, we focus on the variations of interstellar depletions with environment {\it within} the LMC. In an upcoming paper, we will perform a detailed comparison of depletions between the Milky Way, LMC, and SMC. The paper is organized as follows. We describe the observations and abundance measurements in Sections \ref{observation_section} and \ref{measurements_section}. In Section \ref{ci_section}, we describe the derivation of volume density and radiation fields from the \cis fine structure line. The correlations between depletions of different elements and with environment are examined in Sections \ref{correlation_elements} and \ref{correlations_environment}. Combining depletions of different dust constituents, the variations of the dust-to-gas ratio with environment are derived in Section \ref{dg_section}. Section \ref{summary_section} provides a summary of this paper.

\section{Observations}\label{observation_section}

\indent The details of the observing strategy, sample properties, and survey parameters were covered in the METAL Survey paper \citep[][; hereafter Paper I]{RD2019}.  For convenience we provide a brief summary and explanation of the sample here. In short, this study of interstellar depletions in the LMC is based on STIS and COS medium-resolution spectra of 32 massive stars, obtained predominantly as part of the METAL large HST program \citep[GO-14675, see][]{RD2019}, but also from archival HST spectra of the same target sample (DOI \dataset{10.17909/t9-g6d9-rj76}). The target sample used in this analysis is listed in Table 1, along with the LMC \his and H$_2$ column densities toward each star, and the heliocentric velocity intervals with detectable absorption in the \feiis lines (1608 \AA, 2249 \AA, 2260 \AA). Information in Table 1 is directly taken from the survey paper \citep[][, hereafter Paper I]{RD2019}. \\
\indent The COS spectra were obtained with the G130M and G160M gratings, while the STIS spectra used the E140M and E230M. Two of the targets (SK-70 115 and SK-68 73) had archival E230H spectra as well, which were preferred since they minimized the effects of unresolved saturation. The complete list of medium-resolution spectra used in this analysis, including the program IDs of archival data, is included in Paper I (Tables A1 and A2). 

\begin{deluxetable*}{cccccccc}
\centering
\tabletypesize{\scriptsize}
\tablecolumns{7}
\tablewidth{\textwidth}
\tablecaption{Spectroscopic targets and their interstellar parameters}
\tablenum{1}
 
 \tablehead{Target & Ra & Dec & $E(B-V)$  & $\log N($\hi)$_{\mathrm{LMC}}$\tablenotemark{a}& $\log{N(\mathrm{H}_2})$\tablenotemark{b}&  $V_{\mathrm{helio}}$}\\
 
 \startdata
 & $h$ & $deg$ &   mag & cm$^{-2}$ &  cm$^{-2}$ &  km s$^{-1}$ \\
 \hline
 &&&&&&\\
SK-67 2 & 04h47m04.451s & -67d06m53.12s & 0.26 & 21.04 $\pm$  0.12 & 20.95  & 220---310  \\
SK-67 5 & 04h50m18.918s & -67d39m38.10s & 0.14 & 21.02 $\pm$ 0.04 & 19.46  & 240---320 \\
SK-69 279 & 05h41m44.655s & -69d35m14.90s& 0.21 & 21.59 $\pm$ 0.05 & 20.31  & 210---335 \\
SK-67 14 & 04h54m31.889s & -67d15m24.58s & 0.08 & 20.24 $\pm$ 0.06 & 15.01   & 230---335  \\
SK-66 19 & 04h55m53.951s & -66d24m59.35s & 0.36 & 21.85 $\pm$ 0.07 & 20.2 & 225---330  \\
PGMW 3120 & 04h56m46.812s & -66d24m46.72s & 0.25 & 21.48 $\pm$ 0.03 & 18.3  & 220---330\\
PGMW 3223 & 04h57m00.859s & -66d24m25.12s & 0.19 & 21.4 $\pm$ 0.06 & 18.69  & 220---330  \\
SK-66 35 & 04h57m04.440s & -66d34m38.45s & 0.11 & 20.83 $\pm$ 0.04 & 19.3 & 235---320 \\
SK-65 22 & 05h01m23.070s & -65d52m33.40s & 0.11 & 20.66 $\pm$ 0.03 & 14.93   & 240---340  \\
SK-68 26 & 05h01m32.248s & -68d10m42.93s & 0.29 & 21.6 $\pm$ 0.06 & 20.38  & 225---320 \\
SK-70 79 & 05h06m37.262s & -70d29m24.16s & 0.24 & 21.26 $\pm$ 0.04 & 20.26 & 180---270 \\
SK-68 52 & 05h07m20.423s & -68d32m08.59s & 0.18 & 21.3 $\pm$ 0.06 & 19.47  & 195---350  \\
SK-69 104 & 05h18m59.501s & -69d12m54.82s & 0.1 & 19.57 $\pm$ 0.68 & 14.03 & 200---290 \\
SK-68 73 & 05h22m59.781s & -68d01m46.62s & 0.4 & 21.66 $\pm$ 0.02 & 20.09  & 215---330  \\
SK-67 101 & 05h25m56.221s & -67d30m28.67s & 0.08 & 20.2 $\pm$ 0.04 & 14.14  & 230---360 \\
SK-67 105 & 05h26m06.192s & -67d10m56.79s & 0.17 & 21.25 $\pm$ 0.04 & 19.13 & 235---350 \\
BI 173 & 05h27m09.941s & -69d07m56.46s & 0.15 & 21.25 $\pm$ 0.05 & 15.64  & 195---280  \\
BI 184 & 05h30m30.657s & -71d02m31.60s & 0.2 & 21.12 $\pm$ 0.04 & 19.65  & 200---330 \\
SK-71 45 & 05h31m15.654s & -71d04m09.69s & 0.16 & 21.11 $\pm$ 0.03 & 18.63  & 200---345 \\
SK-69 175 & 05h31m25.520s & -69d05m38.59s & 0.17 & 20.64 $\pm$ 0.03 & 14.28  & 190---370  \\
SK-67 191 & 05h33m34.028s & -67d30m19.72s & 0.1 & 20.78 $\pm$ 0.03 & 14.28 & 235---340 \\
SK-67 211 & 05h35m13.905s & -67d33m27.51s & 0.1 & 20.81 $\pm$ 0.04 & 13.98 & 240---355 \\
BI 237 & 05h36m14.628s & -67d39m19.18s & 0.2 & 21.63 $\pm$ 0.03 & 20.05 & 220---350 \\
SK-68 129 & 05h36m26.768s & -68d57m31.90s & 0.22 & 21.59 $\pm$ 0.14 & 20.2  & 230---330 & \\
SK-66 172 & 05h37m05.394s & -66d21m35.18s & 0.18 & 21.27 $\pm$ 0.03 & 18.21  & 240---320  \\
BI 253 & 05h37m34.461s & -69d01m10.20s & 0.23 & 21.67 $\pm$ 0.03 & 19.76  & 200---320 \\
SK-68 135 & 05h37m49.112s & -68d55m01.69s & 0.26 & 21.46 $\pm$ 0.02 & 19.87   & 225---340 \\
SK-69 246 & 05h38m53.384s & -69d02m00.93s & 0.18 & 21.47 $\pm$ 0.02 & 19.71  & 215---320\\
SK-68 140 & 05h38m57.18s & -68d56m53.1s & 0.29 & 21.47 $\pm$ 0.11 & 20.11 & 220---340 \\
SK-71 50 & 05h40m43.192s & -71d29m00.65s & 0.2 & 21.17 $\pm$ 0.05 & 20.13  & 190---320  \\
SK-68 155 & 05h42m54.93s & -68d56m54.5s & 0.28 & 21.44 $\pm$ 0.09 & 19.99 & 220---360  \\
SK-70 115 & 05h48m49.654s & -70d03m57.82s & 0.2 & 21.13 $\pm$ 0.08 & 19.94  &  195---360   \\

\enddata
\tablenotetext{a}{The \his column densities are from \citet{RD2019}}
\tablenotetext{b}{The H$_2$ column densities are from \citet{welty2012}} 
\end{deluxetable*}

\section{Gas-phase column density and abundance measurements}\label{measurements_section}

\indent The quality of the METAL spectra allows us to derive neutral gas abundances for Fe, Mg, Si, Ni, Cu, Cr, Zn, S, in all METAL targets listed in Table 1 (i.e., all METAL targets except SK-69 220, which is an LBV with a very complex stellar spectrum precluding accurate measurements of interstellar features).  Gas-phase column densities for transitions observed in the STIS/E140M ($R$ $=$ 45,000) and STIS/E230M ($R$ $=$ 30,000)  spectra were derived using the apparent optical depth method \citep[AOD, ][]{savage1991, jenkins1996}, as outlined in Paper I for the Si abundances. The methodology for the STIS-based AOD measurements is described in Section \ref{stis_aod}. For transitions in the COS G130M and G160M spectra ($R$ $=$ 15,000-20,000), we used profile fitting to derive gas-phase column densities, following the method described in Section \ref{cos_profile_fitting}. In some instances of targets and ions covered by both instruments, we chose the STIS-based abundances for this analysis given the higher resolution. We compare the column densities derived by these two methods at different spectral resolutions in Section \ref{compare_cos_stis}. Lastly, ionization corrections for singly ionized column densities are negligible in the column density range of our sight-lines \citep[log N(H) = 20-22][]{tchernyshyov2015, jenkins2017}, so the abundance of an element is taken to be that of the measured low ion.\\
\indent The resulting equivalent widths and column densities for each individual spectral line in the STIS spectra are listed in Table 3, while the combined column density and abundance measurements for each element (combining different spectral lines) in the METAL survey are included in Table 5.

\begin{deluxetable}{cccc}
\centering
\tabletypesize{\scriptsize}
\tablecolumns{4}
\tablewidth{8cm}
\tablecaption{Spectral lines and oscillator strengths used for abundance and depletion measurements}
\tablenum{2}
\tablehead{Element/ion & 12 + $\log$(X/H)$_{\mathrm{LMC, tot}}$ &  Wavelength & $\log$ $\lambda f_{\lambda}$} \\
 
 \startdata
& & \AA & \AA  \\
\hline
& &&\\
\mgiis & 7.26$\pm$0.08 & 1239.925 & -0.106  \\
 & & 1240.395  & -0.355 \\
 &&&\\
\siiis & 7.35 $\pm$0.10 & 1808.013 & 0.575 \\
&&&\\
\ois & 8.50 $\pm$ 0.11 & 1355.598 & -2.805 \\
\piis  & 5.10$\pm$0.1 & 1152.818 & 2.451 \\
&&&\\
\siis & 6.94 $\pm$0.04& 1250.578 &0.809 \\
 & & 1253.805& 1.113\\
 &&&\\
\criis & 5.37$\pm$0.07&  2056.254 & 2.326 \\
& &  2066.161 & 2.024 \\
&&&\\
\feiis &  7.32$\pm$0.08 & 1608.451 & 1.968 \\
&& 1611.201 & 0.347\\
& & 2249.877 & 0.612 \\
& &  2260.780 & 0.742\\
&&&\\
\niiis & 5.92$\pm$0.07 & 1317.217  & 1.876  \\
&& 1370.132 &  1.906   \\
&& 1454.842   & 1.672 \\
& & 1709.600 &  1.743  \\
 && 1741.549 &  1.871  \\
&& 1751.910 &   1.686  \\        
&&&\\
\cuiis & 3.89$\pm$0.04 & 1358.773 & 2.569 \\
&&&\\
\zniis & 4.31$\pm$0.15 & 2026.13 & 3.106 \\
&& 2062.664 & 2.804\\
&&&\\
\enddata
\tablecomments{Mg, Si, O, P, Cr, Fe, Ni, Zn LMC total (gas $+$ dust) abundances are from \citet{tchernyshyov2015}; S, Cu abundances are from \citet{asplund2009} scaled by a factor 0.5. Oscillator strengths are from \citet{morton2003}, except for \zniis \citep{kisielius2015}, \siis \citep{kisielius2014}, and \niiis \citep{jenkins2006}}
\end{deluxetable}

\subsection{Column densities with the AOD method in the STIS spectra}\label{stis_aod}

\begin{figure*}
\centering
\includegraphics[width=8cm]{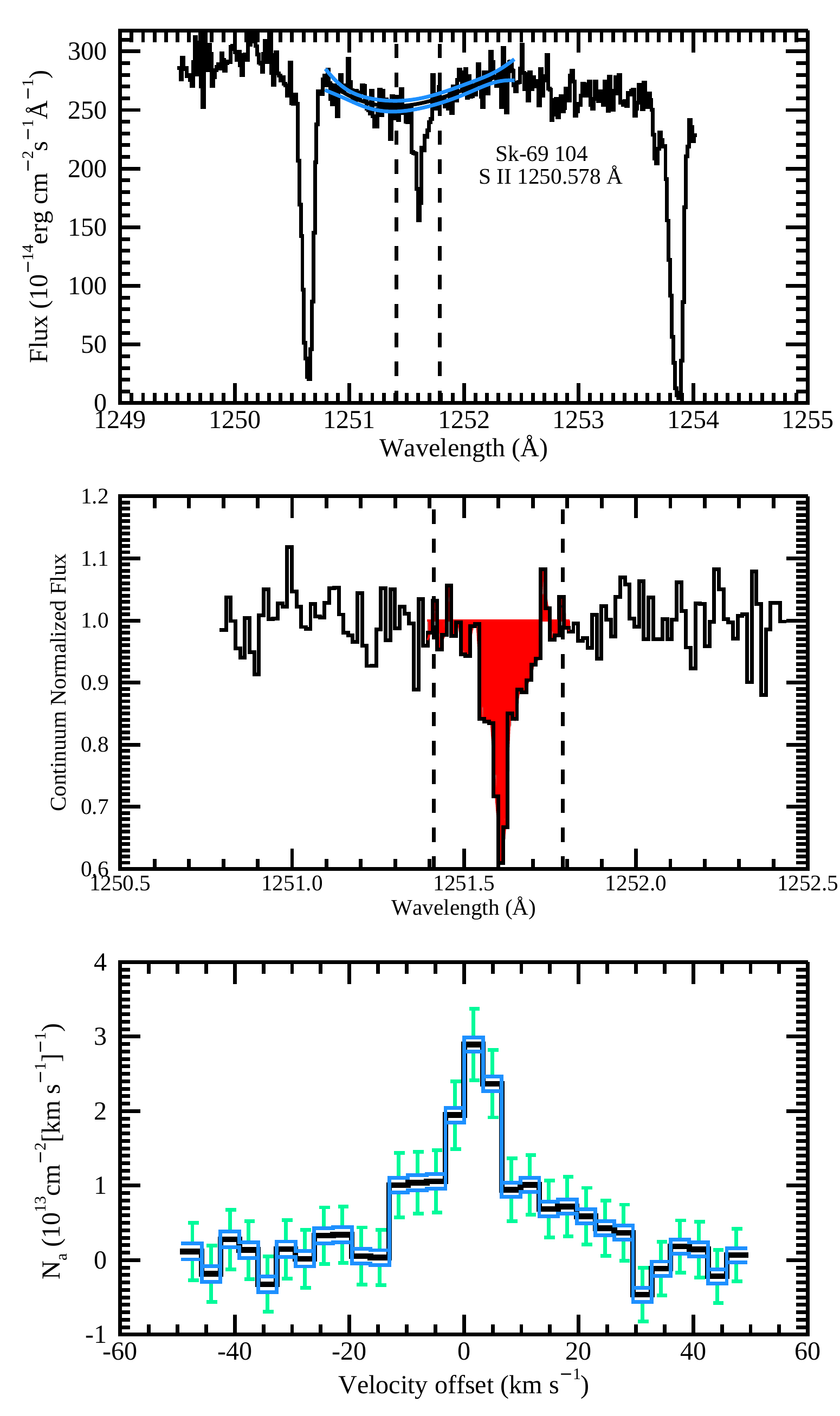}
\includegraphics[width=8cm]{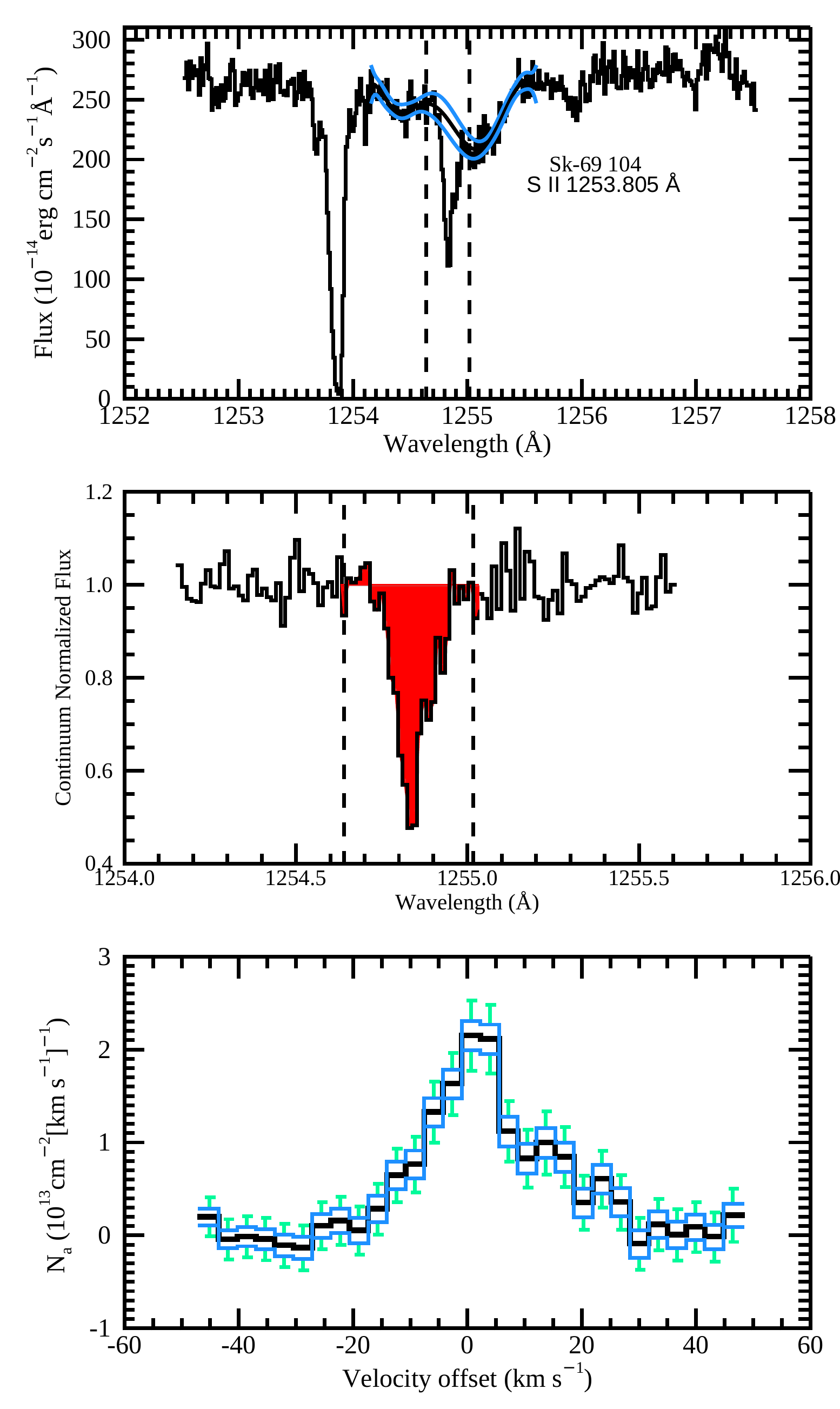}
\caption{Demonstrations of various aspects of deriving AOD outcomes and their uncertainties for Sk-69 104 and \siis ($\lambda$1250 in the left panels, $\lambda$1253 for the right panels). The top panels show the raw fluxes and the adopted continuum curve (black line) and its possible deviations (blue lines). The middle panels show the fluxes normalized to the continuum, with the red fill showing the part of the absorption feature that was included in the derivation of an overall column density. The bottom panels show the apparent column density $N_a(v)$, with the black trace showing the preferred values, the blue traces on either side showing deviations that arise from the upper and lower limits for the adopted continuum, and the green error bars illustrating the random uncertainties arising from photon counting noise. The vertical dashed lines indicate the velocity limits (see Table 1).}
\label{aod_example_sk-69104}
\end{figure*}

\begin{figure}
\centering
\includegraphics[width=8cm]{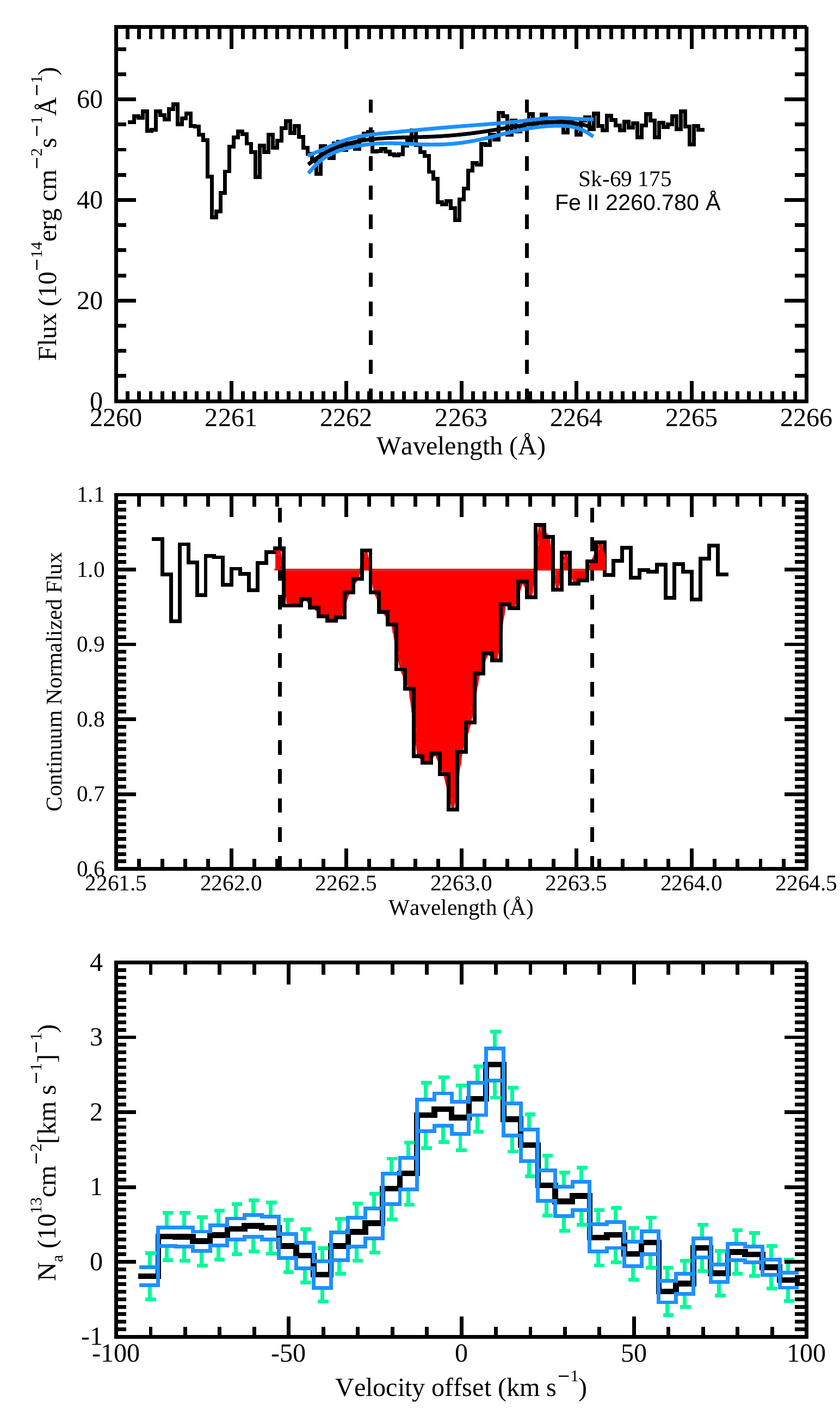}
\caption{Same as Figure \ref{aod_example_sk-69104}, but for Sk-69 175 and \feiis $\lambda$ 2261 \AA.}
\label{aod_example_sk-69175}
\end{figure}

\begin{figure}
\centering
\includegraphics[width=8cm]{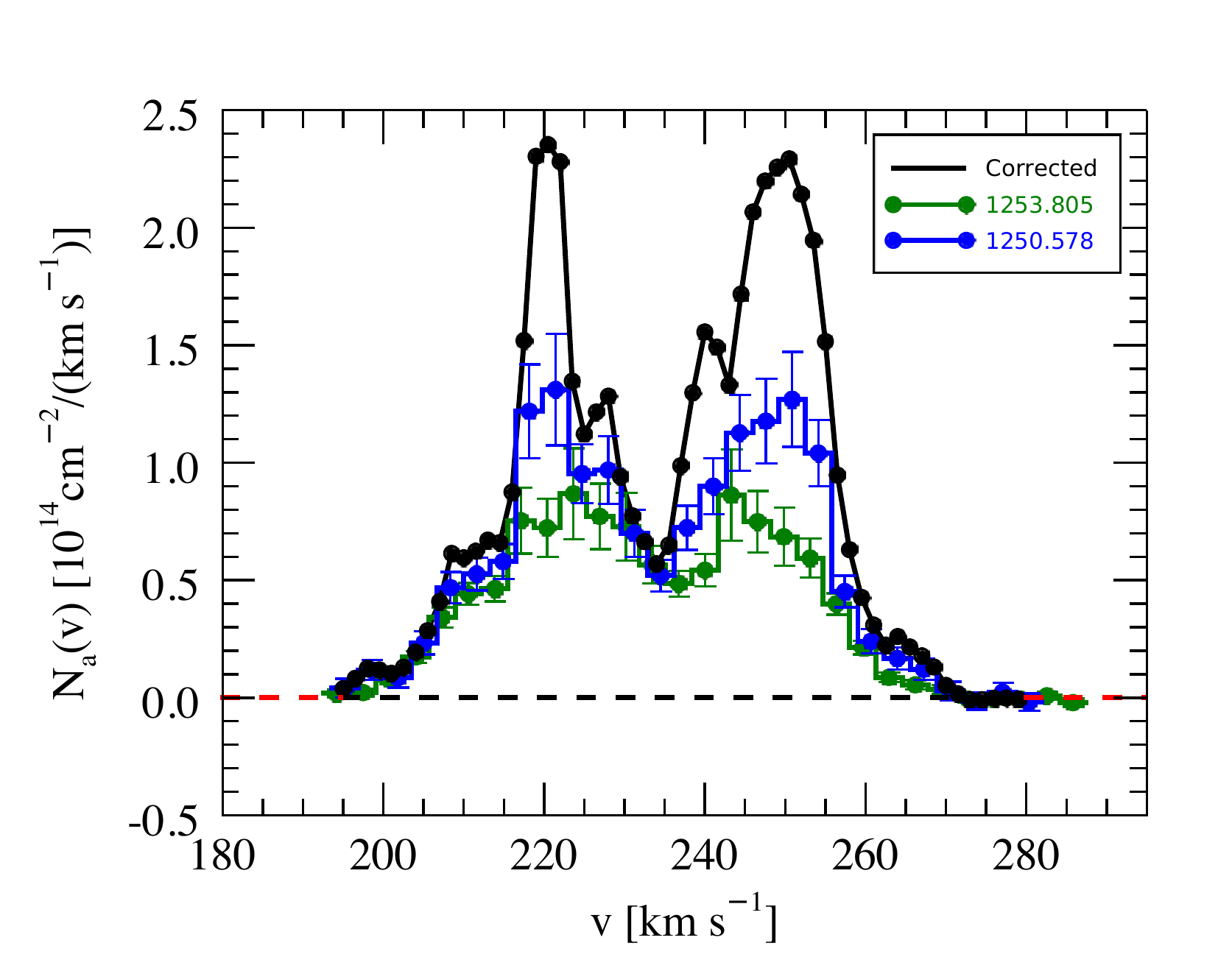}
\caption{Apparent column density of \siis toward BI 173 as measured from the $\lambda$1250 \mAA (blue) and $\lambda$1253 \mAA (green) lines, and corrected for the effects of unresolved saturation following the method described in \citet[][, black]{jenkins1996}. }
\label{example_sii_correction}
\end{figure}

\indent We first determined the continuum levels by best-fitting Legendre polynomials \citep{sembach1992} to fluxes on either side of the absorption profiles (see the top panels of Figures \ref{aod_example_sk-69104} and \ref{aod_example_sk-69175} for some examples). We then applied the AOD method to each feature and determined the apparent column density $N_a$ as a function of velocity. The velocity ranges used to integrate the AOD-based apparent column densities in the STIS spectra are listed in Table 1. The upper portions of the integration velocity ranges are well defined by the sharp, long wavelength edges of strong transitions, such as the \feiis 2344.2 \mAA line \citep[see Figure 6 in][]{RD2019}. The low velocity edge for gas associated with the LMC is however poorly defined, owing to the presence of intermediate and high-velocity gas. Therefore, we define the low velocity edge of AOD integration using observations of \his 21 cm emission in the GASS III survey \citep{mccluregriffiths2009, kalberla2010, kalberla2015} in the directions of our target stars, which reveal a well defined feature arising from gas in the LMC. Specifically, the lower velocity limit of AOD integration corresponds to the  edge of the \his 21 cm emission, where the brightness temperature rises above 0.2 K ($\sim$4$\sigma$). Our basic premise is therefore that much of the foreground gas at lower velocities contains fully ionized hydrogen and thus should not be included. \\
\indent To convert the integrated apparent optical depths to column densities, we assume the oscillator strengths listed in Table 2 for each element/transition. For elements with several transitions from which column densities can be measured, we perform a weighted average of the different column densities, where the weight is given by the inverse square of the error. \\
\indent In some cases the lines are saturated, or nearly so, and when this saturation is not properly resolved, a straight evaluation of the AOD will underestimate the true column density.  If the weakest (or only) line has a central intensity relative to the continuum level $I_0/I_c < 0.05$, we simply declare a lower limit to the column density.  For milder cases of saturation with $I_0/I_c\lesssim 0.3$, we either significantly increased the upper error bound for the derived column density or, when two lines of differing strengths were available, we applied a correction for saturation taking the following approach. In cases where the saturation in the lines of \siis $\lambda\lambda$1250, 1253 \mAA and \zniis $\lambda\lambda$2062, 2026 \mAA did not appear to be too severe and both lines could be measured, the sulfur and zinc column densities were corrected for the effects of unresolved saturation using the method outlined in \citet{jenkins1996}.  Such unresolved saturation effects usually are apparent when the strong transition indicates a lower column density than the weaker one.  The outcome of correcting for this saturation yields a column density that is larger than that from the weak line.  Figure \ref{example_sii_correction} shows an example of this procedure for \sii, where AOD determinations at specific velocities are compared with each other and treated in the same manner as equivalent widths in a standard curve-of-growth analysis. The corrected outcomes for $\log$ N(\sii) and N(\znii) are can be identified in Table 3 because they show no specifications for wavelengths, oscillator strengths, or equivalent widths (as opposed to the measurement of the strong and weak lines). The magnitudes of the corrections are typically $+$0.15 dex for \siis and $+$0.2 for \znii.  \\
\indent The lines of \zniis presented special challenges arising from the interference of other lines at similar wavelengths, which can be troublesome because the velocity ranges of absorption features from the LMC are large.  The left-hand side of the \zniis line at 2062.6604 \mAA may have an overlap from the \criis line at 2062.2361 \AA, which is situated a relative velocity of $-$61.7 \kms, and the right-hand side of the \zniis line at 2026.1370 \mAA can suffer from interference by the Mg I line at 2026.4768 \mAA (at $+$50.3 \kms).  To compensate for such interfering features, we use apparent optical depths of other \criis and Mg I lines, multiply them by the appropriate ratios of $f_{\lambda} \lambda$, and then subtract them with their respective velocity offsets from the \zniis features.  This procedure works well when using the \criis line at 2056.2569 \mAA because its value of $f_{\lambda} \lambda$ is not much different from that of the 2062.2361\mAA line.  The Mg I comparison line at 1827.9351\mAA is considerably weaker than that of the interfering 2026.4768 \mAA transition, and if the latter has unresolved saturation, the compensation will be larger than warranted. When it was clearly evident that this was happening, we disregarded the (usually extreme) right-hand portion of the \zniis 2026 \mAA line and used only the information from the other \zniis feature when we calculated the effects of saturation by comparing the AOD outcomes for the two lines. For the individual results that we obtained for the 2062 \mAA line of \zniis in Table 3, we list the AOD column density outcomes after subtracting off the contribution from \crii.  For the 2026 \mAA line, the corrections were often larger than they should have been, for the reason discussed above.  Hence, for this line we listed the uncorrected column densities.\\
\indent At the opposite extreme, we considered a measurement to be marginal if the equivalent width outcome was less than the 2$\sigma$ level of uncertainty  from noise and continuum placement. For weak lines below this uncertainty threshold, we specified an upper limit for the column density based on a completely unsaturated line having a strength at the measurement value plus a 1$\sigma$ positive excursion, but with an allowance for the fact that negative real line strengths are not allowed even though we occasionally obtained negative measurement outcomes caused by downward noise  fluctuations (or a continuum placement that was too low). Details of how we calculated these 1$\sigma$ upper limits are given in Appendix D of \citet{bowen2008}. Such a calculation avoids the unphysical conclusion that an upper limit for a column density can be nearly zero or negative when the measurement yields an outcome that is $<-1\sigma$. It also yields a smooth transition to a conventional expression of an upper limit as a value plus 1$\sigma$ when the value is larger than twice the noise level.\\
\indent Errors on the column densities stem from the effects of three different sources: (1) noise in the absorption profile, (2) errors in defining the continuum level, and (3) uncertainties in the transition oscillator strengths, $f_{\lambda}$, all of which were combined in quadrature in our error estimation. We evaluate the expected deviations produced by continuum definition by remeasuring the AODs at the lower and upper bounds for the continua, which are derived from the expected formal uncertainties in the polynomial coefficients of the fits as described by \citet{sembach1992}. We multiply these coefficient uncertainties by 2 in order to make approximate allowances for additional deviations that might arise from some freedom in assigning the most appropriate order for the polynomial.\\
\indent When two or more, non-saturated lines - as shown by the weak line strengths and the consistency of the derived column densities between lines of different oscillator strengths - were available for a given element (e.g., Fe II, Ni II, Mg II), we derived the final column density value as average of the column densities derived from each line, weighted by the inverse of their squared errors.

\begin{deluxetable*}{ccccccc}
\centering
\tabletypesize{\scriptsize}
\tablecolumns{7}
\tablewidth{\textwidth}
\tablecaption{Equivalent widths and individual AOD and column density measurements for the STIS observations}
\tablenum{3}
\tablehead{Sight-line & grating & Element & Wavelength & $\log$ $\lambda f_{\lambda}$ & EW & $\log(NX)$ \\}
 \startdata
&  & & \AA & \AA & m\AA & cm$^{-2}$ \\
\hline
&&&&&&\\
BI 173 & STIS-M & OI & 1355.598 & -2.805 & 10.4$\pm$11.9 & $<$18.10 \\
BI 173 & STIS-M & MgII & 1239.925 & -0.106 & 60.1$\pm$6.9 & 15.96$^{+0.13}_{-0.13}$ \\
BI 173 & STIS-M & MgII & 1240.395 & -0.355 & 44.7$\pm$7.2 & 16.06$^{+0.14}_{-0.14}$ \\
BI 173 & STIS-M & SiII & 1808.013 & 0.575 & 236.6$\pm$11.7 & 15.98$^{+0.20}_{-0.20}$ \\
BI 173 & STIS-M & PII & 1152.818 & 2.451 & 166.4$\pm$24.9 & $>$13.60 \\
BI 173 & STIS-M & SII & 1253.805 & 1.113 & 213.3$\pm$6.5 & 15.54$^{+0.04}_{-0.04}$ \\
BI 173 & STIS-M & SII & 1250.578 & 0.809 & 183.2$\pm$8.0 & 15.69$^{+0.04}_{-0.05}$ \\
BI 173 & STIS-M & SII & \nodata & \nodata &  \nodata  & 15.86$^{+0.06}_{-0.06}$ \\
BI 173 & STIS-M & CrII & 2056.254 & 2.326 & 118.2$\pm$8.2 & 13.57$^{+0.04}_{-0.04}$ \\
BI 173 & STIS-M & CrII & 2066.161 & 2.024 & 77.1$\pm$9.1 & 13.65$^{+0.06}_{-0.06}$ \\
BI 173 & STIS-M & FeII & 2260.780 & 0.742 & 157.2$\pm$8.9 & 15.26$^{+0.04}_{-0.04}$ \\
BI 173 & STIS-M & FeII & 2249.877 & 0.612 & 114.2$\pm$8.8 & 15.22$^{+0.04}_{-0.05}$ \\
BI 173 & STIS-M & NiII & 1370.132 & 1.906 & 86.0$\pm$5.4 & 14.03$^{+0.05}_{-0.05}$ \\
BI 173 & STIS-M & NiII & 1317.217 & 1.876 & 53.9$\pm$7.1 & 13.87$^{+0.06}_{-0.07}$ \\
BI 173 & STIS-M & NiII & 1741.549 & 1.871 & 79.4$\pm$12.0 & 13.92$^{+0.07}_{-0.08}$ \\
BI 173 & STIS-M & NiII & 1709.600 & 1.743 & 66.1$\pm$18.3 & 13.96$^{+0.12}_{-0.15}$ \\
BI 173 & STIS-M & NiII & 1751.910 & 1.686 & 79.1$\pm$42.6 & $<$14.21 \\
BI 173 & STIS-M & NiII & 1454.842 & 1.672 & 49.8$\pm$9.6 & 13.97$^{+0.14}_{-0.15}$ \\
BI 173 & STIS-M & CuII & 1358.773 & 2.569 & 17.9$\pm$8.0 & 12.63$^{+0.18}_{-0.27}$ \\
BI 173 & STIS-M & ZnII & 2026.136 & 3.106 & 153.8$\pm$8.5 & 12.95$^{+0.04}_{-0.04}$ \\
BI 173 & STIS-M & ZnII & 2062.664 & 2.804 & 119.8$\pm$8.1 & 13.08$^{+0.04}_{-0.04}$ \\
BI 173 & STIS-M & ZnII & \nodata & \nodata &  \nodata & 13.22$^{+0.04}_{-0.04}$ \\
BI 173 & STIS-M & GeII & 1237.059 & 3.033 & 8.1$\pm$8.6 & $<$12.18 \\
SK-66 19 & STIS-M & SiII & 1808.013 & 0.575 & 254.8$\pm$36.7 & $>$15.91 \\
SK-66 19 & STIS-M & CrII & 2056.254 & 2.326 & 81.7$\pm$30.1 & 13.45$^{+0.12}_{-0.17}$ \\
SK-66 19 & STIS-M & CrII & 2066.161 & 2.024 & 94.7$\pm$24.7 & 13.78$^{+0.10}_{-0.12}$ \\
SK-66 19 & STIS-M & FeII & 2260.780 & 0.742 & 161.0$\pm$14.9 & 15.32$^{+0.05}_{-0.05}$ \\
SK-66 19 & STIS-M & FeII & 2249.877 & 0.612 & 127.0$\pm$13.1 & 15.32$^{+0.05}_{-0.06}$ \\
SK-66 19 & STIS-M & NiII & 1741.549 & 1.871 & 37.5$\pm$63.4 & $<$14.00 \\
SK-66 19 & STIS-M & ZnII & 2026.136 & 3.106 & 250.0$\pm$17.2 & 13.36$^{+0.09}_{-0.11}$ \\
SK-66 19 & STIS-M & ZnII & 2062.664 & 2.804 & 270.6$\pm$70.8 & 13.61$^{+0.10}_{-0.13}$ \\
SK-66 19 & STIS-M & ZnII &  \nodata  &  \nodata &  \nodata  & 13.72$^{+0.10}_{-0.13}$ \\
SK-68 73 & STIS-H & OI & 1355.598 & -2.805 & 16.9$\pm$5.2 & 18.02$^{+0.11}_{-0.15}$ \\
SK-68 73 & STIS-H & MgII & 1240.395 & -0.355 & 118.2$\pm$32.9 & $>$16.85 \\
SK-68 73 & STIS-H & SII & 1250.578 & 0.809 & 182.7$\pm$6.9 & $>$15.92 \\
SK-68 73 & STIS-H & NiII & 1370.132 & 1.906 & 84.6$\pm$8.7 & 14.06$^{+0.06}_{-0.06}$ \\
SK-68 73 & STIS-H & NiII & 1317.217 & 1.876 & 77.8$\pm$6.6 & 14.07$^{+0.05}_{-0.05}$ \\
SK-68 73 & STIS-H & CuII & 1358.773 & 2.569 & 22.4$\pm$5.6 & 12.74$^{+0.13}_{-0.15}$ \\
SK-68 73 & STIS-M & SiII & 1808.013 & 0.575 & 234.5$\pm$12.9 & $>$15.92 \\
SK-68 73 & STIS-M & CrII & 2056.254 & 2.326 & 136.4$\pm$6.6 & 13.68$^{+0.03}_{-0.03}$ \\
SK-68 73 & STIS-M & CrII & 2066.161 & 2.024 & 69.4$\pm$8.1 & 13.63$^{+0.05}_{-0.06}$ \\
SK-68 73 & STIS-M & FeII & 2260.780 & 0.742 & 163.4$\pm$4.8 & 15.32$^{+0.03}_{-0.03}$ \\
SK-68 73 & STIS-M & FeII & 2249.877 & 0.612 & 132.8$\pm$7.3 & 15.32$^{+0.04}_{-0.04}$ \\
SK-68 73 & STIS-M & NiII & 1741.549 & 1.871 & 95.9$\pm$9.5 & 14.00$^{+0.06}_{-0.06}$ \\
SK-68 73 & STIS-M & NiII & 1709.600 & 1.743 & 45.8$\pm$19.4 & 13.83$^{+0.15}_{-0.20}$ \\
SK-68 73 & STIS-M & NiII & 1751.910 & 1.686 & 58.6$\pm$25.5 & 13.96$^{+0.16}_{-0.23}$ \\
SK-68 73 & STIS-M & ZnII & 2026.136 & 3.106 & 199.0$\pm$3.9 & 13.26$^{+0.04}_{-0.04}$ \\
SK-68 73 & STIS-M & ZnII & 2062.664 & 2.804 & 264.2$\pm$8.4 & 13.47$^{+0.03}_{-0.03}$ \\
SK-68 73 & STIS-M & ZnII & \nodata & \nodata  & \nodata  & 13.71$^{+0.20}_{-0.05}$ \\

\enddata

\tablecomments{The entirety of this table is available online in machine-readable format}
\end{deluxetable*}

\subsection{The problematic case of the single Si II $\lambda$1808 transition}\label{section_siii}

\begin{figure*}
\centering
\includegraphics[width=\textwidth]{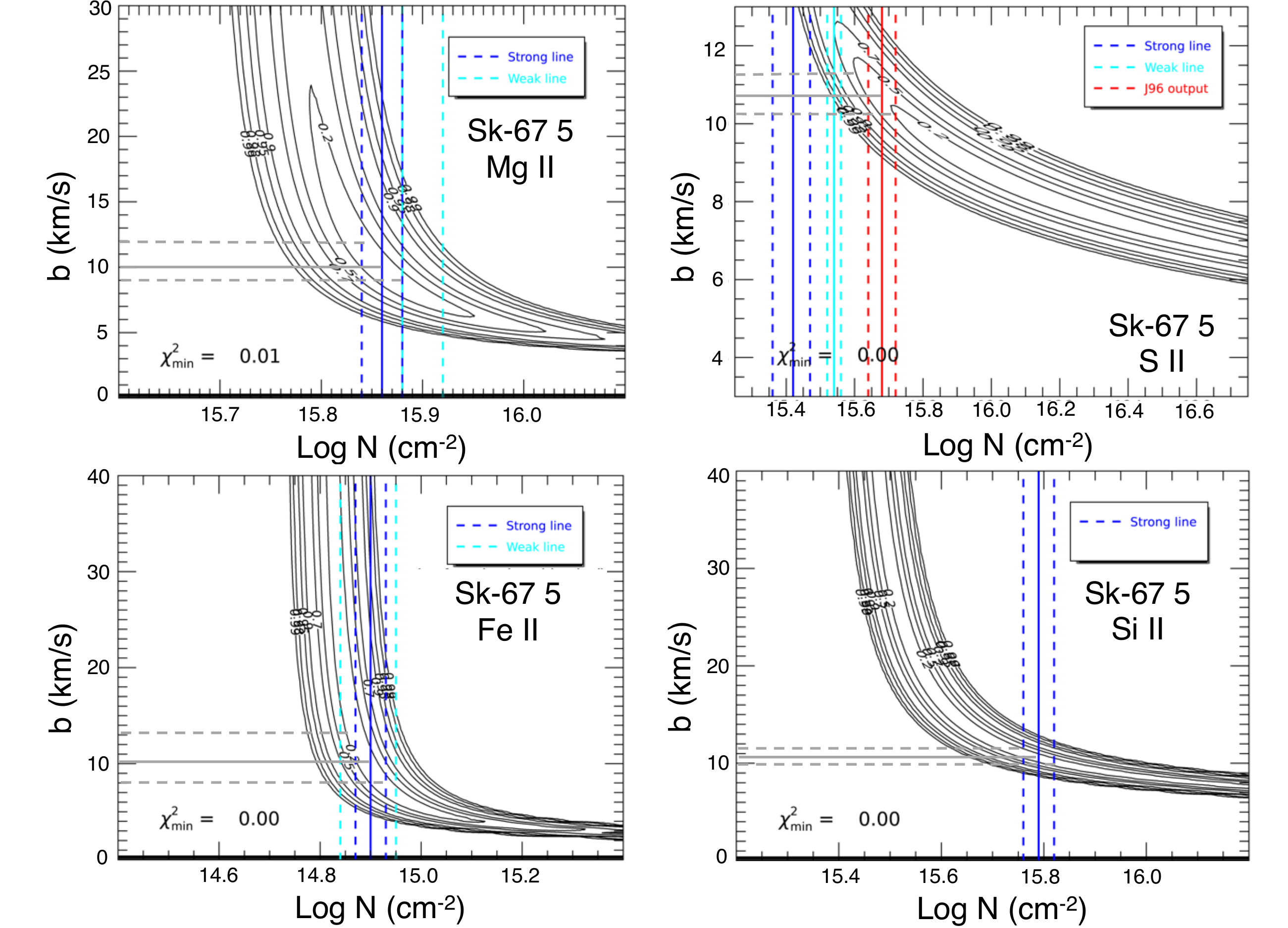}
\caption{Outcome of the AOD and COG approaches to determine column densities of \mgiis (top left), \siis (top right), \feiis (bottom left) and \siiis (bottom right) toward Sk-67 5. Each panel shows the probability contours based on a $\chi^2$ distribution with two degrees of freedom for various combinations of $\log N$ (x-axis) and $b$ (y-axis) for a standard COG, given the measured values of equivalent widths and their uncertainties for each of the two transitions used (e.g., $\lambda \lambda$ 2249, 2260 \mAA for \feii, $\lambda \lambda$ 1250, 1253 \mAA for \sii, $\lambda \lambda$ 1239, 1240 \mAA for \mgii). The column density determined from the AOD method is also overlaid on the contours in blue (strong line) and cyan (weak line). For \feii, \mgii, and \sii, two transitions are available to evaluate and correct for the effects of saturation, and the column densities determined from the AOD and COG are in agreement, and the intersection of the AOD-determined $N$ and the COG contours provides a well-constrained range of $b$ values, shown as gray lines. This value of $b$ can then be applied to the COG contours obtained from the single \siiis transition in order to evaluate and correct for the effects of saturation. For this Sk-67 5 sight-line, the \siiis COG contours and AOD-determined column density are in agreement for the $b$ value derived from the other element. Therefore, saturation is not an issue for this sight-line and no adjustment to the AOD outcome is necessary.}
\label{fig_aod_cog_sk-675}
\end{figure*}

\begin{figure*}
\centering
\includegraphics[width=\textwidth]{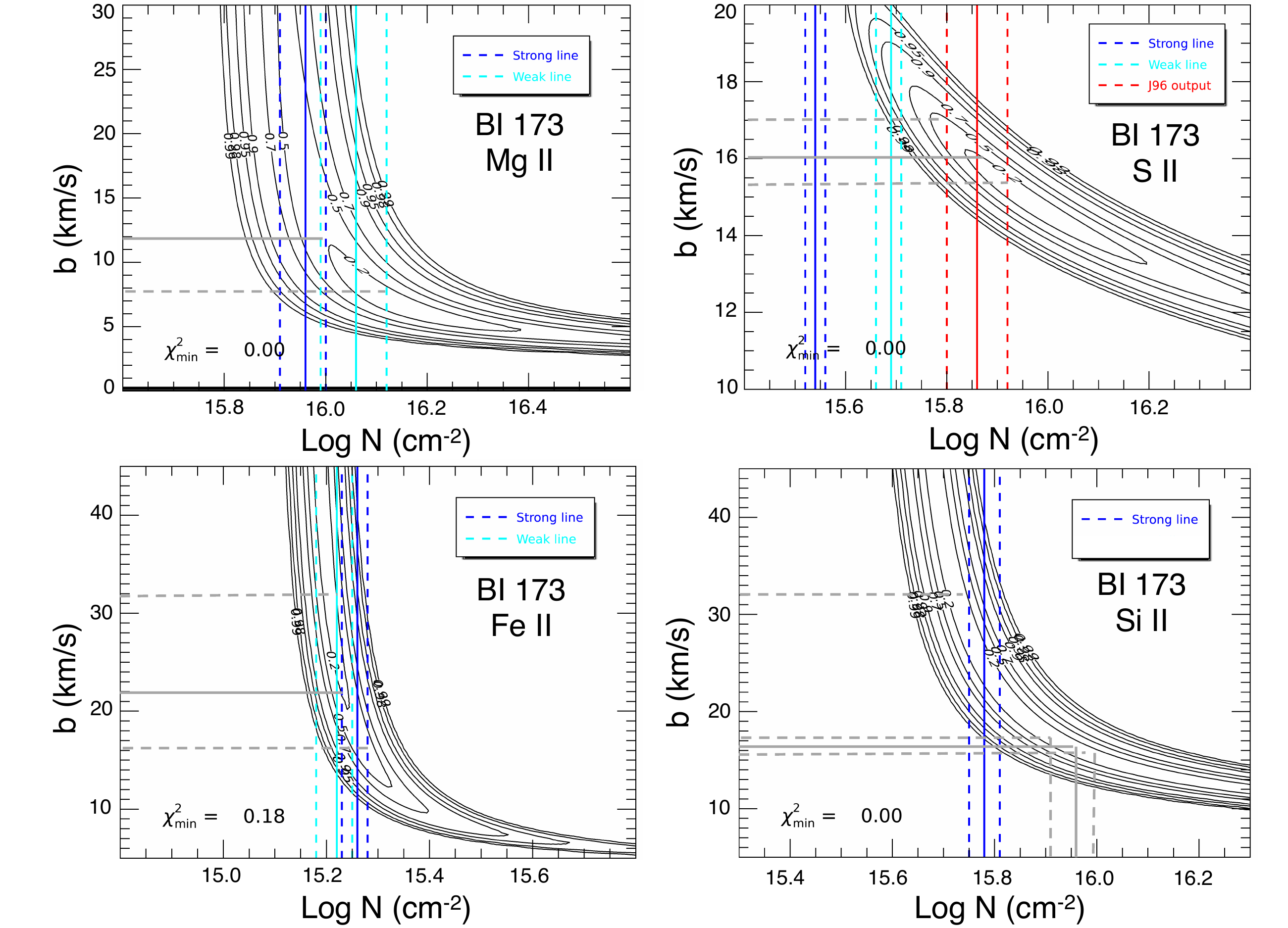}
\caption{Same as Figure \ref{fig_aod_cog_sk-675}, but for the sight-line toward BI 173. In this case, the $b$ value derived from the intersection of the AOD and COG contours for \mgii, \feii, and \siis suggests that the column density of \siiis derived from the AOD should be increased by $+$0.2 dex (solid gray line) and the error bars increased to $-$0.2 and $+$0.2 dex (dashed gray lines). }
\label{fig_aod_cog_bi173}
\end{figure*}

\indent For \siii, the only line that is not always badly saturated is the $\lambda$1808 \mAA transition. This means that we cannot evaluate and correct for the effects of mild saturation on the \siiis column density determination using different transitions of varying oscillator strengths, as is typically done with the AOD method (see previous Section). Nonetheless, the $\lambda$1808 \mAA line is seldom completely saturated, and \siiis being a key component of dust, it is important to measure its abundance. Therefore, we have devised a hybrid method to constrain the column density and abundance of \siii. The method uses both the AOD and curve-of-growth (COG) analysis performed on multiple non- or very mildly saturated transitions for other elements (\feii, \mgii, \sii) with similar equivalent widths to that of the single \siiis $\lambda$ 1808 \mAA transition in order to determine the Doppler parameter, $b$. The approach then applies this $b$ value to the \siiis COG to adjust the AOD-determined \siiis column density and associated errors, which can be impacted by saturation. \\
\indent The methodology for evaluating and correcting the effects of saturation on the \siiis column density determination from the AOD method is illustrated in Figures \ref{fig_aod_cog_sk-675} (for Sk-67 5) and \ref{fig_aod_cog_bi173} (for BI 173). These two figures show probability contours based on a $\chi^2$ distribution with two degrees of freedom for various combinations of $\log N$ (x-axis) and $b$ (y-axis) for a standard COG, given the measured values of equivalent widths and their uncertainties for each of the two transitions used (e.g., $\lambda \lambda$ 2249, 2260 \mAA for \feii, $\lambda \lambda$ 1250, 1253 \mAA for \sii, $\lambda \lambda$ 1239, 1240 \mAA for \mgii). The column density determined from the AOD method is also overlaid on the contours. The COG and AOD results are shown for \mgii, \sii, \feii, and \siii. For \mgii, \sii, \feii, which have two measured transitions, the contours are closed or half-closed and $\log N$ and $b$ are constrained, while for \siii, we have only one measurement but two parameters, and hence the contours are open ended and only indicate unacceptable combinations of $\log N$ and $b$. Our objective is to determine a plausible range of $b$ values for \siiis relying on the combined COG and AOD analyses for \mgii, \feii, \sii, so that $\log N$(\siii) can be constrained given the \siiis COG contours.\\
\indent For transitions that do not suffer from saturation (\mgii, \feii), the column densities determined from the AOD method applied to lines with different oscillator strengths are within errors, and the column density determination is consistent with the COG contours. For \sii, saturation effects can be evaluated and corrected for using the \citet{jenkins1996} method. In this case, the intersection of the corrected column density is generally in agreement with the COG contours. Thus, the intersection of the AOD-determined $N$ and the COG contours provides a well-constrained range of $b$ values, shown as gray lines in Figures \ref{fig_aod_cog_sk-675} and \ref{fig_aod_cog_bi173}. \\
\indent The range of allowed $b$ values depends on the strength of the transition, on the measurement errors on the AOD-derived column density, and on the tightness of the COG contours, which in turn depends on the error bars on the measured equivalent widths. Thus, different elements provide constraints on $b$ that are generally consistent, but can differ between different elements.  One important criterion in prioritizing the $b$-values derived from different elements is that the two features have transition probabilities that differ enough to give good indications of the COG behavior. The lines in the \mgiis doublet have strengths that differ by 1.77, and the depletion behavior of \mgiis is similar to that of \siiis in the MW. However, both of the lines in the doublet are considerably weaker than the \siiis feature, and effective $b$ values can change with increasing line strength as weaker non-Gaussian wings start to become more influential.  \siis has lines that differ in strength by a factor 2.01, and the equivalent widths are more similar to those of \siiis (with the benefit that saturation effects can be corrected for using the \citet{jenkins1996} method).  One possible disadvantage with \siis is that its depletion behavior may differ from that of \siii, which may drive a difference in the velocity behaviors.  The two \feiis lines that we investigated also have equivalent widths similar to that of \siii, but the strengths of the two lines differ by only a factor of 1.35, which weakens the COG test. Additionally, \feiis is substantially more depleted than \siii. In our analysis, we therefore determine the most plausible range of $b$ values by prioritizing the constraints on $b$ from \siis owing to its benefits outlined above, and the fact that is generally offers the tightest constraints (narrowest $b$ range). When \siis is not available, we prefer \feiis over \mgii.\\
\indent In the example of Sk-675 (Figure \ref{fig_aod_cog_sk-675}), \mgiis and \feiis indicate $b$ = 10$\pm$2 km s$^{-1}$, while the strongest constraints comes from \siis with $b$ = 10.75$\pm$0.5 km s$^{-1}$. For \siiis, the intersection of the innermost COG contour and the AOD column density measurement are consistent with $b$ $=$ 10.75 km s$^{-1}$, and the error bars on the AOD measurement are also consistent with the range of $b$ values determined from \sii, \mgii, and \feii.  From the location of this intersection below where the contours become curved, it is clear that there is some saturation of the line, but the AOD analysis seems to have handled it well.  \\
\indent In the second example of BI 173 (Figure \ref{fig_aod_cog_bi173}), the weighted average of the strong and weak \feiis lines gives a most plausible $b$ value of 22 km s$^{-1}$, with a possible range b $>$ 16 km s$^{-1}$. \mgiis yields $b$ $>$ 8 km s$^{-1}$. The weak \mgiis lines show smaller $b$ values than the stronger \feiis lines for the reasons outlined above. This can be understood in terms of a velocity profile that is narrow in the central portion but then has broad wings in the lower portions.  Such a behavior will cause a shift to higher $b$ values for stronger lines. The best constraints on $b$ again arise from \siis for this target, with a most plausible value of 16$\pm$1 km s$^{-1}$. In this case, $b$ = 16$\pm$1 km s$^{-1}$ in the \siiis COG corresponds to $N$ $=$ 15.98$\pm$0.05 cm$^{-2}$, while the column determined from the AOD is $N$ $=$ 15.78$\pm$0.03 cm$^{-2}$. In this case, saturation did therefore impact the \siiis column density determination. In Table 3, we report $N$(\siii) determined from the hybrid AOD/COG method, with upper and lower error bars of $\pm$0.2 dex to capture the possibility of the original measurement and account for possible further effects of saturation in this horizontal part of the COG contours where $N$ is insensitive to changes in $b$. \\
\indent We proceed with this analysis for all sight-lines and find that 14 out of 32 targets in the sample need adjustments to their AOD-derived \siiis column densities. This includes 5 targets for which $I_0/I_c > 0.05$ and therefore $N$(\siii) was derived from the AOD, but the examination of the COG contours revealed that only a lower limit could be estimated. The results of these corrections are listed in Table 4.

\begin{deluxetable}{ccc}
\centering
\tabletypesize{\scriptsize}
\tablecolumns{3}
\tablewidth{8cm}
\tablecaption{Column density adjustments for \siiis based on the hybrid AOD and COG approach}
\tablenum{4}
\tablehead{Target & $N$(\siii)$_{\mathrm{AOD}}$ & $N$(\siii)$_{\mathrm{AOD+COG}}$ \\}
 \startdata
& cm$^{-2}$ & cm$^{-2}$\\
\hline
&&\\
SK-67 2 & 15.63$^{+0.20}_{-0.06}$ & $>$15.63 \\
SK-66 19 & 15.91$^{+0.20}_{-0.16}$ & $>$15.91 \\
PGMW 3120 & 15.91$^{+0.20}_{-0.22}$ & 16.00$^{+0.30}_{-0.22}$\\
SK-66 35 & 15.75$^{+0.20}_{-0.05}$ & 15.80$^{+0.15}_{-0.10}$ \\
SK-65 22 & 15.58$^{+0.20}_{-0.05}$ & 15.65$^{+0.13}_{-0.12}$ \\
SK-68 26 & 15.73$^{+0.20}_{-0.07}$ & $>$15.73 \\
SK-70 79 & 15.75$^{+0.20}_{-0.06}$ & $>$15.75 \\
BI 173 & 15.78$^{+0.05}_{-0.05}$ & 15.98$^{+0.2}_{-0.2}$ \\
BI 184 & 15.80$^{+0.16}_{-0.06}$ & 15.85$^{+0.21}_{-0.11}$ \\
SK-67 191 & 15.61$^{+0.05}_{-0.05}$ & 15.68$^{+0.12}_{-0.12}$ \\
SK-67 211 & 15.74$^{+0.20}_{-0.04}$ & 15.87$^{+0.20}_{-0.17}$ \\
SK-66 172 & 15.71$^{+0.20}_{-0.06}$ & $>$15.71 \\
SK-68 140 & 16.0$7^{+0.27}_{-0.10}$ & 16.27$^{+0.27}_{-0.21}$ \\
SK-70 115 & 15.96$^{+0.04}_{-0.04}$ & 16.06$^{+0.12}_{-0.12}$ \\
\enddata
\end{deluxetable}

\subsection{Column densities with profile fitting in the COS spectra}\label{cos_profile_fitting}

\begin{figure*}
\centering
\includegraphics[width=\textwidth]{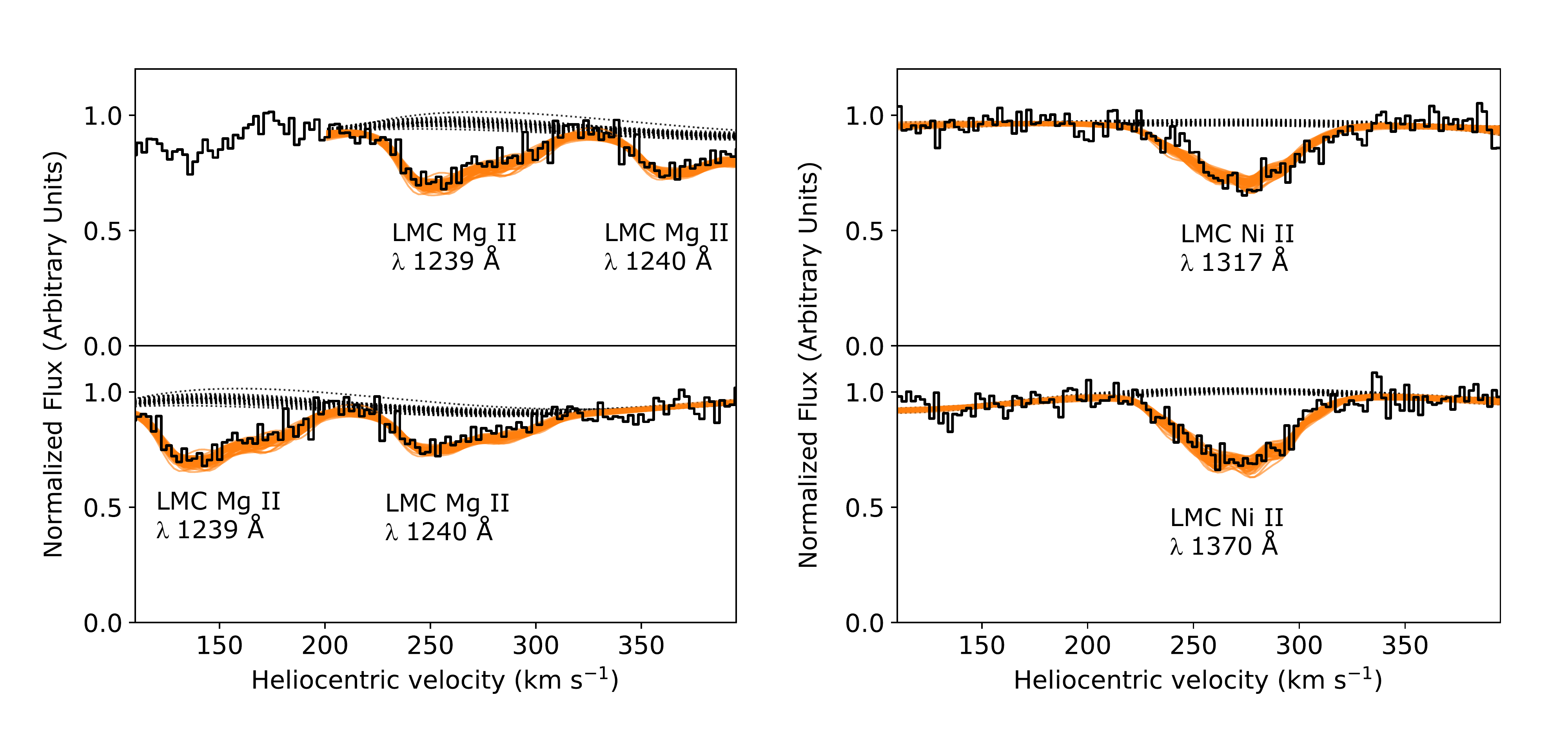}
\includegraphics[width=14cm]{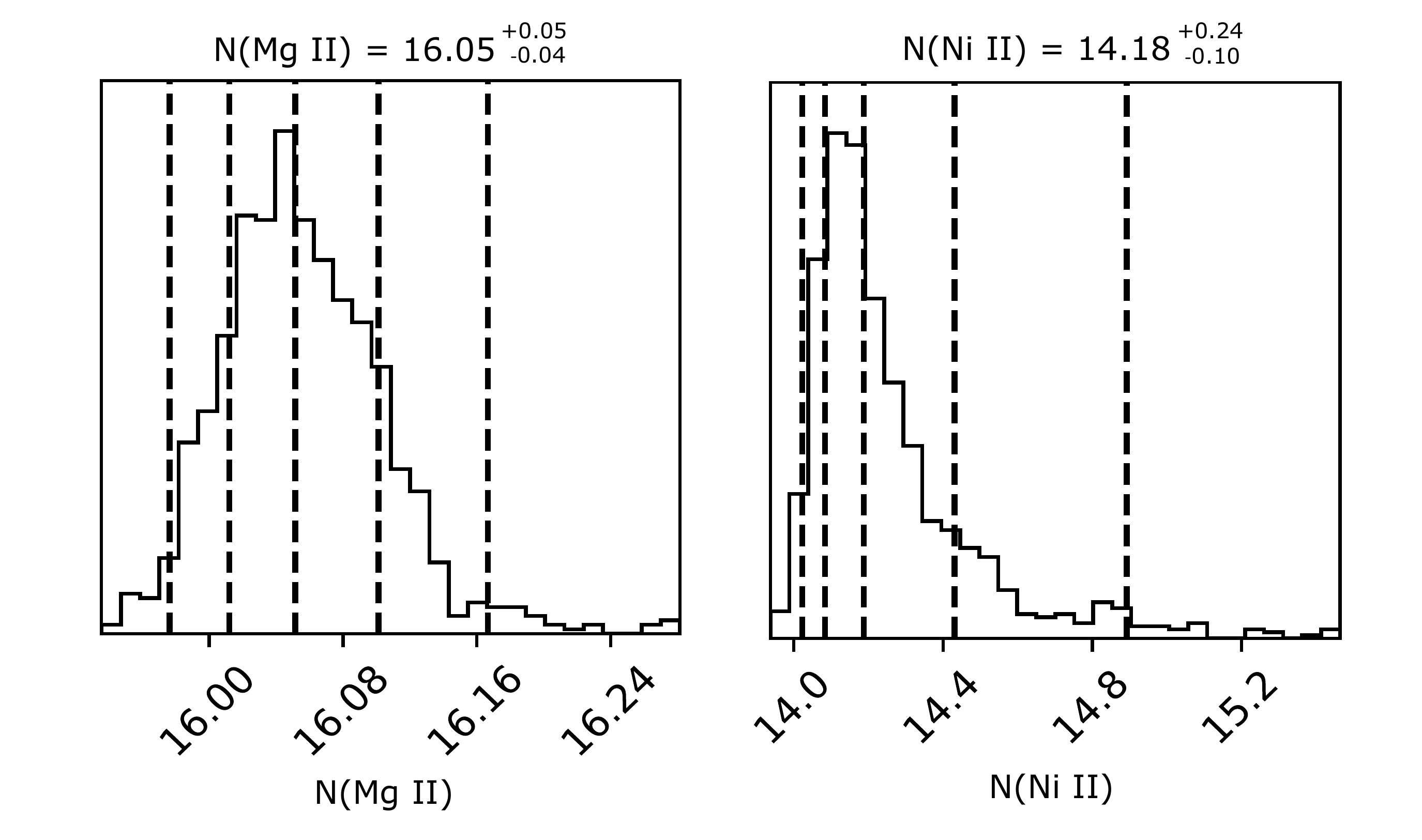}
\caption{Example of profile fitting of the \mgiis ($\lambda \lambda$1239, 1240\AA, left) and \niiis ($\lambda \lambda$1317, 1370 \AA, right) lines in the COS spectrum of BI 184. The top panels show the spectrum near the lines of interest (scaled by a constant), and different realizations of the continuum (gray) and absorption (orange) models sampling the posterior probability density functions shown in the bottom panels.  }
\label{profile_fitting_example}
\end{figure*}

\indent The lower resolution of COS dictated the need for profile fitting, in order to overcome the effects of additional smoothing of the absorption profiles. We performed such fitting to the COS G130M spectra and derive column densities of \mgiis ($\lambda$1239.9253 \mAA,$\lambda$1240.3947 \mAA) and \niiis ($\lambda$1317.217 \mAA, $\lambda$1370.132 \mAA). The approach is similar to that of \citet{tchernyshyov2015}: it simultaneously fits for the continuum and line profiles, which allows us to propagate the total uncertainty (including the uncertainty on the continuum fitting). The approach uses a Gaussian Process \citep{rasmussen2006} to capture the large and short wavelength scale fluctuations in the continuum level, respectively, and uses Voigt profile fitting \citep[e.g.][]{carswell2014} to model the absorption features. \\
\indent For each element (including all available lines), the absorption model consists of one Voigt profile component in every 10 km s$^{-1}$ interval over the range in which absorption can be detected. Each component has a column density, central velocity (within its 10 km s$^{-1}$ interval), and width. One might anticipate that reducing the component spacing could lead to larger column densities since smaller $b$ values would be associated with more tightly spaced components. In Appendix A, we show that the column density measurements obtained from profile fitting do not change beyond their uncertainties when a shorter component spacing of 5 km s$^{-1}$ is used, although a handful of sight-lines do have higher (by $+$0.5-0.7 dex) column densities of \mgii, with correspondingly larger uncertainties, with the tighter component spacing. \\
\indent The modeled absorption features are multiplied by the continuum and convolved with the COS instrumental line spread functions (LSF) in order to forward-model the observations, and allow for the correct propagation of uncertainties. We marginalize over the individual component parameters using a custom implementation of the simplified Manifold Metropolis-adjusted Langevin algorithm \citep{girolami2011} and sum the column densities of all of the components. The inferences of the \mgii, \niii, and \siis column densities are performed independently. For each element, different spectral lines observed in different exposures (e.g., different FP-POS dithers) are fit simultaneously.\\
\indent We use samples drawn from the posterior probabilities using MCMC to build posterior probability distributions for each species column density and the model flux at each wavelength. The reported column densities correspond to the 50$^{th}$ percentile of the posterior distribution. The resulting uncertainties, taken as the difference between the 16$^{th}$ and 50$^{th}$ percentile (lower uncertainties) and between the 50$^{th}$ and 84$^{th}$ percentile (upper uncertainties) of the posterior distribution, include uncertainties on the measurement (noise), continuum estimation, and the possibility of observationally similar but physically different velocity component structures. Figure \ref{profile_fitting_example} provides an example of profile fitting of the \mgiis ($\lambda \lambda$1239, 1240 \AA) and \niiis ($\lambda \lambda$ 1317, 1370 \AA) lines in the COS G130M/1291 spectrum of BI 184. \\

\begin{deluxetable*}{ccccccc}
\centering
\tabletypesize{\scriptsize}
\tablecolumns{7}
\tablewidth{\textwidth}
\tablecaption{Column densities and depletions}
\tablenum{5}
\tablehead{Target & $\log$ N(H) & Element & Grating & $\log$ N(X) & 12 + $\log(X/H)_{\mathrm{LMC,gas}}$ & Depletion $\delta(X)$\tablenotemark{a} \\}
 \startdata
 & cm$^{-2}$ & && cm$^{-2}$ & & \\
\hline
&&&&& &\\

BI 173 & 21.25$\pm$0.05 & CrII & STIS-M & 13.59$\pm$0.03 & 4.34$\pm$0.06 & -1.03$\pm$0.06 \\
BI 173 & 21.25$\pm$0.05 & CuII & STIS-M & 12.63$\pm$0.22 & 3.38$\pm$0.23 & -0.41$\pm$0.23 \\
BI 173 & 21.25$\pm$0.05 & FeII & STIS-M & 15.24$\pm$0.03 & 5.99$\pm$0.06 & -1.33$\pm$0.06 \\
BI 173 & 21.25$\pm$0.05 & GeII & STIS-M & $<$12.18 & $<$2.93 & $<$-0.42 \\
BI 173 & 21.25$\pm$0.05 & MgII & STIS-M & 16.01$\pm$0.10 & 6.76$\pm$0.11 & -0.50$\pm$0.11 \\
BI 173 & 21.25$\pm$0.05 & NiII & STIS-M & 13.96$\pm$0.03 & 4.71$\pm$0.06 & -1.21$\pm$0.06 \\
BI 173 & 21.25$\pm$0.05 & OI & STIS-M & $<$18.10 & $<$8.85 & $<$0.35 \\
BI 173 & 21.25$\pm$0.05 & PII & STIS-M & $>$13.60 & $>$4.35 & $>$-0.75 \\
BI 173 & 21.25$\pm$0.05 & SII & STIS-M & 15.86$\pm$0.06 & 6.61$\pm$0.08 & -0.52$\pm$0.08 \\
BI 173 & 21.25$\pm$0.05 & SiII & STIS-M & 15.98$\pm$0.20 & 6.73$\pm$0.21 & -0.62$\pm$0.21 \\
BI 173 & 21.25$\pm$0.05 & ZnII & STIS-M & 13.22$\pm$0.04 & 3.97$\pm$0.06 & -0.34$\pm$0.06 \\
SK-66 19 & 21.87$\pm$0.07 & CrII & STIS-M & 13.66$\pm$0.09 & 3.79$\pm$0.11 & -1.58$\pm$0.11 \\
SK-66 19 & 21.87$\pm$0.07 & FeII & STIS-M & 15.32$\pm$0.04 & 5.45$\pm$0.08 & -1.87$\pm$0.08 \\
SK-66 19 & 21.87$\pm$0.07 & MgII & COS-M & 16.49$\pm$0.16 & 6.62$\pm$0.17 & -0.64$\pm$0.17 \\
SK-66 19 & 21.87$\pm$0.07 & NiII & COS-M & 14.00$\pm$0.06 & 4.13$\pm$0.10 & -1.79$\pm$0.10 \\
SK-66 19 & 21.87$\pm$0.07 & SiII & STIS-M & $>$15.91 & $>$6.04 & $>$-1.31 \\
SK-66 19 & 21.87$\pm$0.07 & ZnII & STIS-M & 13.72$\pm$0.12 & 3.85$\pm$0.13 & -0.46$\pm$0.13 \\
SK-68 73 & 21.68$\pm$0.02 & CrII & STIS-M & 13.67$\pm$0.03 & 3.99$\pm$0.03 & -1.38$\pm$0.03 \\
SK-68 73 & 21.68$\pm$0.02 & CuII & STIS-H & 12.74$\pm$0.14 & 3.06$\pm$0.14 & -0.73$\pm$0.14 \\
SK-68 73 & 21.68$\pm$0.02 & FeII & STIS-M & 15.32$\pm$0.02 & 5.64$\pm$0.03 & -1.68$\pm$0.03 \\
SK-68 73 & 21.68$\pm$0.02 & MgII & COS-M & 16.49$\pm$0.14 & 6.81$\pm$0.14 & -0.45$\pm$0.14 \\
SK-68 73 & 21.68$\pm$0.02 & NiII & STIS-H & 14.07$\pm$0.04 & 4.38$\pm$0.04 & -1.54$\pm$0.04 \\
SK-68 73 & 21.68$\pm$0.02 & OI & STIS-H & 18.02$\pm$0.13 & 8.34$\pm$0.13 & -0.16$\pm$0.13 \\
SK-68 73 & 21.68$\pm$0.02 & SII & STIS-H & $>$15.92 & $>$6.24 & $>$-0.89 \\
SK-68 73 & 21.68$\pm$0.02 & SiII & STIS-M & $>$15.92 & $>$6.24 & $>$-1.11 \\
SK-68 73 & 21.68$\pm$0.02 & TiII & \nodata. & 12.52$\pm$0.03 & 2.84$\pm$0.04 & -1.92$\pm$0.04 \\
SK-68 73 & 21.68$\pm$0.02 & ZnII & STIS-M & 13.71$\pm$0.12 & 4.03$\pm$0.13 & -0.28$\pm$0.13 \\

\enddata
\tablecomments{This entirety of this table is available online in machine-readable format}
\tablenotetext{a}{The statistical uncertainty is listed here. Systematic errors on the depletions due to uncertainties on the photospheric abundances are not included, because they do not affect the relative trends examined here (e.g., environmental parameters). An estimate of these systematic errors can be found in Table 4 of \citet{tchernyshyov2015}. }
\end{deluxetable*}

\subsection{Gas--phase abundances and depletions}

\indent Gas-phase abundances are derived by taking the ratio of the measured column densities to the total hydrogen column density, N(H) = N(\hi) + 2N(H$_2$), where N(H) is listed in Table 1. The \his column densities are determined from the METAL spectra (see Paper I), while the H$_2$ column densities are from \citet{welty2012}. The depletion (logarithm of the fraction of element X in the gas-phase) is then calculated assuming that the total (gas and dust) neutral ISM abundance of X is equal to the photospheric abundance of X in young stars. Because young stars recently formed out of the ISM and have not yet undergone self-enrichment, they are good proxies for the present-day ISM composition. A number of studies have spectroscopically investigated the composition of luminous young stars in the LMC. However, no single study includes all the elements for which we wish to compute interstellar depletions. \citet{tchernyshyov2015} carefully pooled measurements of young star abundances across studies using a multilevel linear model to account for differences between studies and missing uncertainty information. In this work, we assume the LMC stellar abundances compiled in \citet{tchernyshyov2015} for the total ISM abundances. These reference abundances are listed in Table 2 for each element. The measurement errors on the depletions are obtained by summing the errors on the logarithms of the column densities of X and H in quadrature. Systematic errors on the depletions due to uncertainties on the photospheric abundances are not included in Table 5, because they do not affect the relative trends examined here (e.g., environmental parameters). An estimate of these systematic errors can be found in Table 4 of \citet{tchernyshyov2015}.  \\

\section{Hydrogen densities and radiation fields from the C I fine structure lines}\label{ci_section}

\indent The C I lines (\cis at $\lambda$1276.483 \mAA, \ci$*$ at $\lambda$1276.749 \mAA, and \ci$**$ at $\lambda$1277.719 \AA) observed in the METAL spectra can be used to obtain more insight on the local environment properties of our sight-lines, in the same pencil-beam volume probed by the abundance and depletion measurements. As described in \citet{jenkins2011} and references therein, the ratios N(\ci$*$)/N(\ci)$_{\mathrm{tot}}$ and N(\ci$**$)/N(\ci)$_{\mathrm{tot}}$ provide an estimate of the number density of hydrogen atoms, which, combined with estimate of the N(\cii)/N(\ci)$_{\mathrm{tot}}$ ratio and an iterative approach, yields an estimate of the intensity of the UV radiation field $I$ normalized to the to a value $I_0$ specified by \citet{mathis1983} for the average intensity of ultraviolet starlight in the solar neighborhood. We applied a line profile fitting method to the \cis lines in the METAL spectra and followed this approach to estimate $n(H)$ and $I/I_0$ in 26 out of 32 METAL sight-lines. 

\begin{deluxetable*}{ccccccccccc}
\centering
\tabletypesize{\scriptsize}
\tablecolumns{11}
\tablewidth{\textwidth}
\tablecaption{Measurements of the \ci, \ci*, and \ci** column densities and ratios, derived radiation fields, volume densities, and electron densities}
\tablenum{6}
\tablehead{Target & Instr. & $T_{01}(H_2)$\tablenotemark{a} & $\log_{10}$ N(\ci)$_{\mathrm{tot}}$ & $\log_{10}$ N(\cii) & $f_1$ & $f_2$ & $I/I_0$ & $n$(H) & $g_{low}$ & $n_e$ \\}
 \startdata
  & & K & cm$^{-2}$ & cm$^{-2}$ & & & & cm$^{-3}$ & &cm$^{-3}$ \\
\hline
&&&&&& &\\

SK-67 2 & STIS  & 46.0 & 14.73$\pm$0.22 & 17.21$\pm$0.23 & 0.19$\pm$0.07 & 0.09$\pm$0.04 & 0.6$\pm$0.6 & 55$\pm$37 & 0.86$\pm$0.08 & 0.008$\pm$0.004 \\
SK-67 5 & STIS & 57.0 & 13.86$\pm$0.01 & 16.83$\pm$0.20 & 0.29$\pm$0.01 & 0.10$\pm$0.01 & 2.3$\pm$1.3 & 101$\pm$6 & 0.90$\pm$0.01 & 0.014$\pm$0.001 \\
SK-69 279 & COS & 64.0 & 14.23$\pm$0.04 & 17.41$\pm$0.20 & 0.36$\pm$0.03 & 0.19$\pm$0.03 & 5.1$\pm$3.5 & 143$\pm$2 & 0.72$\pm$0.09 & 0.020$\pm$0.004 \\
SK-67 14 & \nodata & 270.0 & \nodata & \nodata & \nodata & \nodata & \nodata & \nodata & \nodata & \nodata \\
SK-66 19 & COS & 77.0 & 14.37$\pm$0.08 & 17.62$\pm$0.21 & 0.36$\pm$0.06 & 0.31$\pm$0.07 & 6.5$\pm$7.2 & 97$\pm$74 & 0.43$\pm$0.16 & 0.018$\pm$0.009 \\
PGMW 3120 & STIS & 71.0 & 13.69$\pm$0.03 & 17.26$\pm$0.20 & 0.34$\pm$0.03 & 0.14$\pm$0.04 & 7.5$\pm$3.9 & 108$\pm$31 & 0.84$\pm$0.09 & 0.019$\pm$0.003 \\
PGMW 3223 & STIS & 62.0 & 13.78$\pm$0.03 & 17.19$\pm$0.20 & 0.29$\pm$0.02 & 0.23$\pm$0.02 & 5.2$\pm$2.9 & 60$\pm$19 & 0.59$\pm$0.06 & 0.013$\pm$0.002 \\
SK-66 35 & STIS & 71.0 & 13.64$\pm$0.04 & 16.67$\pm$0.20 & 0.29$\pm$0.03 & 0.21$\pm$0.04 & 2.2$\pm$1.4 & 60$\pm$23 & 0.63$\pm$0.09 & 0.011$\pm$0.003 \\
SK-65 22 & STIS & 117.0 & 13.18$\pm$0.06 & 16.48$\pm$0.20 & 0.33$\pm$0.06 & 0.33$\pm$0.06 & 4.4$\pm$3.7 & 45$\pm$48 & 0.37$\pm$0.20 & 0.014$\pm$0.007 \\
SK-68 26 & COS & 59.0 & 14.12$\pm$0.05 & 17.40$\pm$0.20 & 0.44$\pm$0.04 & 0.12$\pm$0.03 & 10.0$\pm$4.0 & 327$\pm$95 & 1.00$\pm$0.01 & 0.036$\pm$0.008 \\
SK-70 79 & STIS & 86.0 & 14.38$\pm$0.02 & 17.09$\pm$0.20 & 0.39$\pm$0.02 & 0.24$\pm$0.01 & 2.6$\pm$1.5 & 171$\pm$34 & 0.67$\pm$0.04 & 0.020$\pm$0.003 \\
SK-68 52 & STIS & 60.0 & 14.00$\pm$0.06 & 17.11$\pm$0.20 & 0.43$\pm$0.03 & 0.24$\pm$0.03 & 10.7$\pm$5.4 & 395$\pm$180 & 0.75$\pm$0.13 & 0.042$\pm$0.016 \\
SK-69 104 & \nodata & 80.0 & \nodata & \nodata & \nodata & \nodata & \nodata & \nodata & \nodata & \nodata\\
SK-68 73 & STIS & 57.0 & 14.49$\pm$0.02 & 17.45$\pm$0.19 & 0.40$\pm$0.01 & 0.23$\pm$0.01 & 6.7$\pm$4.7 & 268$\pm$44 & 0.70$\pm$0.03 & 0.030$\pm$0.004 \\
SK-67 101 & \nodata & 80.0 & \nodata & \nodata & \nodata & \nodata & \nodata & \nodata & \nodata & \nodata \\
SK-67 105 & STIS & 62.0 & 14.09$\pm$0.11 & 17.02$\pm$0.20 & 0.47$\pm$0.05 & 0.18$\pm$0.03 & 9.0$\pm$4.9 & 562$\pm$129 & 1.00$\pm$0.02 & 0.055$\pm$0.011 \\
BI 173 & COS & 117.0 & 13.65$\pm$0.02 & 17.05$\pm$0.20 & 0.35$\pm$0.02 & 0.28$\pm$0.03 & 5.4$\pm$2.7 & 71$\pm$25 & 0.50$\pm$0.07 & 0.016$\pm$0.003 \\
BI 184 & COS & 55.0 & 14.37$\pm$0.72 & 16.96$\pm$0.20 & 0.36$\pm$0.16 & 0.14$\pm$0.10 & 2.1$\pm$2.1 & 185$\pm$180 & 0.86$\pm$0.15 & 0.020$\pm$0.017 \\
SK-71 45 & STIS & 98.0 & 13.60$\pm$0.03 & 16.93$\pm$0.20 & 0.31$\pm$0.04 & 0.22$\pm$0.03 & 3.7$\pm$2.5 & 58$\pm$24 & 0.62$\pm$0.09 & 0.014$\pm$0.003 \\
SK-69 175 & \nodata & 80.0 & \nodata & \nodata & \nodata & \nodata & \nodata & \nodata & \nodata & \nodata \\
SK-67 191 & \nodata & 80.0 & \nodata & \nodata & \nodata & \nodata & \nodata & \nodata & \nodata & \nodata \\
SK-67 211 & \nodata & 80.0 & \nodata & \nodata & \nodata & \nodata & \nodata & \nodata & \nodata & \nodata\\
BI 237 & COS & 61.0 & 14.04$\pm$0.77 & 17.42$\pm$0.20 & 0.30$\pm$0.18 & 0.06$\pm$0.06 & 4.6$\pm$4.0 & 103$\pm$109 & 0.99$\pm$0.08 & 0.016$\pm$0.013 \\
SK-68 129 & COS & 63.0 & 15.51$\pm$0.64 & 17.40$\pm$0.24 & 0.02$\pm$0.14 & 0.00$\pm$0.03 & 0.0$\pm$0.0 & 4$\pm$1 & 0.99$\pm$0.04 & 0.002$\pm$0.001 \\
SK-66172 & STIS & 110.0 & 13.97$\pm$0.02 & 17.04$\pm$0.20 & 0.47$\pm$0.02 & 0.23$\pm$0.02 & 6.8$\pm$2.2 & 377$\pm$72 & 1.00$\pm$0.05 & 0.040$\pm$0.006 \\
BI 253 & COS & 59.0 & 15.52$\pm$0.79 & 17.46$\pm$0.20 & 0.26$\pm$0.23 & 0.03$\pm$0.10 & 0.3$\pm$0.3 & 94$\pm$138 & 1.00$\pm$0.02 & 0.011$\pm$0.012 \\
SK-68 135 & STIS & 91.0 & 14.44$\pm$0.02 & 17.27$\pm$0.19 & 0.34$\pm$0.02 & 0.13$\pm$0.01 & 1.7$\pm$0.8 & 103$\pm$14 & 0.87$\pm$0.03 & 0.014$\pm$0.002 \\
SK-69 246 & STIS & 74.0 & 14.21$\pm$0.02 & 17.28$\pm$0.19 & 0.36$\pm$0.02 & 0.12$\pm$0.01 & 3.3$\pm$1.7 & 141$\pm$18 & 0.92$\pm$0.04 & 0.019$\pm$0.002 \\
SK-68 140 & COS & 64.0 & 15.56$\pm$0.09 & 17.29$\pm$0.22 & 0.11$\pm$0.12 & 0.01$\pm$0.00 & 0.1$\pm$0.1 & 27$\pm$29 & 1.00$\pm$0.00 & 0.004$\pm$0.003 \\
SK-71 50 & STIS & 52.0 & 13.71$\pm$0.92 & 17.06$\pm$0.20 & 0.22$\pm$0.13 & 0.04$\pm$0.04 & 3.4$\pm$0.8 & 70$\pm$40 & 1.00$\pm$0.07 & 0.013$\pm$0.006 \\
SK-68 155 & COS & 68.0 & 14.09$\pm$0.03 & 17.27$\pm$0.21 & 0.34$\pm$0.02 & 0.24$\pm$0.02 & 4.5$\pm$3.4 & 100$\pm$26 & 0.60$\pm$0.07 & 0.017$\pm$0.003 \\
SK-70 115 & STIS & 53.0 & 13.86$\pm$0.01 & 17.00$\pm$0.21 & 0.42$\pm$0.01 & 0.14$\pm$0.01 & 8.1$\pm$3.3 & 310$\pm$25 & 0.95$\pm$0.02 & 0.034$\pm$0.002 \\

\enddata
\tablenotetext{a}{Rotational temperatures $T_{01}$, computed as in Equation 5 of \citet{tumlinson2002}, are taken from \citet{welty2012} and references therein. For Sk-66 172, the iterative computation of density and radiation field from \cis and \ciis line ratios diverges with the $H_2$ $T_{01}$ rotational temperature given in \citet{welty2012} (41$^{+110}_{-140}$ K) input to the model as the kinetic temperature. The closest temperatures for which the models converge is 110K, which are well within the error bars of the temperature estimation.}
\end{deluxetable*}

\begin{figure*}
\centering
\includegraphics[width=\textwidth]{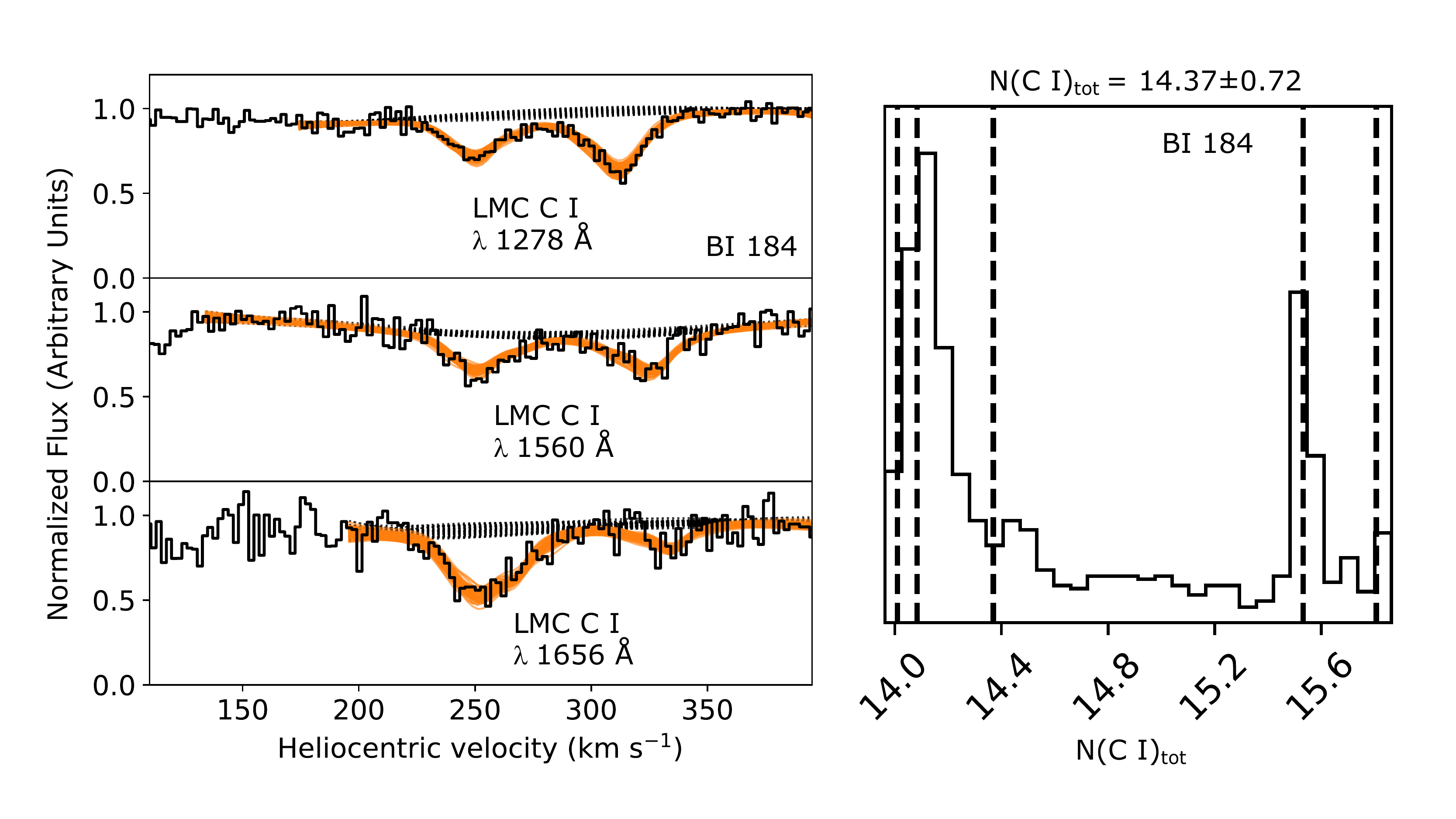}

\caption{Example of profile fitting of the \cis band system in the COS spectrum of BI 184. The left panel shows the spectrum near the lines of interest (scaled by a constant), and different realizations of the continuum (gray) and absorption (orange) models sampling the posterior probability density functions shown in the right panel.}
\label{plot_ci_profile_fit}
\end{figure*}

\begin{figure}
\centering
\includegraphics[width=8cm]{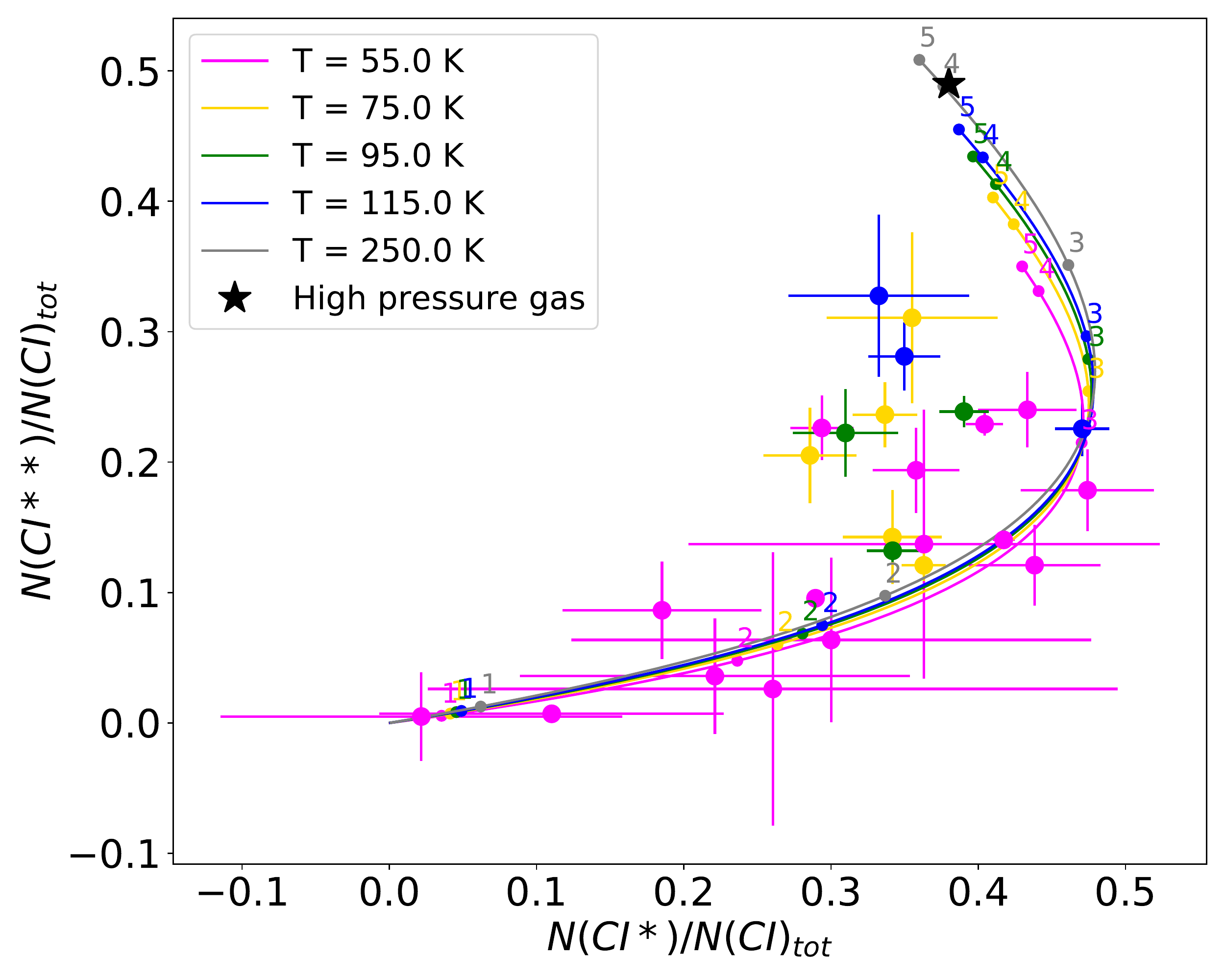}
\caption{Distribution of line ratios $f_2$ $=$ N(\ci$**$)/N(\ci)$_{\mathrm{tot}}$ versus $f_1$ $=$ N(\ci$*$)/N(\ci)$_{\mathrm{tot}}$ for the sight-lines in the METAL program. The model tracks for the low pressure gas are computed for several temperatures indicated in the legend, and the logarithm of the densities along the tracks are indicated on the figure.  }
\label{plot_ci_model_f1_f2}
\end{figure}

\begin{figure}
\centering
\includegraphics[width=8cm]{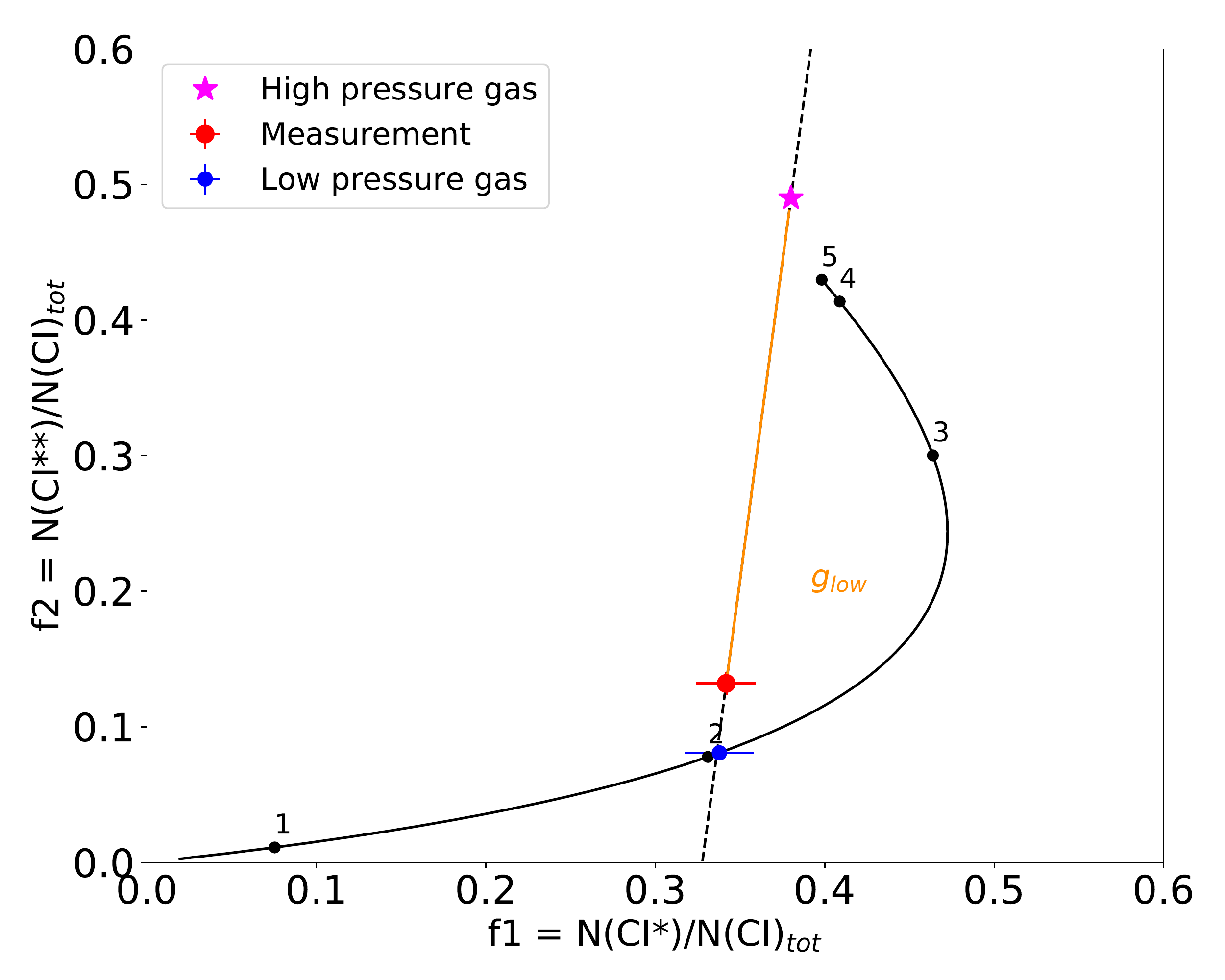}
\caption{Example of modeling and geometrical derivation of the low ($g_{low}$) and high ($1-g_{low}$) pressure gas components fractions from the \cis fine structure line ratios, $f_1$ and $f_2$, for the Sk-68 135 sight-line. The black tracks represent a model of the variations of the ($f_1$, $f_2$) distribution for a fixed temperature (equal to the $H_2$ rotational temperature for this target, or 91 K), as the volume density changes ($\log$ n(H) values are indicated next to each black point). The red point with error bars is the ($f_1$, $f_2$) measurement, which corresponds to the geometric average of the high (pink star at $f_1$, $f_2$ $=$ 0.38, 0.49) and the low pressure component, located at the intersection of the line joining the high pressure component and the measurement, and the model track, thus yielding the density of the low pressure component. }
\label{plot_example_ci}
\end{figure}

\indent We first derive column densities of N(\ci), N(\ci$*$), and N(\ci$**$) (summing up to N(\ci)$_{\mathrm{tot}}$) using the same profile fitting method as described in Section \ref{cos_profile_fitting} (see example in Figure \ref{plot_ci_profile_fit}). We also assume a 10 km s$^{-1}$ component spacing for \ci. As for \mgiis and \niii, we investigate possible systematic differences in the N(\ci), N(\ci$*$), and N(\ci$**$) outcomes for a tighter component spacing of 5 km s$^{1}$ in Appendix A and find results within uncertainties. The resulting column densities are listed in Table 6 and the ratios $f_1$ $=$ N(\ci$*$)/N(\ci)$_{\mathrm{tot}}$ and $f_2$ $=$ N(\ci$**$)/N(\ci)$_{\mathrm{tot}}$ are shown in Figure \ref{plot_ci_model_f1_f2}. \\
\indent As explained in \citet{jenkins2001, jenkins2011} and references therein, for a given kinetic temperature and radiation field intensity, the location of the ($f_1$, $f_2$) measurements follows a track that is dependent on the volume density of hydrogen, $n(H)$ (or equivalently on pressure for a given temperature), and the fraction of low-pressure gas, $g_{\mathrm{low}}$. Geometrically, the composite measurement of ($f_1$, $f_2$) toward a sight-line, which includes contributions from high pressure gas with \cis column density $(1-g_{\mathrm{low}})$N(\ci)$_{\mathrm{tot}}$ and low pressure gas with \cis column density $g_{\mathrm{low}}$N(\ci)$_{\mathrm{tot}}$, corresponds to the center of mass of the points ($f_1^{\mathrm{hp, lp}}$, $f_2^{\mathrm{hp, lp}}$) associated with the high and low pressure contributors, with the weights for each component corresponding to the corresponding fraction of the \cis column density they represent, (1-$g_{\mathrm{low}}$) and $g_{\mathrm{low}}$, respectively. The high pressure component is assumed to have ($f_1^{\mathrm{hp}}$, $f_2^{\mathrm{hp}}$) $=$ (0.38, 0.49), as in \citet{jenkins2011}. We will discuss the effects of different assumptions on the location of the high-pressure component further in this section. The low pressure component follows tracks that depend on density, temperature, and radiation field intensity (responsible for optical pumping of the excited \cis states). The model is identical to the one described in \citet{jenkins2001}, in particular their equations 10--12. This geometrical estimation of the low and high pressure components is illustrated in Figure \ref{plot_example_ci}, showing the ($f_1$, $f_2$) measurement for Sk-68 135. \\
\indent By geometrically matching the ($f_1$, $f_2$) measurements to the model tracks for the low pressure component shown in Figures \ref{plot_ci_model_f1_f2} and \ref{plot_example_ci}, and assuming that the kinetic temperature of the gas is equal to the rotational temperature of $H_2$, $T_{01}$(H$_2$) reported in \citet{welty2012}, $g_{\mathrm{low}}$ and $n(H)$ can be derived, if the radiation field intensity for the low-pressure is known. Fortunately, as described in \citet[][their Equation 1]{jenkins2011}, an estimate of the ionized to neutral carbon ratio, N(\cii)/N(\ci)$_{\mathrm{tot}}$, can provide the necessary constraints on the radiation field intensity. Because the calculation of the radiation field intensity from Equation 1 of \citet{jenkins2011} depends on the local hydrogen volume density $n$(H) (in addition to N(\cii)/N(\ci)$_{\mathrm{tot}}$), and conversely, the determination of $n$(H) from $f_1$ and $f_2$ requires the knowledge of the radiation field intensity, an iterative approach is necessary to solve for both. We start the iterative calculation by assuming the standard radiation field intensity specified by \citet{mathis1983} ($I/I_0$ $=$ 1) in the computation of $n$(H) using the model shown in Figure \ref{plot_ci_model_f1_f2} and the ($f_1$, $f_2$) measurements. We can then apply this density to calculate the radiation field intensity from N(\cii)/N(\ci)$_{\mathrm{tot}}$, following the approach described later in this section. From the density and radiation field, we derive the electron density $n_e$ (the procedure for deriving the electron density is described in the next paragraph). These are the initial values for the radiation field intensity, hydrogen density, and electron density. Next, we proceed with the iterative approach below:
\begin{itemize}
\item Compute an updated radiation field intensity $I/I_0$ using the density $n$(H) and electron density $n_e$ from the previous iteration, as well as fixed measurements (N(\ci)$_{\mathrm{tot}}$, N(\cii), T), as inputs to Equation 1 of \citet{jenkins2011}. \\
\item Using this updated radiation field intensity as input for the model tracks, derive an updated density for the low pressure component, $n$(H), from the observed ($f_1$, $f_2$) line ratios using the geometrical approach described above. \\
\item Using the fixed temperature $T_{01}$(H$_2$) and the updated $I/I_0$ and $n$(H) values as inputs in Equation 24 of \citet{weingartner2001b}, compute an updated electron density $n_e$ (see next paragraph for more details on this step).
\item Repeat this process using the radiation field intensity, density, and electron density output by an iteration as input for the next iteration. Continue iterating on the radiation field intensity, density, and electron density until convergence is reached, i.e. when the difference between successive iterations is less than 5\% on all parameters. 
\end{itemize} 
\indent During the iteration on radiation field and density, a computation of the electron density is required. For this, we solve for the ionization equilibrium electron density following Equation 24 in \citet{weingartner2001b}, given the density and radiation field at each step of the iteration. The calculation also includes physical coefficients such as the ionization from metals of $x_{e,0}$ $=$ 8.7$\times$10$^{-5}$ in the LMC (scaled by a factor of 1/2 to account for the metallicity difference between the LMC and Milky Way), and the cosmic ray ionization rate in the LMC, found to be 30\% \citep{abdo2010} of the Milky Way value 2$\times$10$^{-16}$  s$^{-1}$ \citep{indriolo2007, neufeld2010} (although the results are largely insensitive to the assumed cosmic ray ionization rate within a factor of a few of this value). The radiative plus di-electronic recombination coefficients, $\alpha_r$(H)  and the recombination rates for different elements due to collisions with dust grains, $\alpha_g$(X), are taken from \citet{shull1982} and \citet{weingartner2001b}, respectively. We scale $\alpha_g$(X) down by a factor 1/3 to account for the lower abundance of dust grains in the LMC. While the LMC has half-solar metallicity, \citet{RD2019} showed that the fraction of Si in the dust-phase is a factor 1.5 in the LMC than in the MW, leading to a dust-to-gas ratio 3 times lower in the LMC than in the Milky Way. The results described further in this paper confirm this result for other elements than Si. Since the Equation 24 of \citet{weingartner2001b} used to compute the electron density cannot be solved analytically, we take an iterative approach. \\
\indent As explained above, we use the N(\cii)/N(\ci)$_{\mathrm{tot}}$ ratio to estimate the radiation field intensity, using the density and electron density at each step of the iteration as inputs to \citet[Equation 1 of][]{jenkins2011}. In the METAL sample of sight-lines, the \cii $\lambda$1334 \mAA line is always strongly saturated, and the $\lambda$2325 \mAA line is too weak to be detected. Instead, we estimate the gas-phase \ciis column density with the following procedure. First, we scale the hydrogen column density $N(H)$ for each sight-line by the carbon abundance in the LMC (12 + $\log$(O/H) = 7.94, see Table 2), and obtain an estimate of the total carbon column density in the ISM (gas + dust). Second, we compute the carbon depletion corresponding to the measured iron depletion for each sight-line, using the coefficients presented in \citet{jenkins2009}. We applied this depletion value to the total carbon column density, thus yielding an estimate of the gas-phase carbon column density for each sight-line. Since the \cis column density for the range of hydrogen column densities probed by the METAL sight-lines is negligible compared to the \ciis column density, we can then estimate N(\ci)$_{\mathrm{tot}}$/N(\cii) $\simeq$ N(\ci)$_{\mathrm{tot}}$/N(C). Other ways to estimate N(\cii) from the measured \siis and \mgiis column densities are examined in Appendix A, but yield very similar results to the method used here. Furthermore, N(H) is measured for all targets, unlike N(\mgii) and N(\sii), giving this approach a fundamental advantage. \\
\indent In most cases, the $H_2$ rotational temperature $T_{01}$ was reported in \citet{welty2012}. When this was not the case, we assumed the median value of the \citet{welty2012} sample, or 80 K. For Sk-66 172, the iterative computation of density and radiation field from \cis and \ciis line ratios oscillates between two very different but not very physical models (very low radiation field, very high density and vice-versa), given the $H_2$ $T_{01}$ rotational temperature given in \citet{welty2012} (41$^{+110}_{-140}$ K) input to the model as the kinetic temperature. The closest temperatures for which the models converge is 110 K, which are well within the error bars of the temperature estimation. We assumed these temperatures in the modeling and report these values in Table 6. \\
\indent To propagate the errors on $f_1$ and $f_2$ on the density and radiation field intensity determination, we use a Monte-Carlo approach. We draw sample of ($f_1$, $f_2$) within Gaussian distributions of standard deviation equal to the 1$\sigma$ error on $f_1$ and $f_2$, and repeat the procedure to compute $n(H)$ and $I/I_0$ for each draw. The error on these parameters is then taken as the standard deviation of the resulting distributions. The errors on $n(H)$ and $I/I_0$ are reported in Table 6.\\
\indent We have explored the systematic effect of assuming a different ($f_1^{\mathrm{hp}}$, $f_2^{\mathrm{hp}}$) location for the high-pressure component on the resulting density derivations. Given the geometrical set-up of the approach, the density of the low-pressure component is very weakly dependent on $f_2^{\mathrm{hp}}$ (only $g_{\mathrm{low}}$ would change in this case). Given a displacement $\Delta f_1^{\mathrm{hp}}$ of the assumed location of the high-pressure component, and using the Thales theorem, $\Delta f_1^{\mathrm{lp}}$ $=$ $g_{\mathrm{high}}/g_{\mathrm{low}}$ $\Delta f_1^{\mathrm{hp}}$ $=$ (1-$g_{\mathrm{low}})/g_{\mathrm{low}}$ $\Delta f_1^{\mathrm{hp}}$. For all our targets, $g_{\mathrm{low}}$ $<$ 0.5, and thus, $\Delta f_1^{\mathrm{lp}}$ $<$ $\Delta f_1^{\mathrm{hp}}$. Furthermore, the location of the high-pressure component cannot slide too far off the high-density end of the model tracks. The fiducial value we assume, ($f_1^{\mathrm{hp}}$, $f_2^{\mathrm{hp}}$) $=$ (0.38, 0.49) \citep[as in ][]{jenkins2011}, corresponds to $n(H)$ $\sim$ 10$^4$ cm$^{-3}$ and $T$ $\sim$ 250 K as seen in the gray model tracks in Figure \ref{plot_ci_model_f1_f2}. The model for densities $>$ 10$^5$ cm${-3}$ all converge around the left-most point at n(H) $=$ 10$^5$ cm${-3}$, located at ($f_1$, $f_2$) $=$ (0.36, 0.51). For lower temperature models, the extremity of the model tracks would lie on top of the 10$^5$ cm$^{-3}$ point as well. For $T$ $\sim$ 100 K, this means the high-pressure point is at (0.39, 0.46). Thus, for a reasonable temperature, the high-pressure component is constrained between $f_1$ $=$ 0.36 and 0.39, meaning only a maximum displacement of $\Delta f_1^{\mathrm{hp}}$ $=$ 0.03 is possible. Given the high ratio of $g_{\mathrm{low}}$/$g_{\mathrm{high}}$, this implies that the effects of moving the location of the high-pressure point within reasonable model bounds has a negligible effect on the outcome. We have verified this numerically by recomputing densities using ($f_1^{\mathrm{hp}}$, $f_2^{\mathrm{hp}}$) $=$ (0.36, 0.51) and (0.39, 0.46) and found that the resulting differences in density determinations were typically lower than the 1$\sigma$ statistical error on n(H).\\
\indent Finally, we note that \cis toward 3 of the sight-lines (SK-70 115, SK-68 73, and SK-67 5) was previously analyzed by \citet[][(W16)]{welty2016}. Our measurements of $N$(\ci)$_{\mathrm{tot}}$, $f_1$, and $f_2$ are in reasonable agreement with those in \citet{welty2016}. For SK-67 5, the values compare as follows: $N$(\ci)$_{\mathrm{tot}}$ $=$ 13.75$\pm$0.03 (W16) vs 13.86$\pm$0.01; $f_1$ $=$ 0.28$\pm$0.06 (W16) vs 0.29$\pm$0.01; $f_2$ $=$ 0.09$\pm$ 0.03 (W16) vs 0.10$\pm$0.01. For SK-68 73, the comparison yields: $N$(\ci)$_{\mathrm{tot}}$ $=$ 14.29$\pm$0.03 (W16) vs 14.49$\pm$0.02; $f_1$ $=$ 0.45$\pm$0.06 (W16) vs 0.40$\pm$0.01; $f_2$ $=$ 0.27$\pm$0.04 (W16) vs 0.23$\pm$0.01. Finally, for SK-70 115, we have $N$(\ci)$_{\mathrm{tot}}$ $=$ 13.7$\pm$0.03 (W16) vs 13.86$\pm$0.01; $f_1$ $=$ 0.36$\pm$0.06 (W16) vs 0.42$\pm$0.01; $f_2$ $=$ 0.16$\pm$0.02 (W16) vs 0.14$\pm$0.01. Thus, the column densities of \cis differ beyond the uncertainties, but not by a large amount, while the $f_1$ and $f_2$ values, which are the driving parameters in the density derivation, are within 1$\sigma$. There are differences in the assumptions for the estimation of $N$(\cii), and possibly in the modeling of the \cis line ratios, leading to difference in the output parameters. We derive $n$(H) $=$ 101$\pm$6, 268$\pm$44, and 310$\pm$25 cm$^{-3}$ for the 3 sight-lines while W16 obtain $n$(H) $=$ 119, 1919, and 212 cm$^{-3}$ (no uncertainties reported). Densities for SK-67 5 and SK-70 116 are roughly consistent within our uncertainties between the two studies, but densities toward SK-68 73 differ by a large factor (beyond our uncertainties, but not necessarily beyond the W16 uncertainties, which are not reported). This is due to the location of SK-68 73 in the ($f_1$, $f_2$) diagram, near the curve where change of 5\% in $f_1$ can lead to a large change in density. More generally, this difference indicates densities derived from \cis integrated along the entire line-of-sight may not be physically meaningful when the physical conditions show drastic changes from one component to the next. Finally, for the radiation fields, we find $I/I_0$ $=$ 2.3$\pm$1.3, 6.7$\pm$4.7, 8.1$\pm$3.3, while W16 report 2.4, 5.7, 4.5 (no uncertainties reported). The two sets of measurements are within the uncertainties reported here.

\section{The correlation of depletions between different elements}\label{correlation_elements}

\begin{figure*}
\centering
\includegraphics[width=\textwidth]{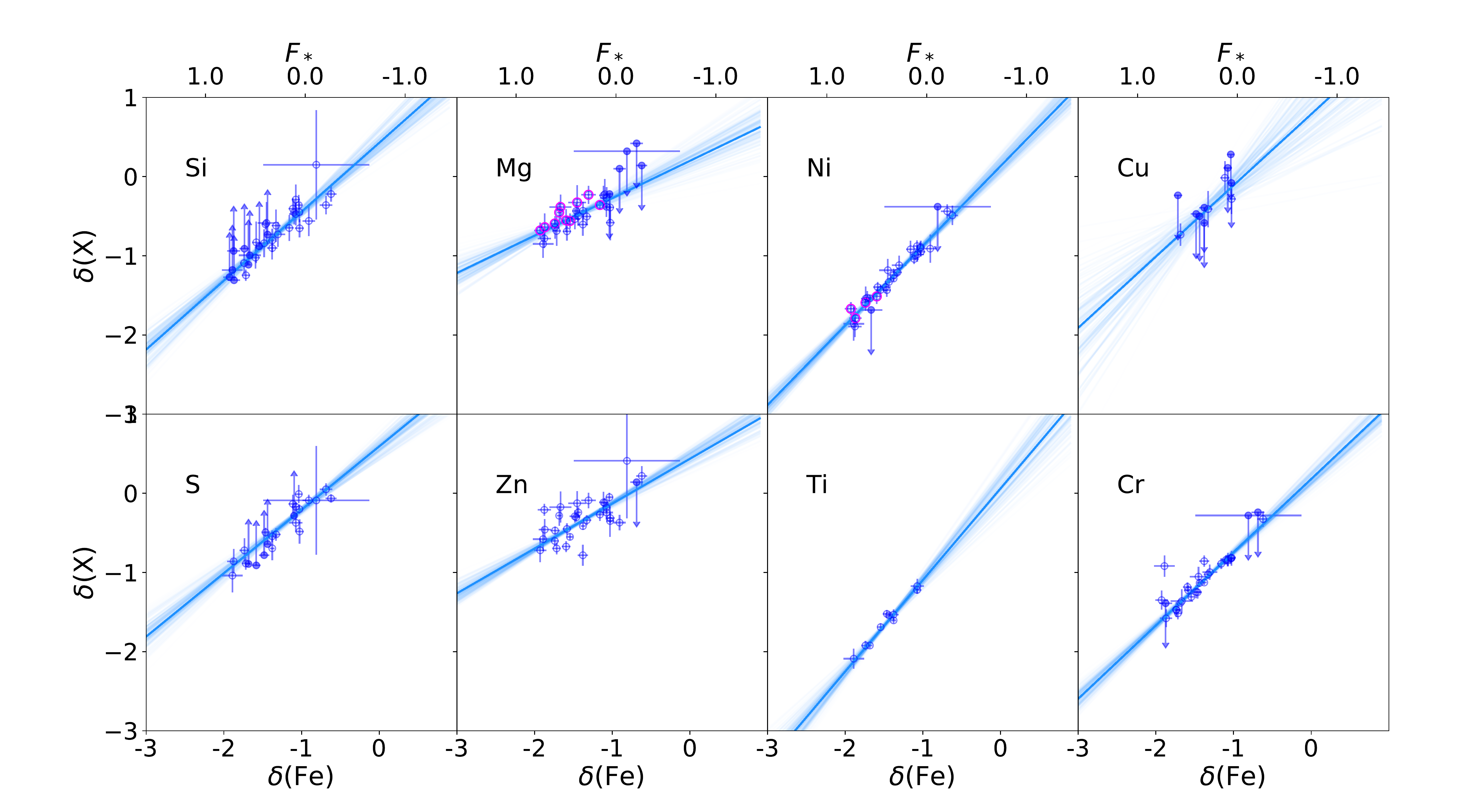}
\caption{Depletions (log fraction in the gas-phase) for Si, Mg, Ni, Cu, S, Zn, Ti, Cr, as a function of iron depletions ($\delta$(Fe)). Measurements made with COS at slightly lower resolution are outlined in magenta. The fits of the depletions are shown by the gray lines, the transparency of which is equal to the square-root of the probability of a given fit. }
\label{plot_deps_fe}
\end{figure*}

\indent As pointed out by \citet{jenkins2009} in the Milky Way, depletions for different elements correlate well with each other, indicating that the depletion process is a collective one. We investigate these correlations in the LMC using the METAL data in Figure \ref{plot_deps_fe}. As expected, we also find tight correlations between the depletions of different elements and that of iron. Following \citet{jenkins2009} and \citet{jenkins2017}, we fit those correlations with linear functions of the form

\begin{equation}
\delta(X) = a(X) \left (\delta(Fe) - z(X) \right ) + b(X)
\end{equation}

\noindent where $a(x)$ and $b(X)$ (slope and intercept of the relation between depletions of element X and Fe depletions) are fitted for, and where

\begin{equation}
z(X) = \cfrac{\sum_{\mathrm{los}}\cfrac{\delta(Fe)}{\sigma(\delta(X))^2}}{\sum_{\mathrm{los}}\cfrac{1}{\sigma(\delta(X))^2}}
\end{equation}

\begin{figure}
\centering
\includegraphics[width=8cm]{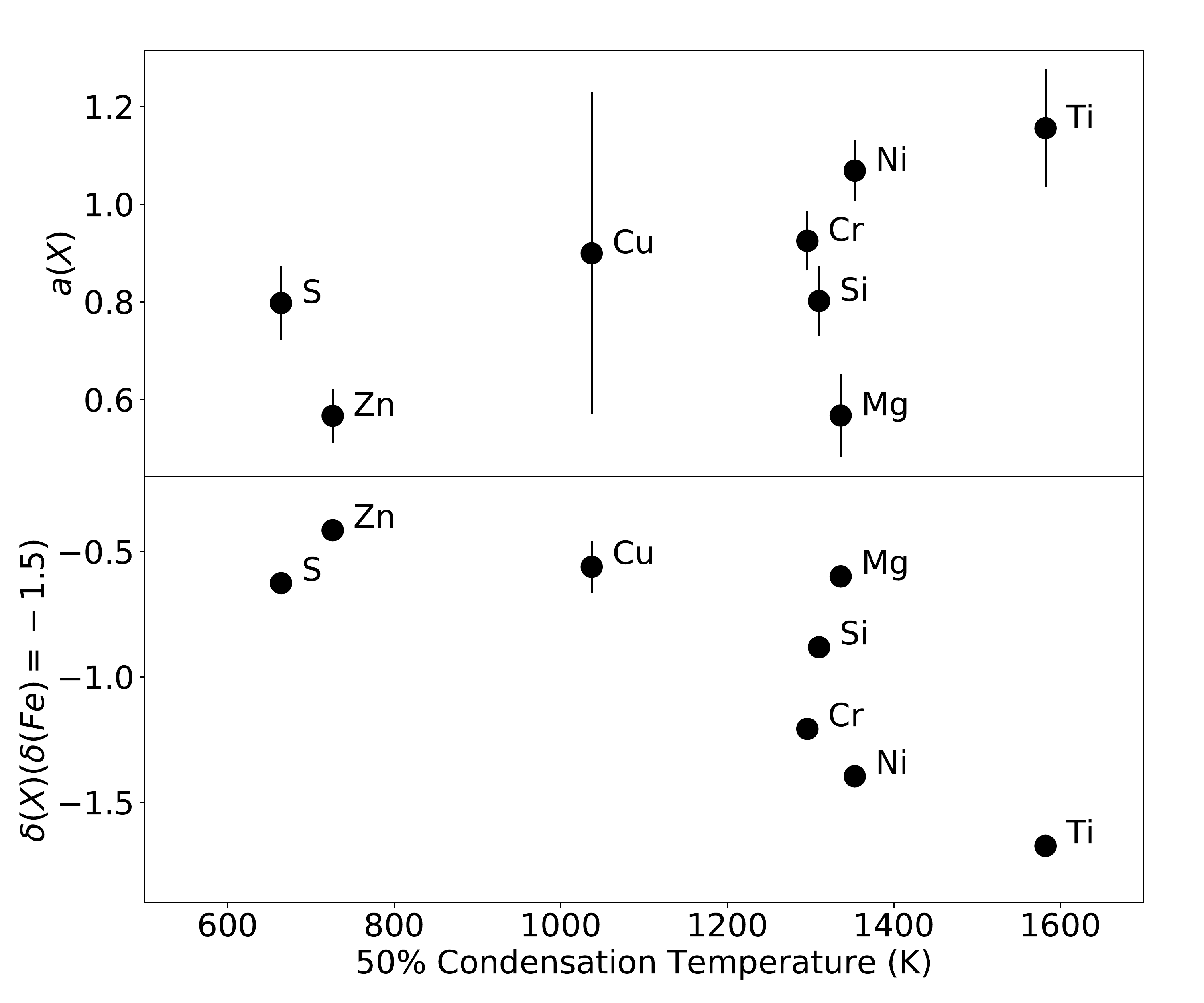}
\caption{Slope $a$ (top) and zero-point (bottom) of the relation between depletion of an element X, $\delta(X)$, and iron depletion, $\delta(Fe)$, as a function of the 50\% condensation temperature. The zero-point of the $\delta(X)$ - $\delta(Fe)$ relation is computed at $\delta(Fe)$ $=$ $-1.5$}
\label{plot_fe_temps}
\end{figure}

\noindent Here $\sigma(\delta)$(X) is the error on $\delta$(X). Introducing the zero-point reference $z(X)$ in $\delta(Fe)$ has the benefit to reduce the covariance between the formal fitting errors in $a(X)$ and $b(X)$ to near zero, as explained in \citet{jenkins2009}. The resulting parameters and their uncertainties are listed in Table 7, where we also list the p-value and correlation coefficients for this relation for each element. The correlation coefficients can be artificially enhanced by the covariant errors on $\delta$(Fe) and $\delta$(X), through common errors on N(H). We account for this following the method described in \citet[Appendix B of][]{jenkins1986b}. The ''corrected'' correlation coefficient is also listed in Table 7. Accounting for covariant errors only marginally reduces the correlation coefficient, which is $>$0.8 for all elements but Zn.\\
\indent As previously observed in the Milky Way \citep{savage1996, jenkins2009, ritchey2018}, the slope of the relation between the depletion of an element and that of iron steepens with increasing condensation temperature. Similarly, the depletion level (the absolute value of the depletion value, i.e, how depleted the gas phase metals are) increases with increasing condensation temperature. This can be seen in Figure \ref{plot_fe_temps}, which shows the slope of the $\delta(X)$ - $\delta(Fe)$ relation as a function of condensation temperature of element X, and the zero-point of this relation (at $\delta(Fe)$ $=$ $-1.5$). The 50\% condensation temperatures, listed in Table 8, are taken from \citet{lodders2003}. The original premise for the relation between depletion and condensation temperatures was in the context of grains being produced in outflows from stars \citep{field1974}. Nevertheless, such a relation is expected in the context of grain formation and destruction in the ISM owing to the correlation between condensation temperature and the strengths of chemical bonds.\\
\indent Zn and Magnesium are the most volatile elements. They deplete at a rate roughly half that of iron and have depletion levels about 10 times smaller than iron. Si, Cr, Cu, Ni, and Ti deplete at a similar rate (slope 0.8-1.2), but only the depletion levels of Cr, Ni, Ti reach that of Fe, with Si and Cu being about 5 and 10 times less depleted than Fe, respectively. S and Zn, which are commonly assumed to suffer little to no depletion, actually present significant depletions. S is 7 times less depleted than iron, but depletes at 80\% the rate of iron, thus reaching depletion levels of $-$1 dex (i.e., 90\% of S in the dust phase). Zn depletes at a lower rate than S and with overall lower levels (about 10 times less than Fe), but nevertheless reaches depletion levels of $-$0.86 dex (87\% of Zn in the dust phase). This effect has important implications for studies of the chemical enrichment of the universe through QSO absorption spectroscopy of damped Lyman-$\alpha$ systems (DLAs), in which S and Zn are often used as metallicity tracers.

\indent Given the tight correlations between depletions of different elements, \citet{jenkins2009} introduced the parameter $F_*$ to describe the collective advancement of the depletion process in the Milky Way. The depletion levels of element X is then modeled by a linear relation with $F_*$, of the form $\delta(X)$ $=$ $A_X \times (F_* - z_X) + B_X$. $F_*$ $=$ 0 corresponds to the most lightly depleted sight-lines in the Milky Way with $\log$ N(H) $>$ 19.5 cm$^2$, where ionization corrections are negligible, and $F_*$ $=$1 corresponds to the well studied heavily depleted velocity component toward $\zeta$ Oph. Since the $F_*$ scale is tied to the particular sight-lines used to anchor the $F_*$ $=$ 0 and 1 extremes, a comparison of depletion patterns using the $F_*$ parameter in other galaxies requires one to use the same normalizations for $F_*$. Therefore, similar to the computation of $F_*$ in SMC by \citet{jenkins2017}, the $F_*$ parameter in the LMC is given by:

\begin{equation}
F_*(LMC) = \frac{\delta(Fe)- B_{Fe}}{A_{Fe}} + z_{Fe}
\label{fit_fe_dep}
\end{equation}

\noindent where $A_{Fe}$ $=$ $-1.285$ , $B_{Fe}$ $=$ $-1.513$ , and $z_{Fe}$ $=$ 0.437 are the coefficients of the linear relation between $\delta(Fe)$ and $F_*$ in the Milky Way given in Table 4 of \citet{jenkins2009}. Thus, $F_*$ in galaxies other than the Milky Way corresponds to a scaling of iron depletions, which makes it easier to compare depletion patterns to those in the Milky Way. The top axis of Figure \ref{plot_deps_fe} shows the $F_*$ scale in the LMC. Iron was chosen as a proxy for $F_*$ due to its abundance of spectral lines of different oscillator strengths, allowing straight-forward measurements of column densities, abundances, and depletions (all METAL sight-lines have an iron depletion determination). \\
\indent

\begin{deluxetable*}{cccccccccc}
\centering
\tabletypesize{\scriptsize}
\tablecolumns{10}
\tablewidth{\textwidth}
\tablecaption{Fits of depletions vs iron depletion $\delta$(Fe)}
\tablenum{7}
\tablehead{Elements & $a_X$ & $\sigma(a_X)$ & $b_X$ & $\sigma(b_X)$\tablenotemark{a} & $z_{X}$  &  $r$\tablenotemark{b} &  $r_c$\tablenotemark{c} & $p-value$ & STD\tablenotemark{d}  \\}
 \startdata
 &&&&&&&&& \\
\hline
&&&&&&&\\

Si & 0.871 & 0.093 & -0.68 & 0.030 & -1.27 & 0.85 & 0.80 & 2.56e-07 & 0.11 \\
Ni & 1.006 & 0.059 & -1.26 & 0.017 & -1.38 & 0.98 & 0.98 & 3.84e-21 & 0.04 \\
Mg & 0.472 & 0.087 & -0.50 & 0.024 & -1.47 & 0.74 & 0.72 & 1.07e-05 & 0.08 \\
Cu & 0.900 & 0.330 & -0.44 & 0.094 & -1.37 & 0.88 & 0.88 & 1.21e-01 & 0.34 \\
Cr & 0.925 & 0.061 & -1.13 & 0.016 & -1.42 & 0.87 & 0.86 & 3.46e-09 & 0.09 \\
Zn & 0.567 & 0.056 & -0.36 & 0.015 & -1.41 & 0.66 & 0.56 & 5.09e-05 & 0.10 \\
S & 0.801 & 0.075 & -0.31 & 0.025 & -1.13 & 0.93 & 0.91 & 3.00e-08 & 0.08 \\
Ti & 1.156 & 0.120 & -1.63 & 0.023 & -1.46 & 0.99 & 0.99 & 2.08e-07 & 0.07 \\

\enddata
\tablenotetext{a}{Systematic errors on $b_X$ due to uncertainties on the photospheric abundances are not included, because they do not affect the relative trends examined here (e.g., environmental parameters). An estimate of these systematic errors can be found in Table 4 of \citet{tchernyshyov2015}. }
\tablenotetext{b}{Correlation coefficient}
\tablenotetext{c}{Correlation coefficient corrected for covariant errors (through the $\log$ N(H) dependence of $\delta$(Fe) and $\delta$(X)) following \citet[Appendix B of][]{jenkins1986b}}
\tablenotetext{d}{Standard deviation of the measurements about the fit}

\end{deluxetable*}

\begin{deluxetable}{cc}
\centering
\tabletypesize{\scriptsize}
\tablecolumns{2}
\tablewidth{3cm}
\tablecaption{50\% Condensation Temperatures}
\tablenum{8}
\tablehead{Elements & $T_c$ \\}
 \startdata
 & \\
O & 182\\
Mg & 1336 \\
Si & 1310\\
S & 664\\
Ti & 1582\\
Cr & 1296\\
Fe & 1334\\
Ni & 1353\\
Cu & 1037\\
Zn & 726\\

\enddata
\tablecomments{50\% condensation temperatures are from \citet{lodders2003}}
\end{deluxetable}

\section{Variations of depletions with local environment}\label{correlations_environment}

\indent In this Section, we explore the environmental parameters driving variations of the interstellar depletions and subsequently dust-to-metal ratio within the LMC.  Thanks to the tight correlation between depletions of iron and other elements, we can explore the variations of depletions as a function of environment using iron as a representative element. We examine the correlations between iron depletions and hydrogen column density (N(H)), H$_2$ fraction (f(H$_2$)), hydrogen volume density (n(H)), radiation field intensity ($I/I_0$), and distance from various landmarks in the LMC, in particular to its center, located about 1 kpc West of the 30 Dor massive star-forming region.

\subsection{Multi-linear regression of iron depletions versus local environment parameters}\label{multilinear_section}

\begin{figure*}
\centering
\includegraphics[width=\textwidth]{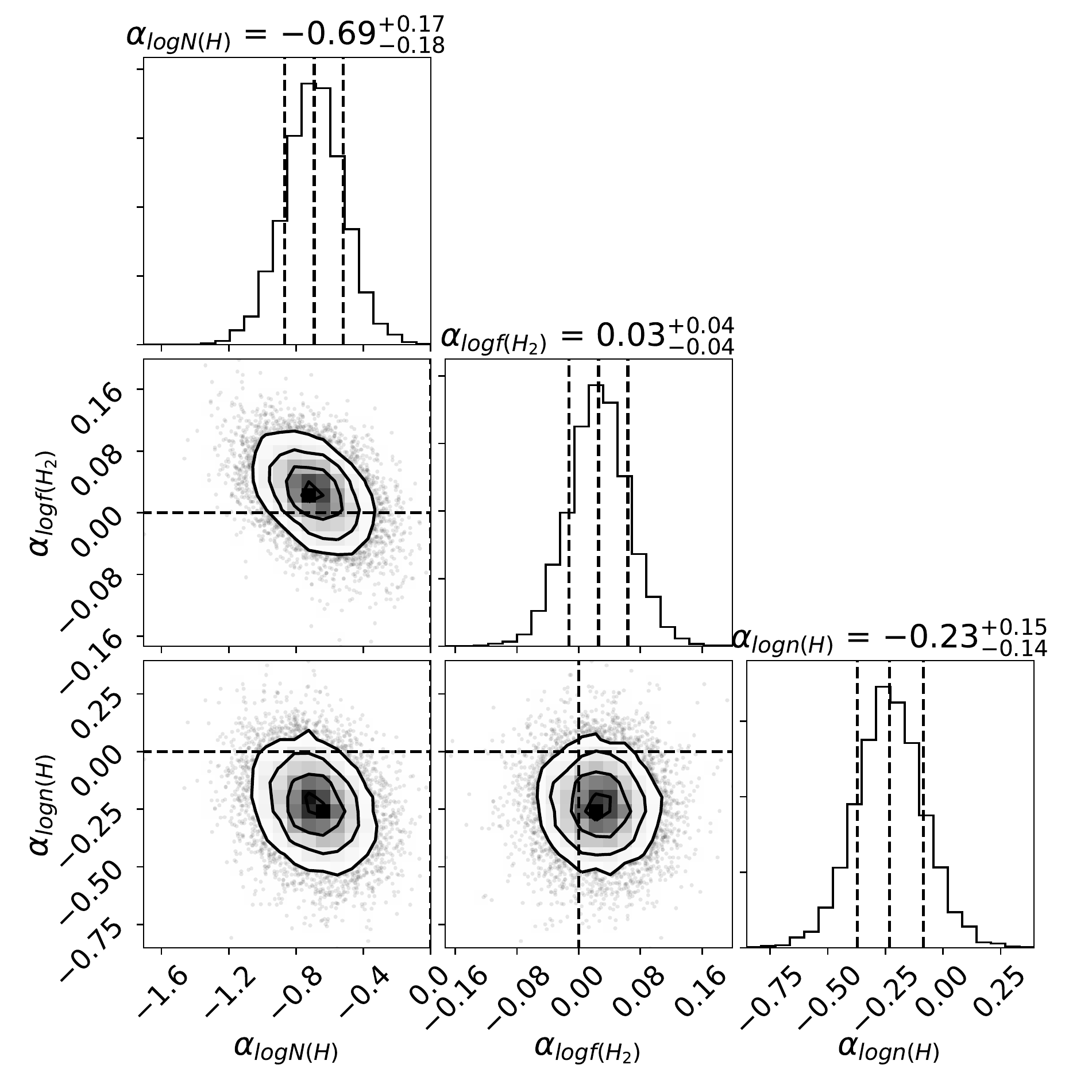}
\caption{Distributions of the slopes of the iron depletions versus the logarithm of the hydrogen column density, $\alpha_{log N(H)}$, the logarithm of the $H_2$ fraction, $\alpha_{f(H2)}$, and the hydrogen volume density, $\alpha_{log n(H)}$, such that $\delta(Fe)$ $=$ $\alpha_{log N(H)}$ $\log N(H)$ $+$ $\alpha_{f(H2)}$ $f(H_2)$  + $\alpha_{n(H)}$ $\log n(H)$ $+$ $\beta$(Fe)}
\label{plot_corner_column_fh2_dens}
\end{figure*}

\begin{figure*}
\centering
\includegraphics[width=\textwidth]{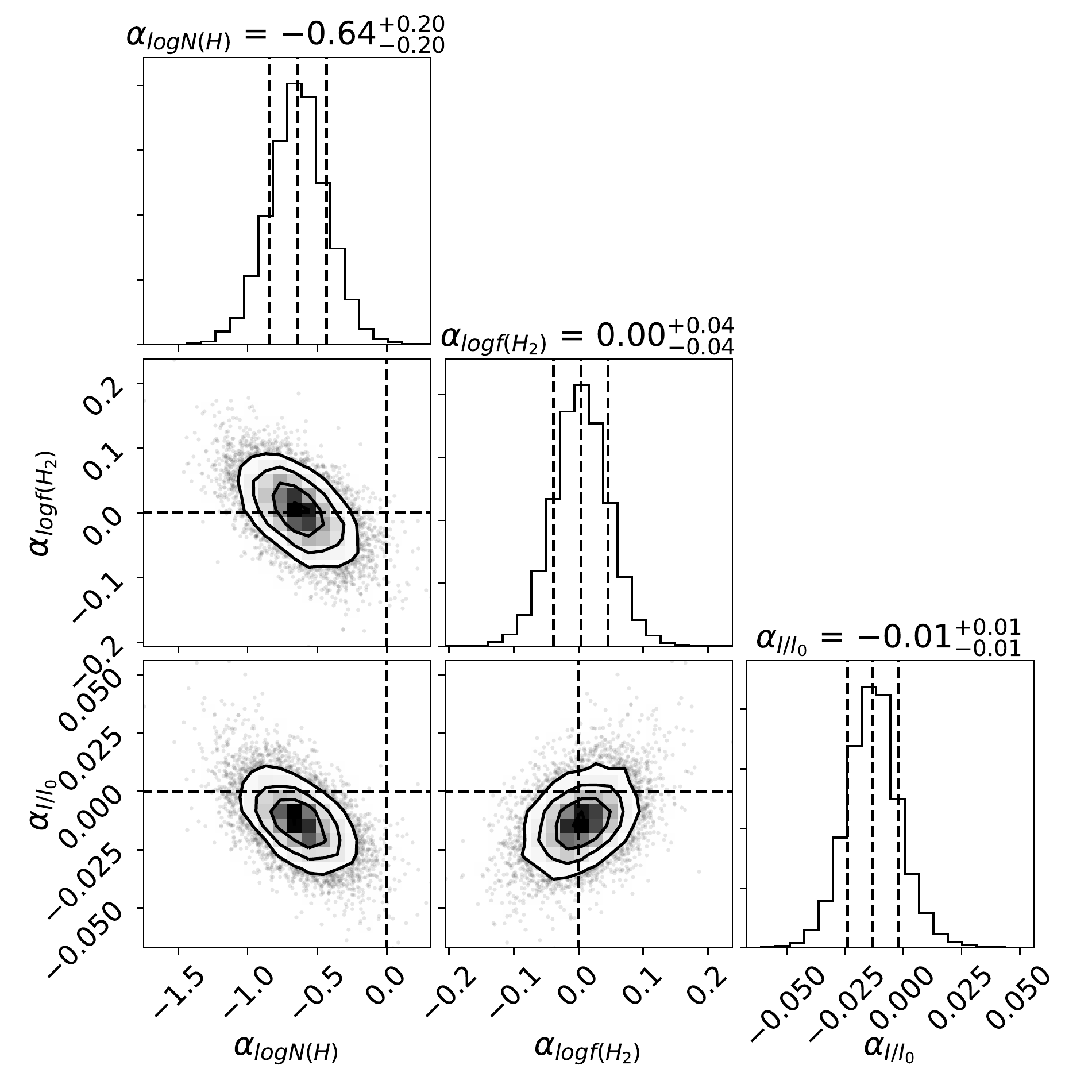}
\caption{Distributions of the slopes of the iron depletions versus the logarithm of the hydrogen column density, $\alpha_{log N(H)}$(Fe), the logarithm of the $H_2$ fraction, $\alpha_{f(H2)}$(Fe), and the radiation field intensity, $\alpha_{I/I_0}$(Fe), such that $\delta(Fe)$ $=$ $\alpha_{log N(H)}$(Fe) $\log N(H)$ $+$ $\alpha_{f(H2)}$(Fe) $f(H_2)$  + $\alpha_{I/I_0}$(Fe) $I/I_0$ $+$ $\beta(Fe)$}
\label{plot_corner_column_fh2_rad}
\end{figure*}

\indent In order to determine which environmental parameters amongst N(H), n(H), $I/I_0$, $f(H_2)$ drive the variations of depletions and dust-to-metal ratio, we first perform a multi-linear regression of the iron depletions ($\delta(Fe)$) as a function of combinations of 3 of these parameters. We cannot perform a robust multi-linear regression analysis on more parameters in one instance with only 32 measurements. The covariance of errors in the depletions (recall $\delta(X)$ $\propto$ $N(X)/N(H)$) with $N(H)$ and $f(H_2)$ can strengthen or weaken the inferred correlation between parameters. To account for this, we use the {\it mlinmix\_err} IDL package developed by \citet{kelly2007}, which uses a Bayesian approach to multi-linear regression, accounting for errors and covariances between dependent and independent variables. \\
\indent We explore the multi-linear correlations between $\delta(Fe)$ and $\log$ N(H), $f(H_2)$ and $n(H)$ in Figure \ref{plot_corner_column_fh2_dens}, which shows the distributions of the slopes of the iron depletions versus each parameter, $\alpha_{\log N(H)}$(Fe), $\alpha_{\log f(H2)}$(Fe), and $\alpha_{\log n(H)}$(Fe), such that:

\begin{multline}
\delta(\mathrm{Fe}) =\alpha_{\log \mathrm{N(H)}}(\mathrm{Fe}) \log \mathrm{N(H)} + \\
\alpha_{\log f(\mathrm{H}_2)}(\mathrm{Fe}) \log f(\mathrm{H}_2)  +\\
 \alpha_{\log \mathrm{n(H)}}(\mathrm{Fe}) \log \mathrm{n(H)} + \beta(\mathrm{Fe})
\end{multline}

\noindent We find $\alpha_{\log N(H)}$(Fe) $=$ $-0.74\pm0.18$, $\alpha_{\log f(H2)}$(Fe) $=$ 0.01$\pm$0.04, and $\alpha_{\log n(H)}$(Fe) $=$ $-$0.14$\pm$0.12. In other words, the clearest and strongest anti-correlation of depletions with environment occurs with hydrogen column density. We find no correlation with the fraction of molecular gas $f(H_2)$ in the regime probed by the METAL sight-lines (3$\times10^{-7}$ --- 0.62 with a median value of 0.03). There is a secondary marginal anti-correlation with hydrogen volume density, as traced by the \cis gas. In theory, one would expect a strong correlation between depletions and volume density, because dust growth timescales are inversely proportional to density. However, since \ciis is the dominant form of carbon in the neutral translucent ISM probed by this spectroscopic program, the \cis gas may represent a small mass and volume fraction of the gas traced by our sight-lines, and thus the density in the \cis gas may not be representative of the mean density along the line of sight, traced by other metals and \hi. Rather, since we are viewing the LMC nearly face on, variations in the path length are effectively driven by any changes in the scale height of the gas perpendicular to the plane of the LMC. The magnitudes of such variations are probably small compared to the variability of N(H) in our sample. Hence, N(H) should be a good proxy for the average n(H) over the entire line of sight to a star embedded near the plane of the LMC, explaining the resulting strong correlation with depletions. \\
\indent In Figure \ref{plot_corner_column_fh2_rad}, we perform a multi-linear regression with the combination of $\log N(H)$, $\log f(H_2)$, and $I/I_0$. In this case, we find $\alpha_{\log N(H)}$(Fe) $=$ $-$0.74$\pm$0.19, $\alpha_{\log f(H2)}$(Fe) $=$ 0.00$\pm$0.04, and $\alpha_{\log I/I_0}$(Fe) $=$ $-$0.05$\pm$0.12. We therefore conclude that there is no correlation between depletions and radiation field intensity or H$_2$ fraction, at least in the parameter space probed by the METAL survey. We note however, that the radiation field probed by the \cis gas, as for the density, may not be representative of the radiation field illuminating the gas along the entire line of sight, since the \cis gas is associated with a small fraction of the \hi.

\subsection{Depletions vs hydrogen column density}\label{section_dep_nh}

\begin{figure*}
\includegraphics[width=\textwidth]{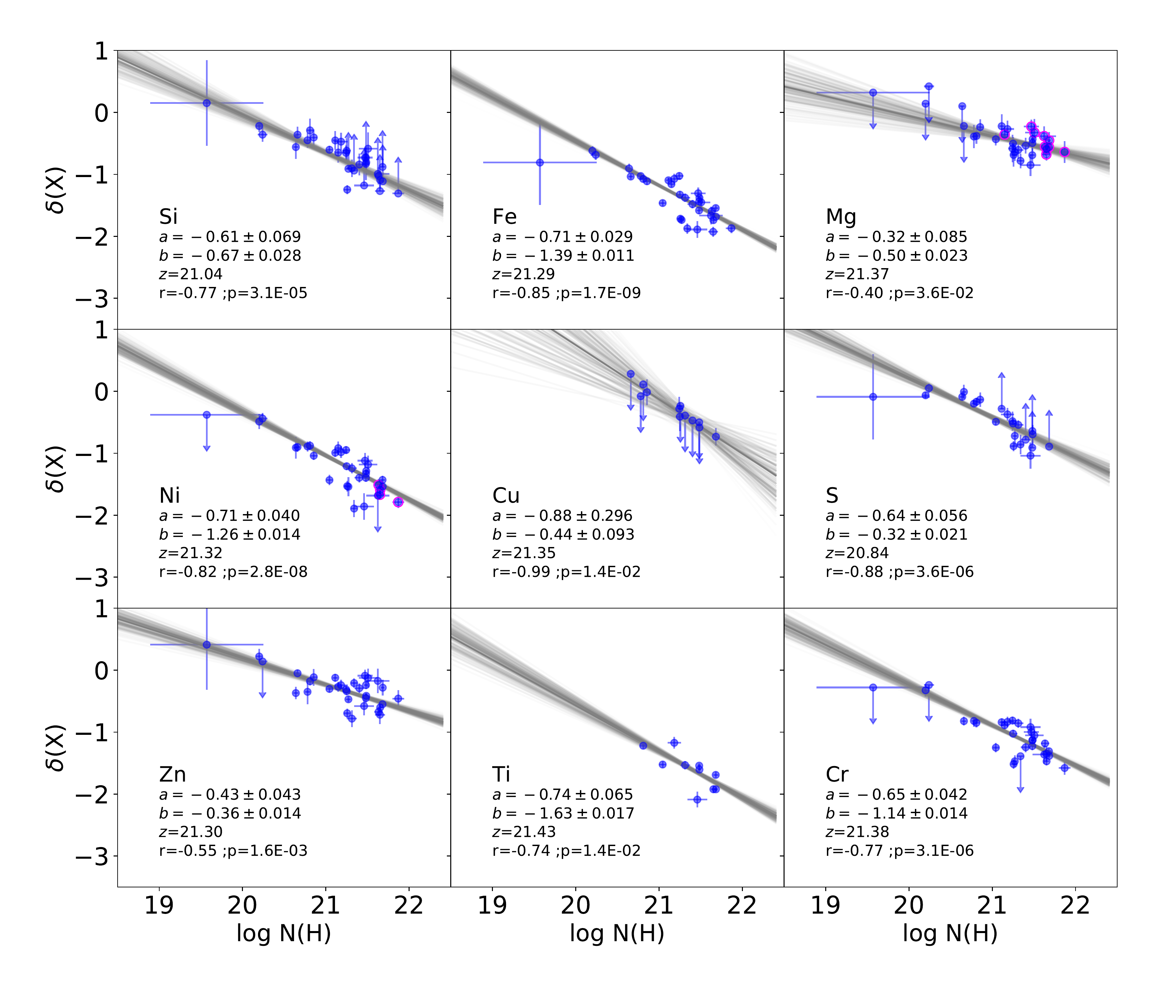}
\caption{Depletions (log fraction in the gas-phase) for Si, Fe, Mg, O, Ni, Cu, S, Zn, Ti, Cr, as a function of $\log$ N(H), where N(H) is from \citet{RD2019} and includes both \his and H$_2$. Measurements made with COS at slightly lower resolution than STIS are outlined in magenta. The $a_{H}$, $b_{H}$, $z_{H}$ parameters of the best-fit correlations, as well as the correlation coefficient $r$ and $p$-value are indicated in each panel.}
\label{plot_deps_nh}
\end{figure*}

\begin{deluxetable*}{ccccccc}
\centering
\tabletypesize{\scriptsize}
\tablecolumns{7}
\tablewidth{\textwidth}
\tablecaption{Fits of depletions vs $\log$ N(H)}
\tablenum{9}
\tablehead{Elements & $a_H$  & $b_H$\tablenotemark{a} & $z_{H}$  & $r$\tablenotemark{b} & $p-value$ & STD\tablenotemark{c} \\}
 \startdata
 &&&&&& \\
\hline
&&&&& &\\

Fe & -0.711$\pm$0.03 & -1.385$\pm$0.01 & 21.29 & -0.85 & 1.72e-09 & 0.08 \\
Si & -0.614$\pm$0.07 & -0.672$\pm$0.03 & 21.04 & -0.77 & 3.06e-05 & 0.09 \\
Ni & -0.709$\pm$0.04 & -1.263$\pm$0.01 & 21.32 & -0.82 & 2.75e-08 & 0.10 \\
Mg & -0.318$\pm$0.09 & -0.496$\pm$0.02 & 21.37 & -0.40 & 3.63e-02 & 0.11 \\
Cu & -0.877$\pm$0.30 & -0.441$\pm$0.09 & 21.35 & -0.99 & 1.39e-02 & 0.10 \\
Cr & -0.648$\pm$0.04 & -1.138$\pm$0.01 & 21.38 & -0.77 & 3.06e-06 & 0.10 \\
Zn & -0.426$\pm$0.04 & -0.362$\pm$0.01 & 21.30 & -0.55 & 1.58e-03 & 0.09 \\
S & -0.637$\pm$0.06 & -0.315$\pm$0.02 & 20.84 & -0.88 & 3.62e-06 & 0.10 \\
Ti & -0.740$\pm$0.07 & -1.627$\pm$0.02 & 21.43 & -0.74 & 1.40e-02 & 0.24 \\

\enddata
\tablenotetext{a}{The first and second uncertainties reported represent the statistical error and systematic error on the photospheric abundances of young stars (see Table 4 in \citet{tchernyshyov2015})}
\tablenotetext{b}{Correlation coefficient}
\tablenotetext{c}{Standard deviation of the measurements about the fit}
\end{deluxetable*}

\indent  Because timescales for accretion of gas-phase metals onto dust grains become shorter as density increases \citep{asano2013, zhukovska2016}, it is expected that the fraction of metals in the gas (i.e., depletion) decreases with increasing hydrogen column density or volume density (they are related). Section \ref{multilinear_section} demonstrates that the strongest correlation between depletions and environment is with hydrogen column density, even accounting for the covariance between depletions and $\log$ N(H). Such a trend has also been observed in the Milky Way \citep{wakker2000, jenkins2009} for the full suite of the major components of the ISM, and Magellanic Clouds \citep{tchernyshyov2015, RD2019}, albeit for a limited range of elements. \\
\indent Here, we quantify the correlation between depletions and $\log$ N(H) for all elements probed by the METAL spectra. In Figure \ref{plot_deps_nh}, the depletions toward the 32 METAL sight-lines in the LMC decrease with increasing hydrogen column density for all elements probed by the survey: Si, Fe, Mg, O, Ni, Cu, S,  Zn, and Cr. The Ti depletions are taken from \citet{welty2010}, an optical spectroscopic study of a large number of LMC sight-lines.  We fit the relation between elemental depletions and $\log N(H)$ taking into account the errors in $\log N(H)$ and depletions, and using a linear function of the form: 

\begin{equation}
\delta(X) = a_H(X) \left (\log N(H) - z_H(X) \right ) + b_H(X)
\label{fit_dep_nh_eq}
\end{equation}

\noindent where $a_H(x)$ and $b_H(X)$ (slope and intercept of the relation between depletions of element X and $\log$ N(H)) are fitted for, and where

\begin{equation}
z_H(X) = \cfrac{\sum_{\mathrm{los}}\cfrac{\log N(H)}{\sigma(\delta(X))^2}}{\sum_{\mathrm{los}}\cfrac{1}{\sigma(\delta(X))^2}}
\end{equation}

\noindent Again, introducing the zero-point reference $z_H(X)$ in $\log$ N(H) has the benefit to reduce the covariance between the formal fitting errors in $a_H(X)$ and $b_H(X)$ to near zero.
\noindent As pointed out by \citet{jenkins2009}, introducing the $z_H(X)$ parameter practically removes the covariance in the errors on the fitted solutions $a_H(X)$ and $b_H(X)$. The best-fit parameters for each element, as well as the $p$ and $r$ values, are listed in Table 9. The tight correlations seen in Figure \ref{plot_deps_nh} are reflected in the $r$ values, which range from $-$0.99 to $-$0.40, with high significance  ($p$ $<$ 3.6 $\times 10^{-2}$). 


\subsection{What other parameter(s) drives depletion variations?}

\begin{figure}
\centering
\includegraphics[width=8cm]{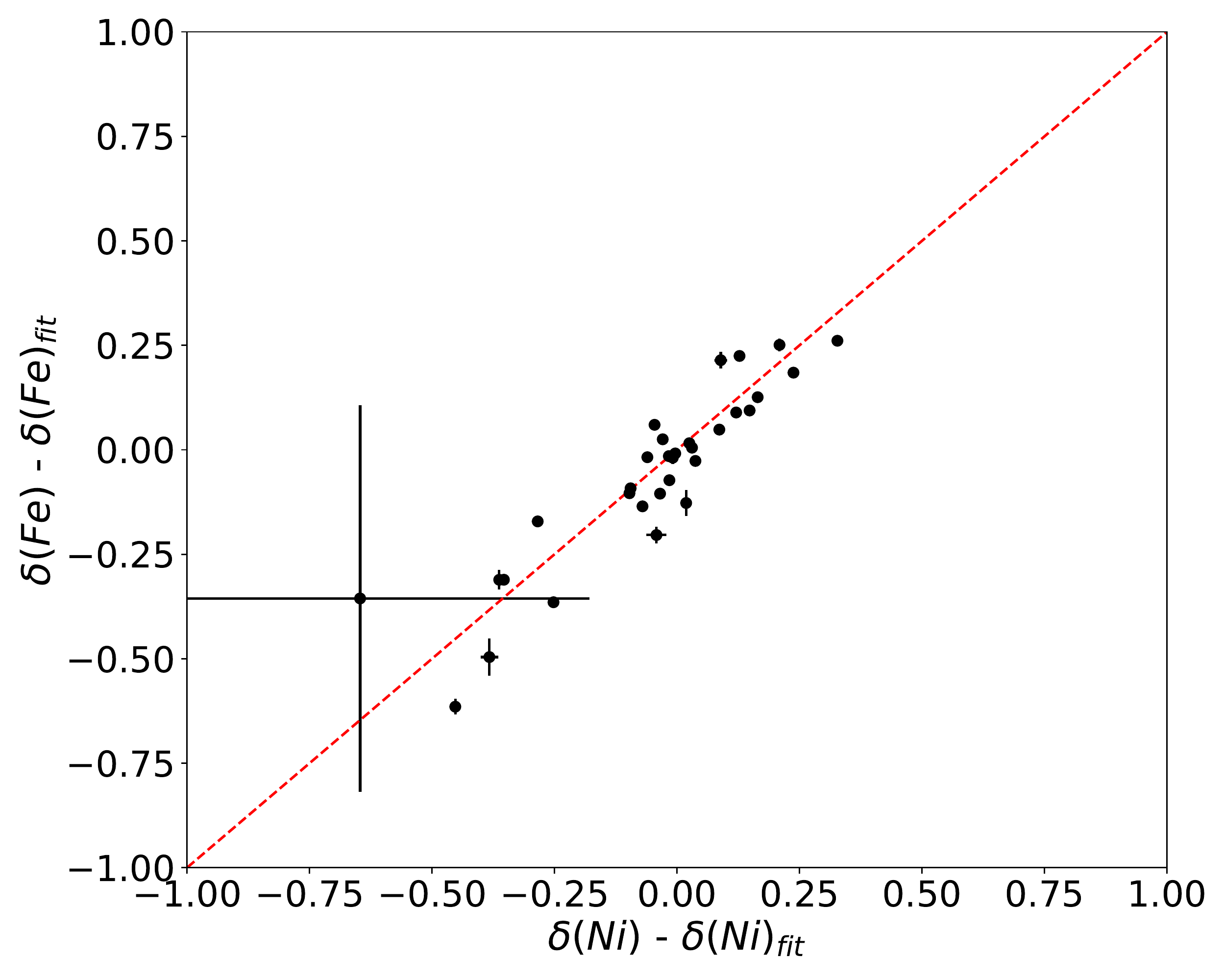}
\caption{Residuals of the fitted correlation between iron depletion and $\log$ N(H) ($\delta$(Fe) - $\delta$(Fe)$_{\mathrm{fit}}$) versus residuals of the fitted correlation between nickel depletions and $\log$N(H) ($\delta$(Ni) - $\delta$(Ni)$_{\mathrm{fit}}$). The residuals are computed as $\delta$(X) - $\delta$(X)$_{\mathrm{fit}}$ $=$ $\delta$(X) $-$ $a_H$(X)($\log$ N(H) $-$ $z_H$(X)) $-$ $b_H$(X). The residuals are not random noise: residuals for all elements correlate with each other, indicating a physical origin to this secondary depletion variation. The red dashed line indicates a 1:1 correlation.}
\label{plot_fe_ni_residuals}
\end{figure}

\begin{figure}
\centering
\includegraphics[width=8cm]{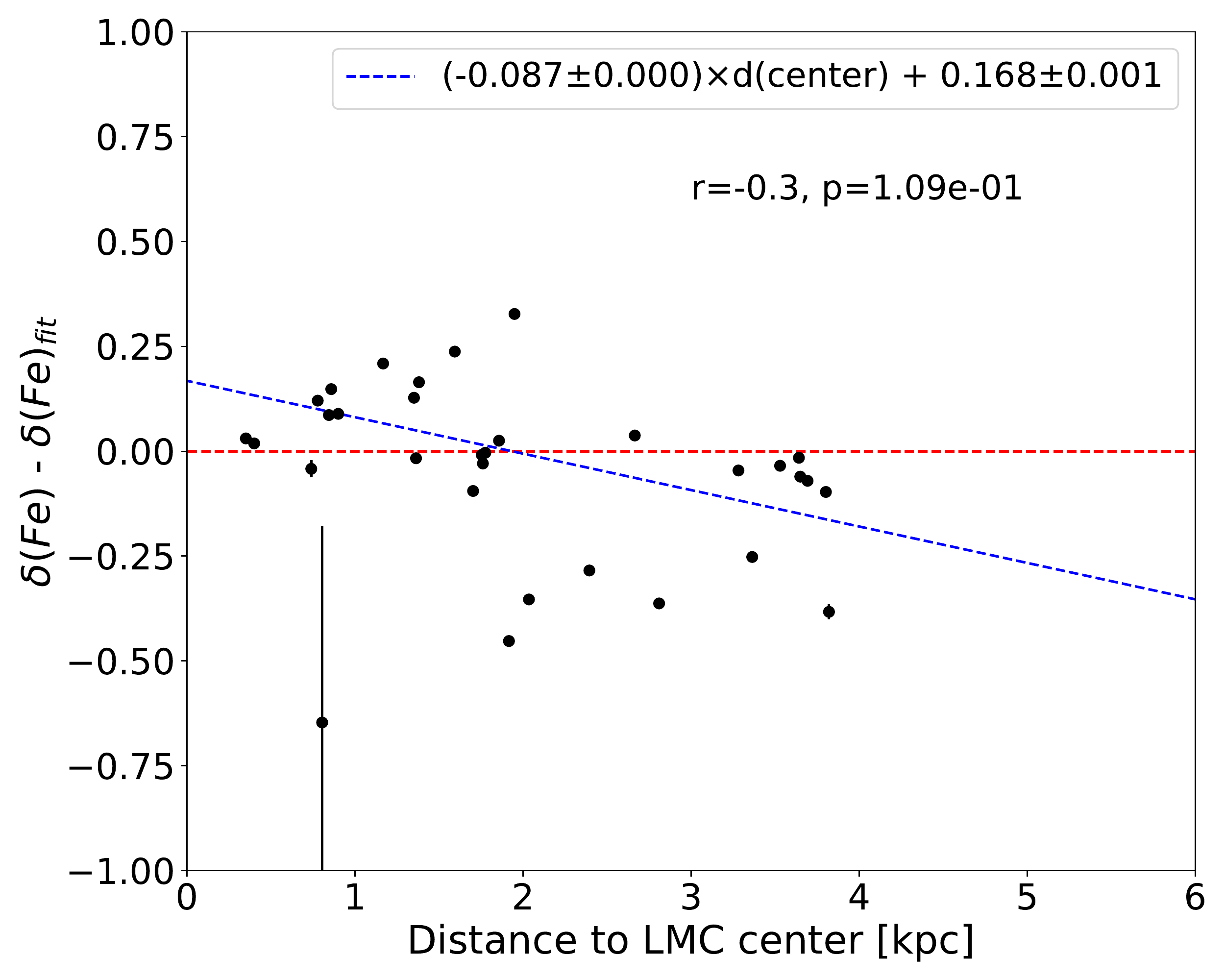}
\caption{Residuals of the correlation between iron depletion and $\log$ N(H) ($\delta$(Fe) - $\delta$(Fe)$_{\mathrm{fit}}$) versus distance to the LMC center. The residuals are computed as $\delta$(Fe) - $\delta$(Fe)$_{\mathrm{fit}}$ $=$ $\delta$(Fe) $-$ $a_H$(Fe)($\log$ N(H) $-$ $z_H$(Fe)) $-$ $b_H$(Fe).}
\label{plot_residuals_fe_dist_center}
\end{figure}

\begin{figure*}
\centering
\includegraphics[width=\textwidth]{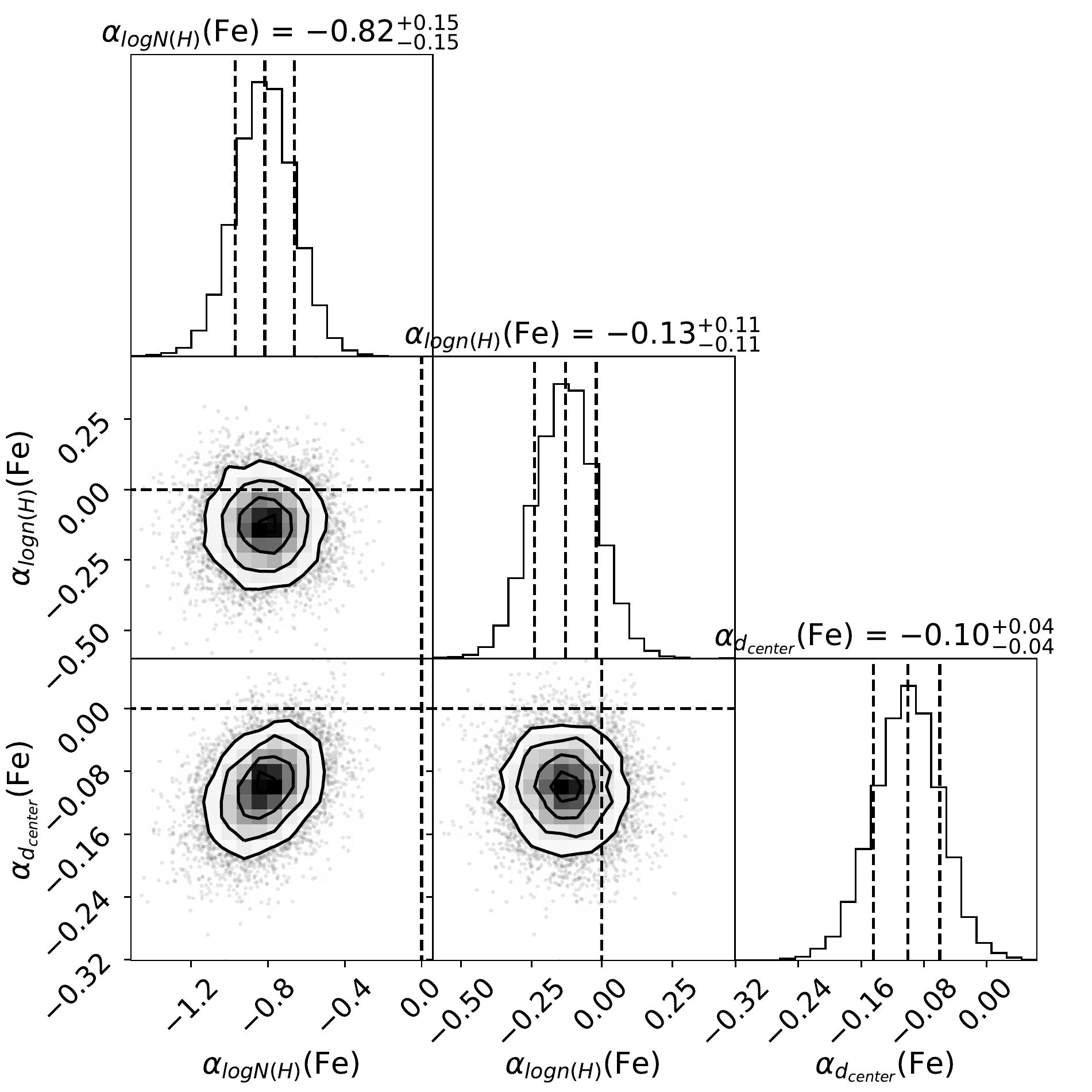}
\caption{Distributions of the slopes of the iron depletions versus the logarithm of the hydrogen column density, $\alpha_{log N(H)}$, the hydrogen volume density, $\alpha_{log n(H)}$, and the distance to the LMC center, $\alpha_{d(center)}$, such that $\delta(Fe)$ $=$ $\alpha_{log N(H)}$ $\log N(H)$ $+$ $\alpha_{n(H)}$ $\log n(H)$ $+$ $\alpha_{d(center}$ $d(center)$ $+$ $\beta$}
\label{plot_corner_column_dens_dcenter}
\end{figure*}

\begin{figure*}
\centering
\includegraphics[width=13cm]{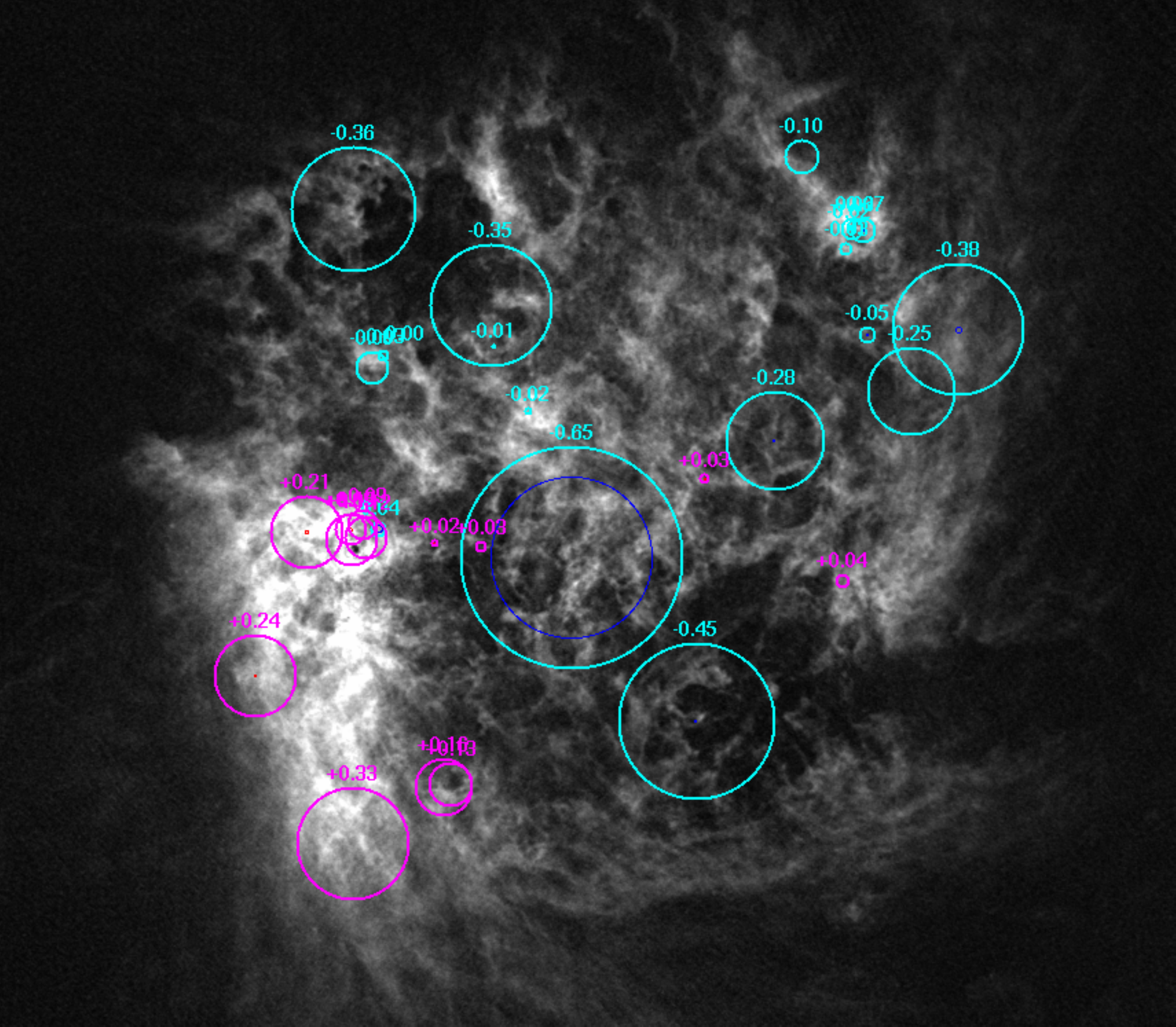}
\caption{Map of the \his column density in the LMC derived from 21 cm emission \citep{kim2003}, with the METAL sight-lines overlaid. Cyan and magenta circles correspond to sight-lines with negative and positive residuals from the main relation between iron depletions and hydrogen column density, respectively. The size of the circles is proportional to the absolute value of the residuals. The blue and red (very small) circles similarly scale with the depletion measurement error, which is negligible compared to the magnitude of the residuals for all sight-lines except Sk-69 104. There is a clear East-West split in the gas-phase abundances and depletions, with sight-lines along the South-East over-dense \his filament \citep{mastropietro2009} having enhanced gas-phase metallicities (up to $+$0.3 dex), and sight-lines on the West side of the LMC center having depressed metallicities (up to $-$0.5 dex).  }
\label{map_residuals}
\end{figure*}

\indent While hydrogen column density appears to be a main driver of the depletion levels, and depletions do not appear to correlate with volume density, radiation field intensity, or $H_2$ fraction, the residuals of the $\delta(X)$---$\log N(H)$ for different elements X correlate strongly with each other, as seen in Figure \ref{plot_fe_ni_residuals}. This indicates a physical origin for these residuals and subsequently a secondary correlation with another parameter other than volume density, $H_2$ fraction, or radiation field intensity. We note that the residuals of the $\delta(X)$---$\log N(H)$ correlation could also be caused by variations of similar magnitude of the true total metallicity of the ISM (gas + dust). \\
\indent We investigated possible correlations of these residuals with tracers of interstellar shocks, star-formation and feedback, such as the distance to the closet supernova (SN) remnant \citep{temim2015}, the H$\alpha$ surface brightness from the SHASSA survey at ~1$'$ (15 pc) resolution \citep{gaustad2001}, 24 $\mu$m surface brightness in the SAGE Spitzer survey of the LMC \citep{meixner2006} at 6$''$ resolution (1.5 pc)), and the dust temperature at 36$''$ (10 pc) resolution \citep{gordon2014}. We found no correlation between the residuals of the $\delta(X)$---$\log N(H)$ correlation and any of these parameters. \\
\indent Ultimately, we uncovered a significant correlation between the residuals of the $\delta(X)$---$\log N(H)$ correlation and distance to the center of the LMC, located at RA $=$ 82.25 \degs and DEC $=$ $-$69.5 \degn, about 1 kpc to the West of 30 Doradus. The relation is shown in Figure \ref{plot_residuals_fe_dist_center} for iron. The correlation coefficient is $-$0.3 with a p-value of 0.1, indicating a weak correlation. Similar values were found for other elements, such as nickel and silicon. Given this newly discovered dependence, and the marginal correlation with volume density discussed in Section \ref{multilinear_section}, we ran the multi-linear regression on hydrogen column density, volume density, and distance to the LMC center, shown in Figure \ref{plot_corner_column_dens_dcenter}. Once the most significant dependences are included, the correlations tighten-up, and we finally find $\alpha_{\log N(H)}$(Fe) $=$ $-$0.82$\pm$0.15, $\alpha_{\log n(H)}$(Fe) $=$ $-$0.13$\pm$0.11, and $\alpha_{d_{center}}$(Fe) $=$ $-$0.10$\pm$0.04. \\
\indent Having determined that the drivers of depletion variations are hydrogen column density, hydrogen volume density (marginal correlation), and distance to the LMC center, we then compute the slopes $\alpha_{\log N(H)}$(X), $\alpha_{\log n(H)}$(X), $\alpha_{d_{center}}$(X) for all elements X, as well as the intercept of the multilinear correlation, $\beta$. The results are listed in Table 10. All elements exhibit similar anti-correlations between their depletions, the hydrogen volume density, and the distance to the LMC center. \\
\indent The anti-correlations between depletions, hydrogen column density and volume density are expected if metals accrete onto dust grains in the ISM, since the timescale for accretion is inversely proportional to density (and subsequently column density). The anti-correlation with distance to the LMC center is a relatively surprising finding, which could result from two effects. The first is a possible metallicity gradient. We assume constant total abundances for the ISM based on the lack of observed gradient in $>$ 1 Gyr old stars \citep[e.g.][and references therein]{cioni2009} and \hiis regions \citep[][and references therein]{ToribioSanCipriano2017}. If a metallicity gradient did exist in young (a few tens of million years old) stars, it would give the appearance of a gradient in the depletions, as observed here. The second effect possibly causing the observed negative radial gradient in depletions is dust processing (formation and destruction).  In this case, metals are less depleted from the gas-phase near the LMC center and 30 Dor, and more depleted into dust grains away from 30 Dor.\\
\indent To get more insight into which effect might be at play, we map out the residuals of the main trend between iron depletions and hydrogen column density relative to the LMC gas disk in Figure \ref{map_residuals}. There is a clear gradient in gas-phase metallicity (and/or depletions) from East to West. Sight-lines on the East (left) side of the LMC center, along the \his filament associated with the South-East \his overdensity \citep{nidever2008, mastropietro2009}, have positive residuals up to $+$0.3 dex, i.e., higher gas-phase metallicities for their \his column density than the fiducial trend. Conversely, sight-lines on the West (right) side of the LMC center has negative residuals (down to $-$0.5 dex), i.e., lower gas-phase metallicities for their \his column density. \\
\indent Stars and \hiis regions in the LMC do no exhibit a similar pattern in metallicity variations. While a shallow ($-$0.047$\pm$0.003 dex kpc${-1}$) radial metallicity gradient, similar to the one we derive, is observed in 1-2 Gyr old AGB stars \citep[][and references therein]{cioni2009}. The metallicity gradient for red giants in clusters in the disk is also negligible \citep{grocholski2006, grocholski2007}. Furthermore, abundances in \hiis regions show either a mild, not statistically significant gradient \citep[$-$0.03 dex kpc$^{-1}$][]{pagel1978} or no gradient at all \citep{ToribioSanCipriano2017}, albeit with sparse samples (11 and 4 \hiis regions respectively). Similarly, measurements in OB stars in N11 and NGC 2004 \citep{trundle2007}, as well as 30 Doradus \citep{markova2020} show very similar abundances, within errors. \\
\indent  On the other hand, one could argue that the increased turbulence and feedback from active star-formation on the East side of the LMC, compressed as a result of the collision with the SMC \citep{tsuge2019}, might result in increased dust destruction rates returning metals to the gas-phase faster than in the quiescent, trailing West side of the LMC. This would result in a gas-phase metallicity gradient, with the East side being more metal rich than the West side, while the total metallicity of the LMC remains uniform. This would be consistent with the finding reported in \citet{tsuge2019} that the gas-to-dust ratio is 30\% higher on the East side than the West side of the LMC. While \citet{tsuge2019} interpret this result as an indication that the East side is more metal poor as a result of the mixing with the lower metallicity SMC gas, the observed reduced dust abundance on the East side might actually result from dust processing. The gas-phase metal enhancement observed in this study on the East side compared to the West side of the LMC would be a direct consequence from metals being returned from the dust to the gas-phase as a result of this increased large-scale shock-induced dust processing.\\
\indent Metallicity gradient or variations in dust processing? The jury is still out. Given the sparse spatial coverage of \hiis regions and OB stars on the face of the LMC, it is possible that metallicity variations such as those shown in Figure \ref{map_residuals} might have been missed. Stellar abundances in the 100+ LMC young massive stars from the ULLYSES HST Director's discretionary program will provide an essential dataset to test whether the metallicity of stars formed out of the ISM in the last few millions years exhibits spatial variations similar to the ones observed in the neutral gas.\\
\indent If we are indeed looking at East-West variations in the metallicity of the ISM (dust and gas), models accounting for the dynamical interaction between the LMC, SMC, and the MW halo in their predictions for the star-formation history and chemical enrichment of the LMC will be required. A possible (albeit admittedly highly speculative) explanation is that intense star formation occurred in the last few tens of millions years ago near the leading edge of the LMC, which is compressed by ram pressure from the MW halo. This star-formation could have induced fountains of chemically enriched gas \citep{bustard2018}, which would then be ejected high enough above the clockwise rotating disk to be swept by the $\sim$250 km s$^{-1}$ motion of the LMC through the MW halo, flowing south along the \his filament, before condensing back into the disk. This process can occur on 50-100 Myr timescales \citep{kim2018}, recently enough to affect the metallicity of young stars and the ISM.

\begin{deluxetable*}{cccccc}
\centering
\tabletypesize{\scriptsize}
\tablecolumns{6}
\tablewidth{\textwidth}
\tablecaption{Multilinear regressions results for depletions as a function of hydrogen column density, hydrogen volume density, and distance to the LMC center}
\tablenum{10}
\tablehead{Elements & \# measurements & $\alpha_{\log N(H)}$  & $\alpha_{\log n(H)}$ & $\alpha_{d(center)}$  & $\beta$  \\}
 \startdata
 &&&&& \\
\hline
&&&&& \\

Fe & 32 & -0.82$^{+0.15}_{-0.15}$ & -0.13$^{+0.11}_{-0.11}$ & -0.10$^{+0.04}_{-0.04}$ & 16.45$^{+3.30}_{-3.34}$ \\
Si & 23 & -0.73$^{+0.26}_{-0.26}$ & -0.45$^{+0.19}_{-0.18}$ & -0.08$^{+0.07}_{-0.07}$ & 15.94$^{+5.64}_{-5.49}$ \\
Mg & 27 & -0.34$^{+0.13}_{-0.12}$ & -0.09$^{+0.08}_{-0.09}$ & -0.07$^{+0.03}_{-0.03}$ & 7.08$^{+2.69}_{-2.71}$ \\
Ni & 30 & -0.82$^{+0.16}_{-0.16}$ & 0.00$^{+0.12}_{-0.12}$ & -0.10$^{+0.04}_{-0.04}$ & 16.38$^{+3.50}_{-3.40}$ \\
S & 18 & -1.38$^{+0.46}_{-0.52}$ & -0.10$^{+0.25}_{-0.22}$ & -0.13$^{+0.09}_{-0.10}$ & 29.19$^{+10.85}_{-9.75}$ \\
Zn & 31 & -0.42$^{+0.13}_{-0.13}$ & -0.26$^{+0.09}_{-0.09}$ & -0.05$^{+0.03}_{-0.03}$ & 9.24$^{+2.80}_{-2.82}$ \\
Ti & 10 & -1.32$^{+0.36}_{-0.46}$ & 0.05$^{+0.29}_{-0.31}$ & -0.22$^{+0.09}_{-0.12}$ & 27.00$^{+10.18}_{-7.95}$ \\
Cr & 27 & -0.62$^{+0.19}_{-0.19}$ & -0.10$^{+0.12}_{-0.12}$ & -0.06$^{+0.05}_{-0.05}$ & 12.45$^{+4.05}_{-4.01}$ \\

\enddata
\end{deluxetable*}

\section{Gas-to-dust ratio from depletions and variations with hydrogen column density}\label{dg_section}

\indent With depletion measurements in hand, we can estimate how the dust-to-gas ratio (D/H) varies with environment in the LMC. Note that we do not include the contribution of helium here, which would increase the gas mass by 36\% from the hydrogen mass (G/D = 1.36 H/D). Since the depletions correlate in the strongest way with hydrogen column density, we focus here on the variations of D/H with N(H).  The dust-to-gas ratio, given by Equation \ref{dg_eq}, can be obtained from the fractions of each element in the gas ($10^{\delta(X)}$) and dust $(1-10^{\delta(X)})$ phases.

\begin{figure*}
\centering
\includegraphics[width=13cm]{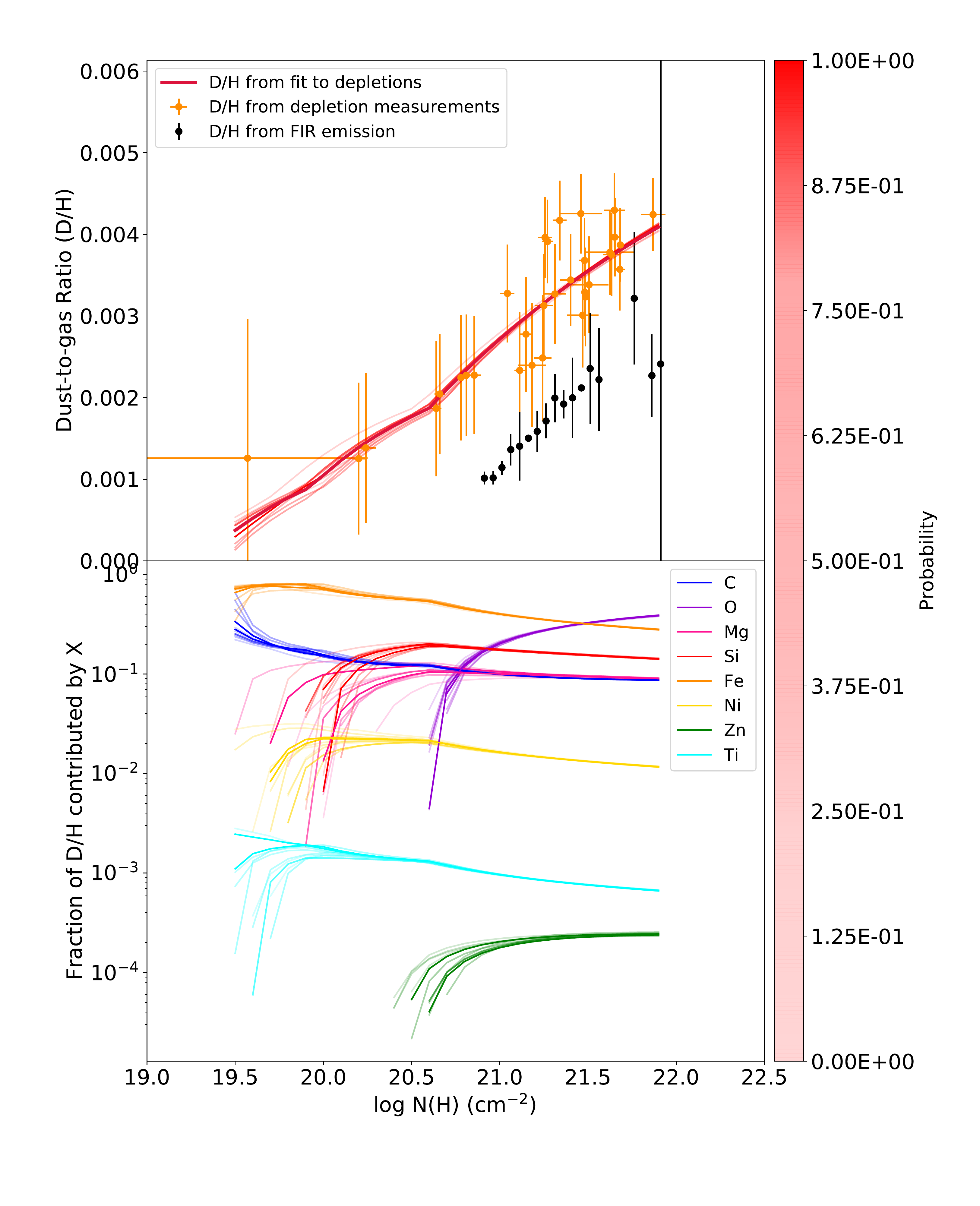}
\caption{(Top) Dust-to-gas ratio (D/H), obtained from the collection of depletions measured by the METAL program, as a function of the logarithm of the hydrogen column density, N(H) (orange points and red lines). The orange points are measurements for each sight-line, while the red lines, the transparency of which scales with the posterior probability of the realization, were obtained from the fits of the individual depletions with $\log$ N(H). For comparison, the D/H measured from FIR, 21 cm, and CO 1-0 in \citet{RD2017} is shown in black. (Bottom) Contributions of different elements X to D/H.}
\label{plot_dg}
\end{figure*}

\begin{equation}
D/H = \sum_{X} \left (1-10^{\delta(X)} \right ) A(X) W(X)
\label{dg_eq}
\end{equation}

\noindent where A(X) is the abundance of element X given in Table 2 and W(X) is the atomic weight of element X. 

\begin{figure*}
\centering
\includegraphics[width=\textwidth]{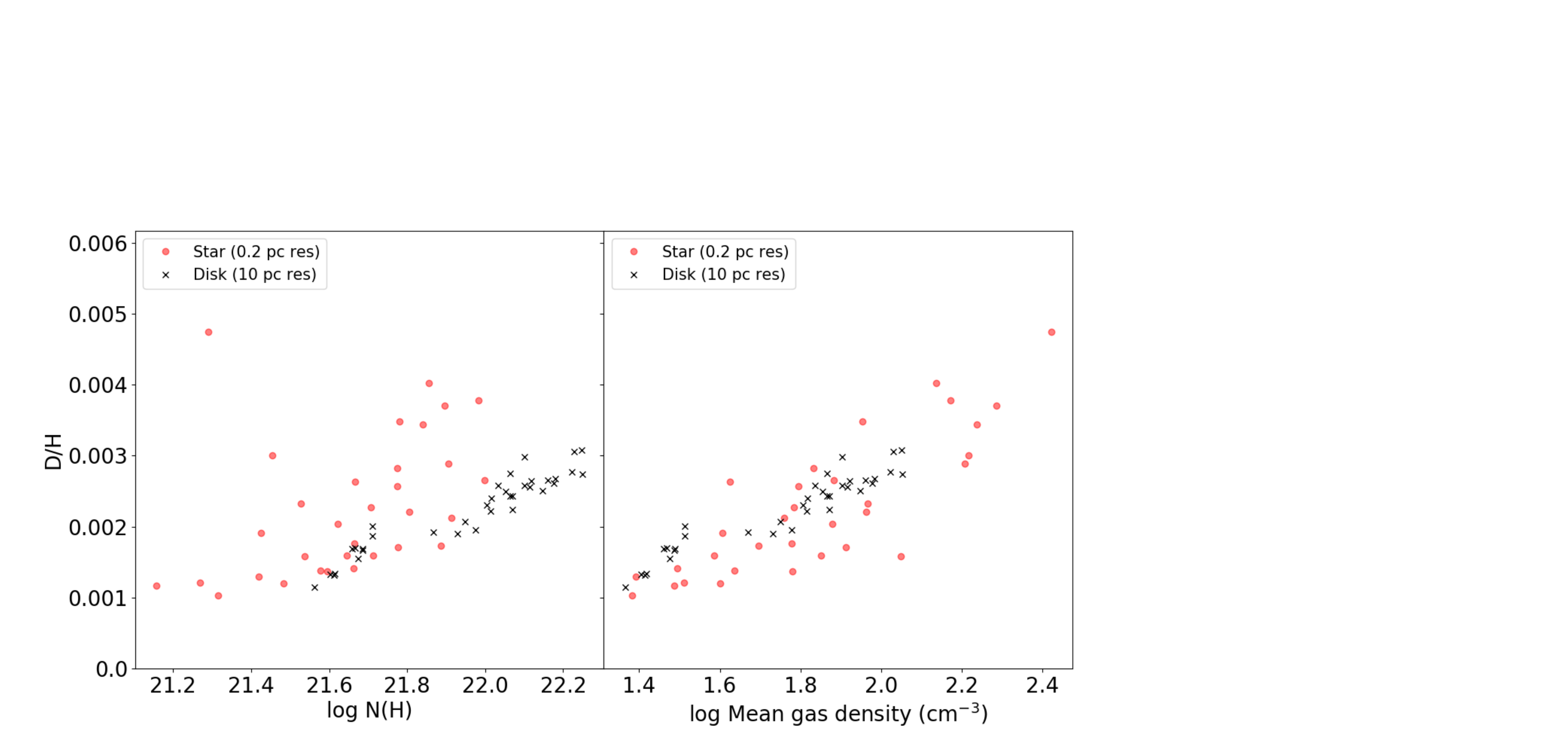}
\caption{Dust-to-gas ratio (D/H) as a function of the logarithm of hydrogen column density (left) and mean density (right), in a toy model based on a hydrodynamic simulation. In the toy model, the local dust-to-gas ratio is assumed to be dependent on the local volume density, via density-dependent timescales for the formation of dust (see Equations \ref{timescale_eq}). The red points, emulating pencil beam absorption spectroscopy measurements, were obtained by computing the integrated D/H (ratio of surface densities of dust to gas integrated along the sight-lines all the way to the depth of the source) toward 32 point sources (1 pixel wide) randomly placed in the 256$^3$, which is assumed to be 50 pc wide (0.2 pc pixels). The black points, emulating D/H measurements at the lower resolution of FIR observations, were computed by integrating the dust and gas surface densities over the entire depth of the box, as well as 50 pixels on each side (10 pc). Owing to the different regions probed by these different types of measurements and the hierarchical density structure of the ISM, the resulting integrated dust-to-gas ratio follows trends with hydrogen column density that are offset from one another. The trends agree between the two types of measurements as a function of mean volume density.}
\label{plot_dg_sim}
\end{figure*}

The METAL survey provided interstellar depletions for Mg, Si, S, Fe, Ni, Cr, Cu, Zn, with Ti from \citet{welty2010}. METAL only yielded 4 measurements of oxygen depletions, which are not sufficient to constrain variations with column density. The carbon $\lambda$1334 is heavily saturated, and the very weak $\lambda$2325 line is not detected in any of the METAL sight-lines. Carbon and oxygen are the two most abundant metals in the ISM, and hence the largest reservoir of metals for dust growth is not included in our accounting of elements in the gas and dust phases. To circumvent this limitation in our estimation of D/H as a function of hydrogen column density, we assume that the depletions of C and O as a function of iron depletions behave similarly to those in the Milky Way, where the weak C and O lines are detected and interstellar depletions can be measured. Figure 3 of \citet{jenkins2017} shows that the relation between iron depletions and those of other elements probed by METAL only very weakly vary between the MW and SMC, so this is the best assumption we can make given the data we have for C and O at the intermediate metallicity of the LMC. Nonetheless, we make a note of caution that a deficiency of carbon relative to other elements in the LMC ($\log$ C/O $=$ $-$0.56 in the LMC vs $-$0.30 in the MW, and $-$0.62 in the SMC) may potentially affect the rate of carbon depletions compared to those of other elements. Indeed, we know that the fraction of carbonaceous dust and polycyclic aromatic hydrocarbons (PAH) is different between the MW, LMC, and SMC, possibly due to the different chemical affinities of dust compounds induced by the lower carbon adundances in the Magellanic Clouds compared to the MW. \\
\indent Knowing the iron depletions for all our sight-lines, we apply the known MW relation between $\delta$(C) or $\delta$(O) and $\delta$(Fe) \citep{jenkins2009} and obtain an estimate of $\delta$(C) and $\delta$(O) for each sight-line. We then compute the D/H according to Equation \ref{dg_eq} for each sight-line. The resulting dust-to-gas ratio values are plotted in the top panel of Figure \ref{plot_dg} as orange points.\\ 
\indent In addition, we perform the same type of estimate of D/H for for each N(H) in a grid of column densities covering the observed range (log N(H) = 20---22 cm$^{-2}$). We determine the depletion of Fe, $\delta(Fe)_{\mathrm{LMC}}$, corresponding to this value of N(H) from Equation \ref{fit_dep_nh_eq} and its best-fit coefficients in Table 9 ($a_H(Fe)_{\mathrm{LMC}}$, $b_H(Fe)_{\mathrm{LMC}}$, and $z_H(Fe)_{\mathrm{LMC}}$). We then use the fitted relations between depletions of different elements, $\delta(X)_{\mathrm{LMC}}$ and that of iron, described by Equation \ref{fit_fe_dep} and its best-fit coefficients listed in Table 7 ($a(X)_{\mathrm{LMC}}$, $b(X)_{\mathrm{LMC}}$, $z(X)_{\mathrm{LMC}}$), to estimate the depletions of each element covered by METAL for each N(H) in the grid. We take this approach instead of using the fits between $\log$ N(H) and $\delta$(X) directly because the fitted correlations between $\log$ N(H) and $\delta$(Fe) on the one hand, and $\delta$(Fe) and $\delta$(X) on the other hand, are less noisy. For C and O, for which we do not have sufficient measurements in the LMC, we proceed as described above, by using the seemingly ''universal'' relationship between the depletion of iron and that of other elements. The resulting trend between D/H and N(H) is plotted as a red line in Figure \ref{plot_dg}.\\
\indent We propagate the uncertainties at each step of the computation of the D/H using a Monte-Carlo approach, by drawing realizations of each parameter ($a_H(Fe)_{\mathrm{LMC}}$, $b_H(Fe)_{\mathrm{LMC}}$, $a(X)_{\mathrm{LMC}}$, $b(X)_{\mathrm{LMC}}$, $a(C,O)_{\mathrm{MW}}$, $b(C,O)_{\mathrm{MW}}$) within their uncertainties. In Figure \ref{plot_dg}, the uncertainties on the D/H measurements for each sight-line are shown as orange error bars, while the fitted trend of D/H vs N(H) for each realization has a color proportional to its posterior probability. \\
\indent D/H in the LMC increases by a factor 4 between $\log$ N(H) $=$ 20 and 22 cm$^{-2}$, from D/H $\simeq$ 0.001 (H/D $=$ 1000) to D/H $\simeq$ 0.004 (H/D $=$ 250). This magnitude of variation is similar to findings by \citet{tchernyshyov2015} in the LMC and SMC based on a more limited sample of depletions. \\
\indent For comparison, \citet{RD2017} found from FIR and \his 21 cm emission that G/D decreases from 1500 (H/D = 1100) at $\Sigma_g$ $=$ 9 \Msu pc$^{-2}$, corresponding to $\log$ N(H) $=$ $8\times10^{20}$ cm$^{-2}$, to 500 (H/D = 370) at $\Sigma_g$ $=$ 100 \Msu pc$^{-2}$ (log N(H) $=$ 10$^{22}$ cm$^{-2}$ (black points in Figure \ref{plot_dg}), albeit with large systematic uncertainties inherent to the lack of constraints on the FIR dust emissivity and the CO-to-H$_2$ conversion factor. Figure \ref{plot_dg} shows that the slope of the relation between $\log$ N(H) and D/H is consistent between the FIR and UV-spectroscopy based depletions, but the zero-point of the FIR-based trend lies a factor of 2 lower than the depletion-based trend. Two possible effects could explain this difference.\\
\indent First, as pointed out earlier in this paper, and also by author studies \citep[e.g.,][]{RD2014, clark2019}, the dust FIR opacity is degenerate with dust surface density: even if the dust temperature is known from fitting the FIR SED, the FIR surface brightness at a given wavelength is still proportional to the product of opacity and dust surface density \citep[see, e.g., Equation 7 of][]{gordon2014}. Furthermore, the dust opacity can vary significantly with environment \citep[e.g.][]{stepnik2003, kohler2012, demyk2017}. A factor of 2 systematic uncertainty on FIR opacity, and therefore on dust surface densities derived from FIR maps, would not be surprising at all. It is therefore possible that the offset seen in Figure \ref{plot_dg} between the D/H derived from depletions and FIR maps is simply due to an uncertain assumption on the FIR opacity.\\
\indent Second, the geometrical set-up is fundamentally different between pencil-beam UV spectroscopy toward massive stars randomly distributed in the dust disk and FIR observations in a 10-15 pc beam probing the interstellar dust through the entire depth of the disk. Already, a comparison of \his column densities derived from fitting the damping wings of the Ly-$\alpha$ line at 1216 \mAA toward massive stars and those derived from 1' resolution (15 pc) 21 cm observation yield vastly different column densities. To evaluate the magnitude of these geometric effects on the D/H vs N(H) trends, we used a prototypical hydrodynamic simulation from \citet{federrath2010} to create a simple toy model. We scale the 256 pixel box to a mean density of 20 cm$^{-3}$, and assume that a pixel in gas density cube is 0.2 pc wide. Following the chemical evolution model by \citet{asano2013}, we furthermore assume that the G/D is given by G/D $=$ (G/D)$_0$ $f_0/f_{\mathrm{dust}}$, where $f_0$ is the fraction of metals locked in dust at (G/D)$_0$ (we assume $f_0$ = 0.15 and (G/D)$_0$ $=$ 1200 in the LMC), and $f_{\mathrm{dust}}$ is given by Equations 16 of \citet{RD2017}, repeated here for clarity:

\begin{equation}
f_{\mathrm{dust}} = f_0  \frac{ \exp{\left (\frac{t}{\tau_{acc}}  \right )} }   {1-f_0 + f_0 \exp{\left ( \frac{t}{\tau_{acc}} \right ) } }
\label{gd_mod_eq}
\end{equation}

\noindent and

\begin{multline}
\left ( \frac{\tau_{acc}}{yr} \right ) = 2\times 10^7 \times \\
\left ( \frac{<a>}{0.1 \mu m} \right ) \left (\frac{n_H}{100 \mathrm{cm}^{-3}} \right )^{-1} \left (\frac{T_d}{50 K} \right)^{-\frac{1}{2}} \left (\frac{Z}{0.02} \right )^{-1}
\label{timescale_eq}
\end{multline}

\noindent  with Z $=$ 0.01 for the LMC (half solar), and we assume $t$ $=$1Gyr. This simple model basically assumes that dust growth with a timescale inversely proportional to density and metallicity, and this allows us to compute the H/D and $n_{\mathrm{dust}}$ $=$ $n_{\mathrm{gas}}$ $\times$ (D/H) in each model cell based on its gas density.\\
\indent We place 32 random sight-lines in the simulation box such that their hydrogen column density is between 10$^{20}$ and 10$^{22}$ cm$^{-2}$ as in our sample, and compute the surface density of gas $\Sigma(H)$, the surface density of dust $\Sigma_{\mathrm{dust}}$ and the D/H $=$ $\Sigma_{\mathrm{dust}}$/$\Sigma_{\mathrm{gas}}$ toward each sight-line. To emulate the nature and lower spatial resolution of FIR observations, we re-bin the simulated cube to a 5$^3$ cube of 10 pc wide pixels, and compute surface densities of gas and dust and the dust-to-gas ratio through the entire depth of this simulated box.  The resulting trends of D/H vs hydrogen column density in this simple toy model are shown in Figure \ref{plot_dg_sim}. Owing to the different regions probed by these different types of measurements (absorption spectroscopy toward point sources in the disk vs FIR emission photometry through the entire disk) and the hierarchical density structure of the ISM, the relation between D/H and $\log$ N(H) recovered from the randomly placed pencil-bean sight-lines and the coarser resolution integrated measurements are offset by a factor $>$1.5. Similar geometric effects could be at play in our comparison of D/H derived from depletions using UV spectroscopy to massive stars and from FIR+21cm+CO emission data, and thus contribute to the zero-point offset in the trends between N(H) and D/H obtained from UV depletions and FIR maps. \\
\indent This toy model furthermore shows that a dependence on volume density of the dust-to-gas ratio will transpire as a column density dependence once integrated along the line-of-sight. The density dependence would be recovered if one could measure the mean density along the line of sight, but that is not the case as the density traced by the \cis fine structure lines is only representative of the \cis gas, which represents a small fraction of the gas along the line of sight.
\indent In the bottom panel of Figure \ref{plot_dg}, we show the fractional contribution of each element to the D/H, as a function of hydrogen column density. This figure provides some constraints about the composition of dust expected from depletion measurements. In particular, the mass contributions of refractory elements, such as Fe, Ti, Ni, and that of carbon, decrease with increasing column density. Conversely, volatile elements such as O, S, Si, Mg, Zn increasingly contribute to the dust mass budget as the column density of ISM increases. In particular, O, S, and Zn only contribute above $\log$ N(H) $>$ 21 cm$^{-2}$.

\section{Summary}\label{summary_section}

\indent In this paper, we presented the analysis of the 32 UV medium-resolution COS and STIS spectra toward massive stars in the half-solar metallicity LMC taken as part of the METAL large program. From the spectra, chemical abundances and interstellar depletions (fraction of an element in the gas-phase) for key constituents of dust in neutral gas (Mg, Si, Fe, Ni, S, Zn, Cr, and Cu) are derived using the apparent optical depth method. For each sight-line, we also derive the volume density, radiation field, and electron density from the \cis fine structure lines and an estimate of the \ci/\ciis line ratio. Combined with previously determined atomic and molecular hydrogen column densities, this allows us to probe the environmental dependence of depletions. \\
\indent We find that the depletions of different elements are tightly correlated with each other (even after accounting for covariant errors), implying that 1) depletions respond to the same environmental parameters causing their variations, and 2) depletions for an element (e.g., Fe for which many lines of different oscillator strengths are accessible throughout the UV) can be used as a proxy for depletions for another element. For all elements, depletions decrease with increasing hydrogen column density, at rates that do not systematically depend on condensation temperature.\\
\indent We find that hydrogen column density is the main driver for depletion variations, even after accounting for the covariant errors between depletions and hydrogen column density. Within the parameter space probed by METAL, no correlations with molecular fraction or radiation field intensity are found. There are, however, secondary, weaker correlations with volume density, derived from the \cis fine structure lines, and distance to the LMC center ($-$0.1 dex kpc$^{-1}$), located about 1 kpc West of the massive star-forming region 30 Doradus. Theoretically, one would expect the dust-to-gas ratio to correlate mainly with volume density. However, the density derived by the \cis fine structure lines only traces a small fraction of the gas along the line of sight (dominated by \ciis, not \ci), while the mean density is not observable. Hence, it is not too surprising that depletions are not well correlated with the volume density derived from the \cis lines, but better correlated with hydrogen column density, which is a good proxy for the average volume density along the line of sight for a galaxy seen face-on. A simple simulation furthermore shows that a purely volume density-dependent dust-to-gas ratio will transpire as correlated with column density as well, once integrated along the line of sight. 
\indent The two-dimensional distribution of the gas metallicity reveals that the gradient in depletions with distance to the LMC center results from an East-West pattern in metallicity, with gas in the \hi-rich East side being more enriched than on West side. The split in metallicity occurs roughly along the line perpendicular to the leading edge of the LMC. On the compressed side of the LMC traced by the Southeast \his overdensity \citep[SEHO][]{nidever2008}, where intense star-formation is occurring (in 30 Dor for example), and has occurred in the last few ten millions years (as evidences by the super-shells), the gas is more enriched in metals (by up to $+$0.3 dex), while gas on the western side of the LMC reaches gas-phase metallicities lower by up to $-$0.5 dex given their hydrogen column density. This could be explained by enhanced depletion effects on the East side due to the increased feedback from ongoing star-formation, and/or metallicity differences between the two sides of the LMC, perhaps due to enhanced star-formation on the compressed side combined with the gas motions in the halo. Either way, explanations for this distinct pattern will require measurements of abundances in young massive stars probed by the ULLYSES HST Director's Discretionary program and through modeling of gas flows in and around the LMC. \\
\indent Combining the depletions for different elements, we compute the dust-to-gas ratio (hydrogen only, D/H) as a function of hydrogen column density. D/H in the LMC increases by a factor 4, from  D/H $\simeq$ 0.001 (H/D $=$ 1000) at $\log$ N(H) $=$ 20, to D/H $\simeq$ 0.004 (H/D $=$ 250) at $\log$ N(H) $=$ 22 cm$^{-2}$. We therefore confirm the factor 3-4 increase in dust-to-metal and dust-to-gas ratios between the diffuse and translucent/molecular ISM observed from far-infrared, 21 cm, and CO 1-0 observations in the LMC. While the slope of the relation between hydrogen column density and D/H is consistent between depletions and FIR measurements, the zero point of the relation based on UV depletions lies a factor of 2 higher compared to the FIR measurements. This discrepancy can be explained by systematic uncertainties in the FIR opacity of dust assumed to convert the surface brightness to surface density measurements, and/or by geometrical effects combining the hierarchical nature of the ISM and the different volumes probed by the FIR measurements (corresponding to $\sim$ 10 pc pixels on the sky probing the entire depth of the gas $+$ dust disk) and the D/H measurements based on UV spectroscopy (pencil-beam sight-lines scattered with the depth of the disk).\\
\indent Future work based on the parallel imaging obtained as part of METAL to produce extinction maps in the LMC and compare them to FIR emission based obtain with the {\it Spitzer} and {\it Herschel} space telescopes will allow us to characterize the FIR opacity of dust and its variations in this galaxy. This work will in turn substantially reduce the systematic uncertainties on D/H measured from FIR, and disentangle opacity and geometric effects responsible for the factor 2 discrepancy between D/H measured from FIR and from UV-spectroscopy-based depletions. \\
\indent Additionally, an upcoming paper will examine how the relation between dust-to-metal ratio and environment changes at different metallicities, based on similar depletion samples in the Milky Way \citep{jenkins2009} and Small Magellanic Cloud \citep{jenkins2017}. The characterization of depletion patterns at a range of metallicities will provide key information for interpreting metallicity measurements in DLAs, reduce systematic uncertainties on dust-based gas mass estimates in galaxies in the nearby and high-redshift universe, and constrain chemical evolution models. 

\acknowledgments{We thank the referee for providing insightful comments and suggestions that contributed to validating the robustness of the results. We thank Chris Evans and Danny Lennon for discussing and providing literature references for abundances of OB stars in the LMC. Edward B. Jenkins was supported by grant HST-GO-14675.004-A to Princeton University.  Benjamin Williams acknowledges support from grant HST-GO-14675.009-A to the University of Washington. Lea Hagen and Karl Gordon were supported by grant HST-GO-14675.002-A to the Space Telescope Science Institute. Karin Sandstrom and Petia Yanchulova Merica-Jones acknowledge support from grant HST-GO-14675.008-A to the University of California, San Diego. This work is based on observations with the NASA/ESA Hubble Space Telescope obtained at the Space Telescope Science Institute, which is operated by the Associations of Universities for Research in Astronomy, Incorporated, under NASA contract NAS5-26555. These observations are associated with program 14675. Support for Program number 14675  was provided by NASA through a grant from the Space Telescope Science Institute, which is operated by the Association of Universities for Research in Astronomy, Incorporated, under NASA contract NAS5-26555. This paper makes uses of the {\it corner} python code \citep{corner}.
}

\bibliographystyle{/users/duval/stsci_research/bibtex/apj11}
\bibliography{/Users/duval/stsci_research/biblio_all}

\section{Appendix A: Validation of the measurements}

\subsection{Impact of the component spacing in the profile fitting}

\begin{figure*}
\centering
\includegraphics[width=8cm]{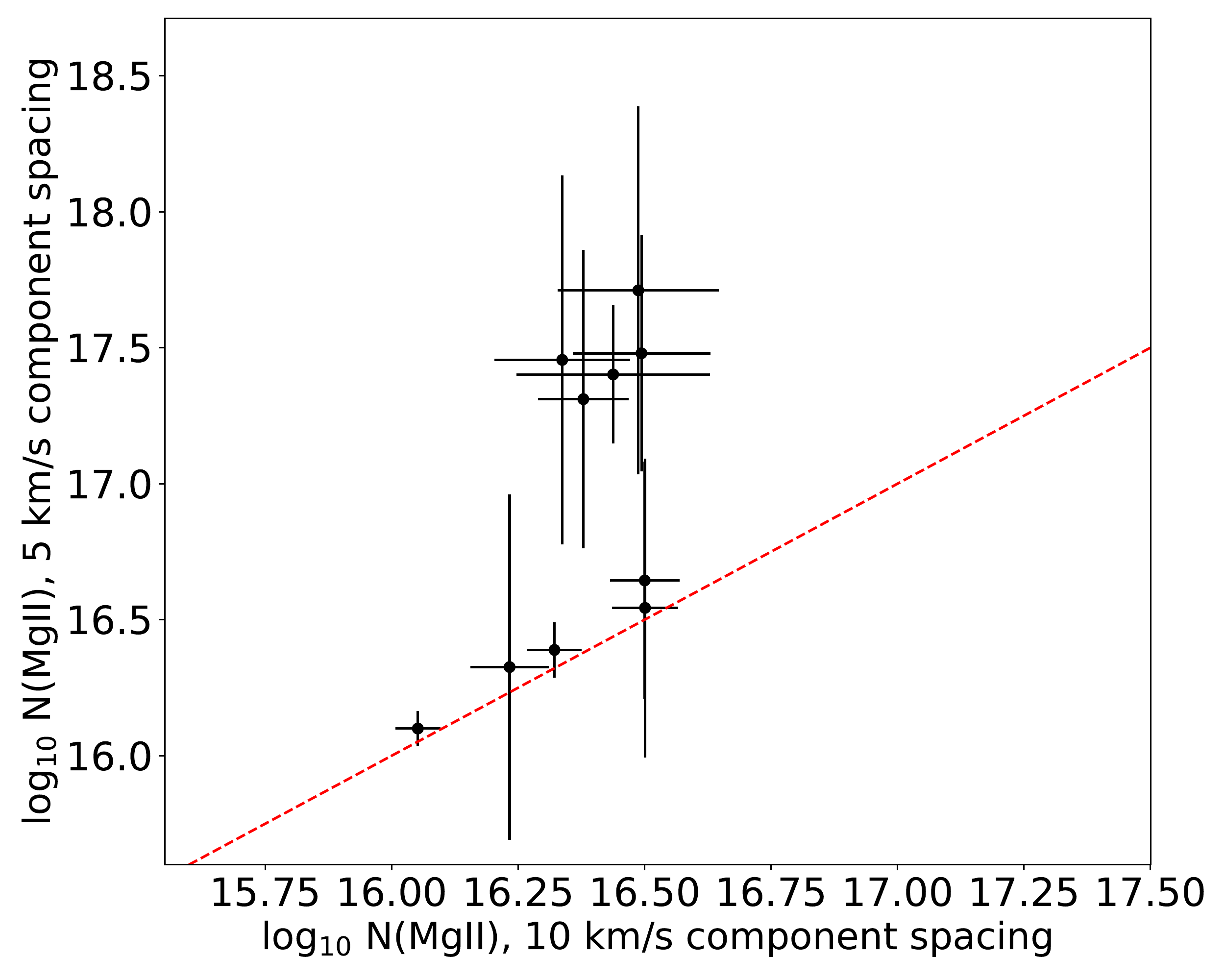}
\includegraphics[width=8cm]{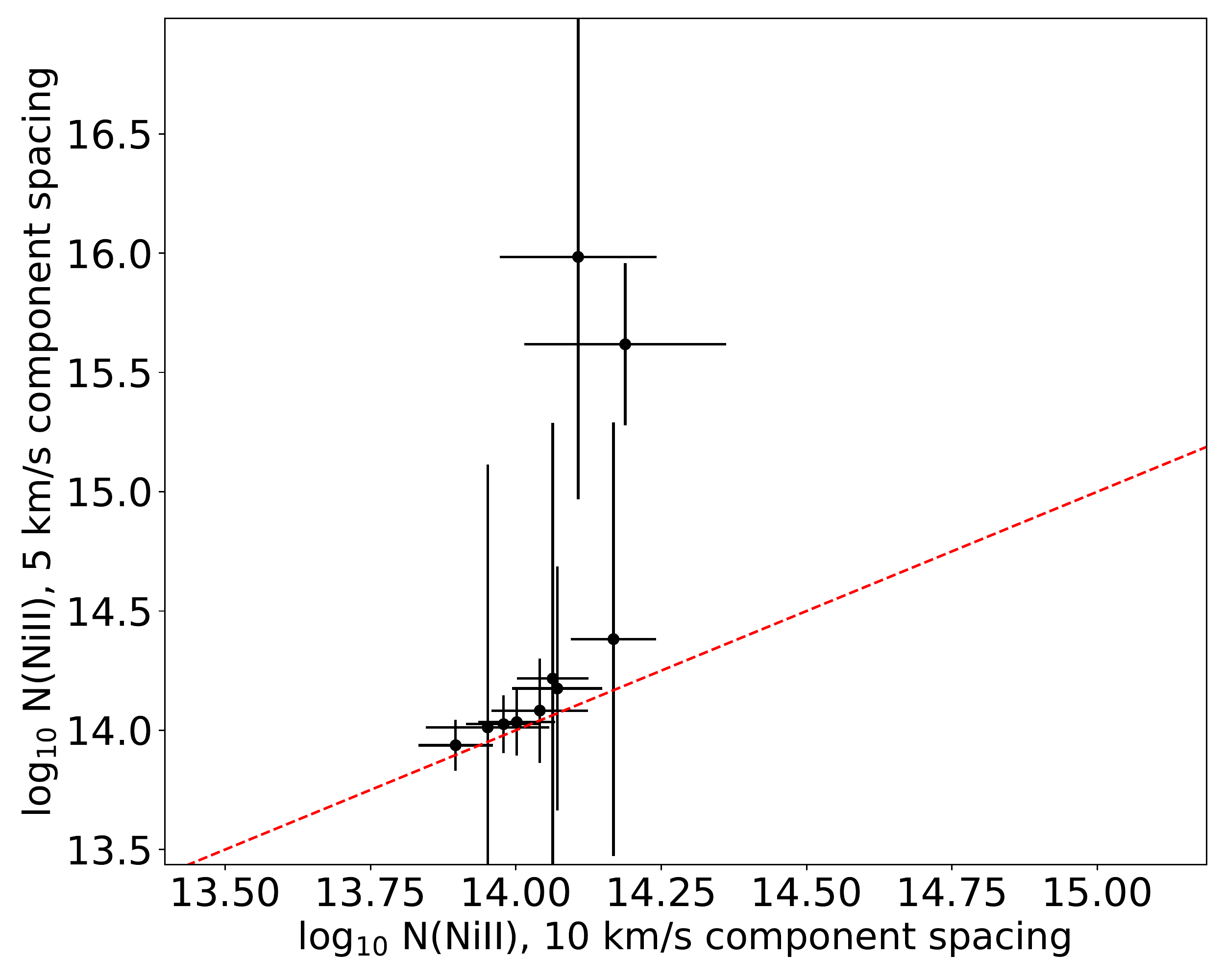}
\includegraphics[width=8cm]{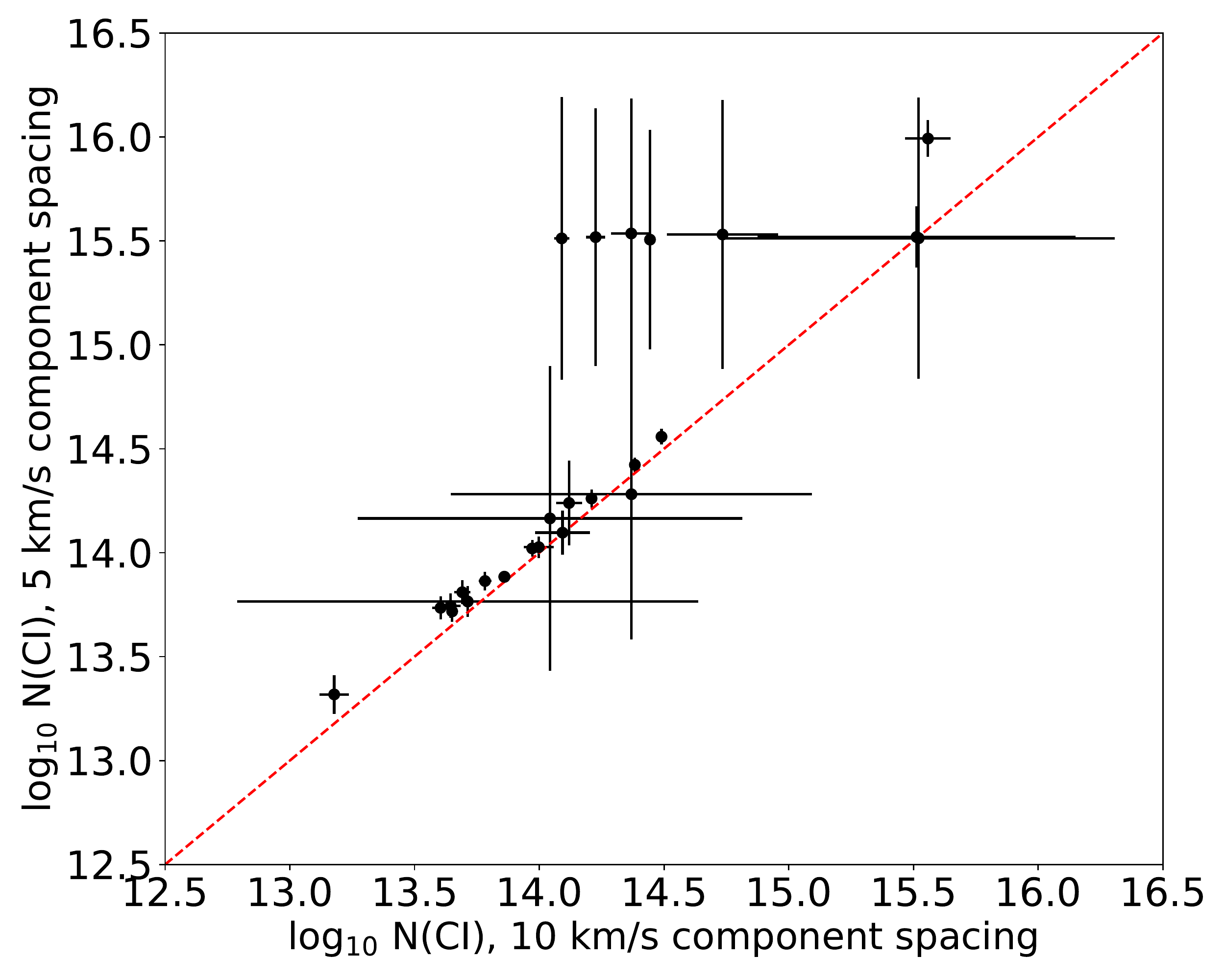}
\includegraphics[width=8cm]{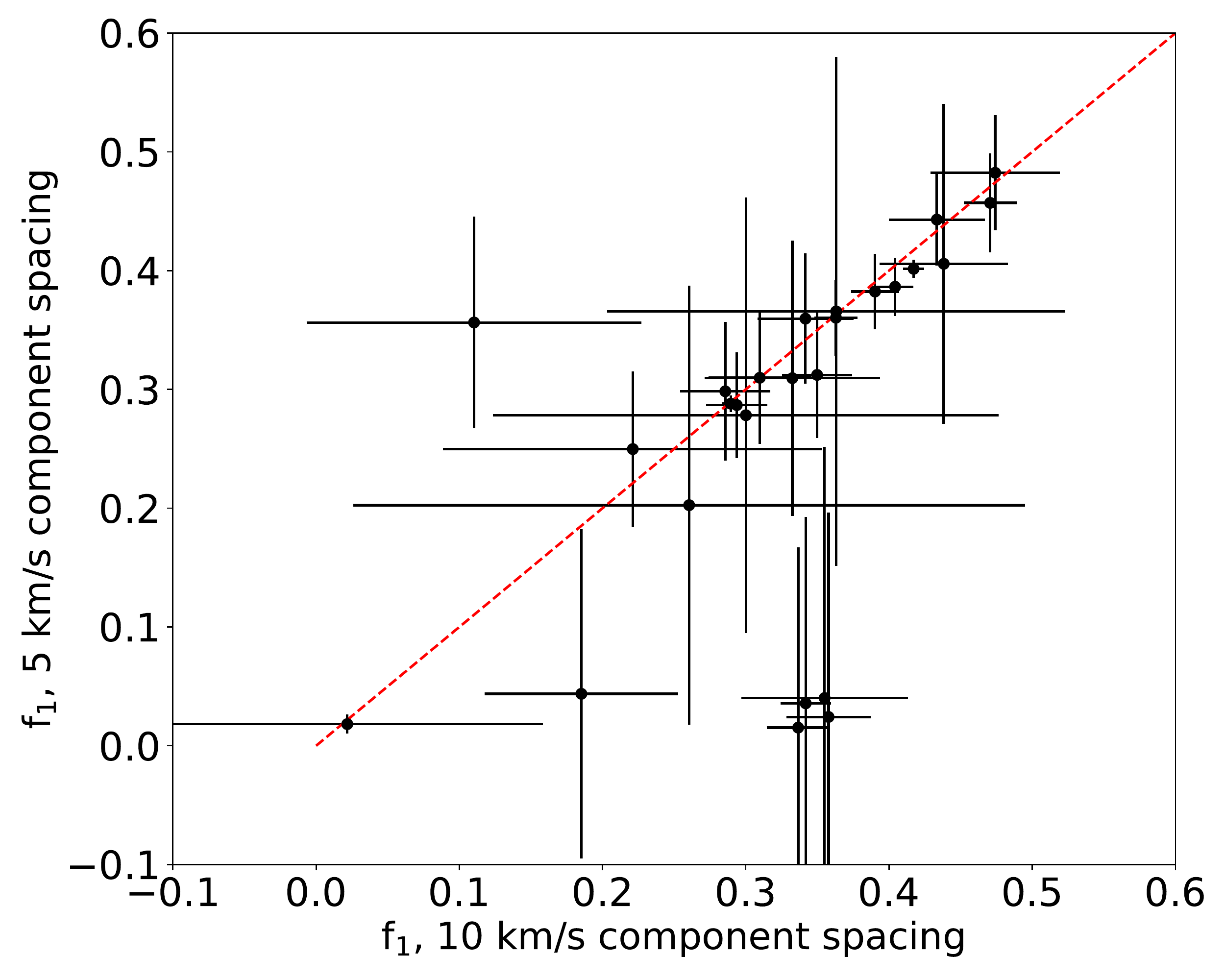}
\includegraphics[width=8cm]{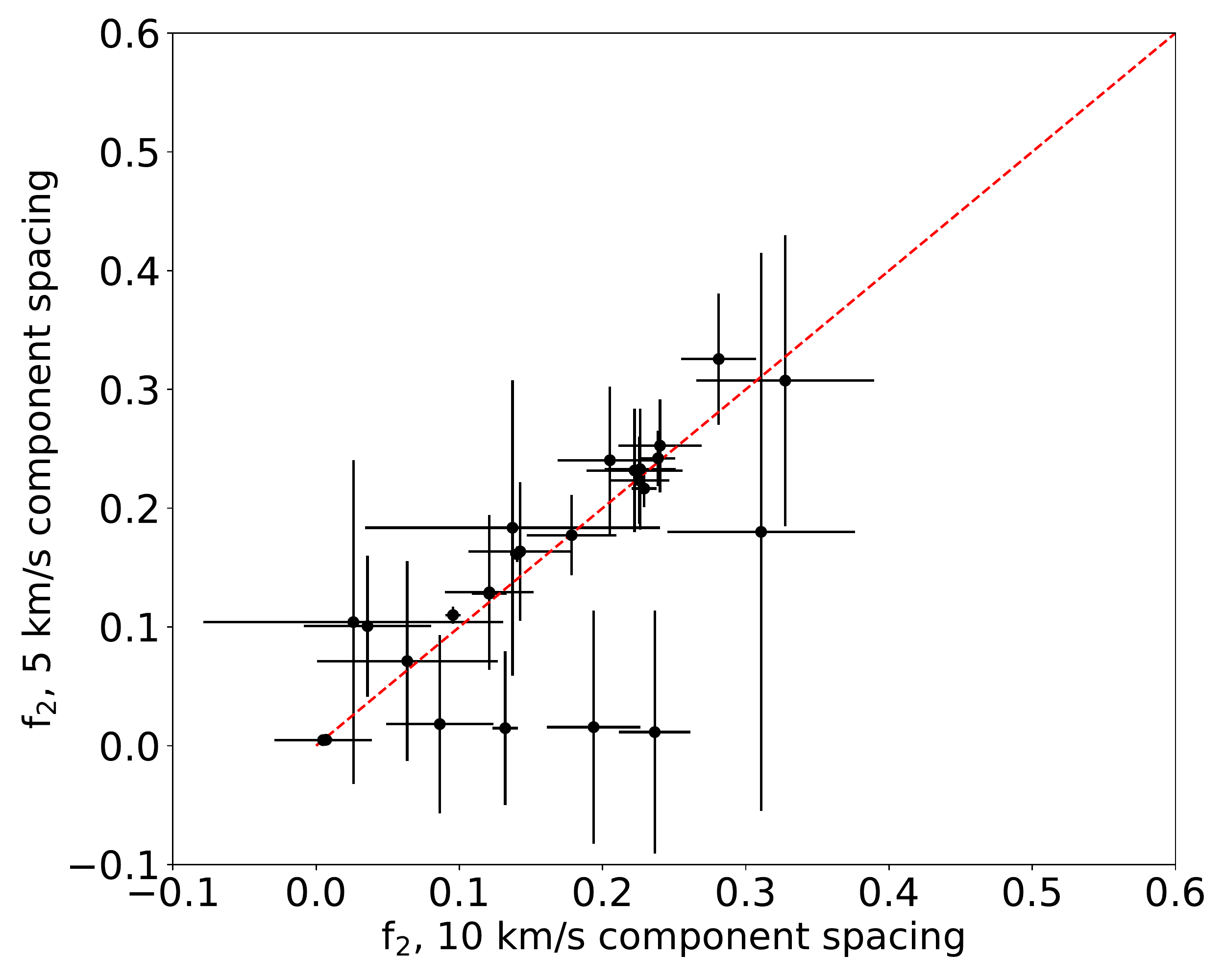}

\caption{Comparison of column densities derived from the profile fitting method applied to COS spectra with 10 km s$^{-1}$ component spacing (x-axis) and 5 km s$^{-1}$ spacing (y-axis) for \mgiis (top left), \niiis (top right), \cis (middle left), $f_1$ $=$ $N$(\ci$^*$)/$N$(\ci)$_{\mathrm{tot}}$ (middle right) and $f_2$ $=$ $N$(\ci$^{**}$/$N$(\ci)$_{\mathrm{tot}}$ (bottom). The red dashed line indicates a 1:1 comparison.}
\label{plot_5_10kms}
\end{figure*}

\indent In fitting the COS line profiles (\mgii, \niii, \ci), we nominally use a component spacing of 10 km s$^{-1}$. One might anticipate that reducing the component spacing could lead to larger column densities since smaller $b$ values would be associated with more tightly spaced components. To evaluate the impact of the choice of component spacing on the column density outcomes, we have performed the profile fitting on the COS spectra for the \mgii, \niii, and \cis lines with 5 km $^{-1}$ component spacing. The comparison of the results for 10 km $^{-1}$ and 5 km $^{-1}$ component spacing is shown in Figure \ref{plot_5_10kms}.\\
\indent   For \niiis, the column density measurements with 5 and 10 km s$^{-1}$ component spacing are in excellent agreement for all but 2 sight-lines (BI 184 and BI 253), for which the column density derived with a 5 km s$^{-1}$ component spacing exceeds the column density derived from 10 km s$^{-1}$ spacing by 1-1.5 dex. The uncertainties are nonetheless much larger in the 5 km s$^{-1}$ case, and thus both results are within uncertainties. Furthermore, the AOD method applied to the \niiis lines in the STIS NUV spectra (1709, 1741, and 1751 \AA) for both sight-lines yields column density outcomes that are in near perfect agreement (within 0.1 dex) of the column densities derived from the profile fitting applied to the COS FUV spectra (1317, 1370 \AA)  with the more realistic 10 km s$^{-1}$ component spacing. \\
\indent For \mgii, 5 out of 10 sight-lines show excellent agreement between the results obtained with 5 and 10 km s$^{-1}$ component spacings, while for the other half (Sk-69 279, Sk-66 19, Sk-68 73, BI 253, Sk-68 140), the tighter component spacing yields column densities 0.1 dex in excess of the ones derived from the 10 km s$^{-1}$ spacing (albeit still within uncertainties given the larger errors associated with the 5 km s$^{-1}$ spacing). Thus, it is possible that our \mgiis values derived from profile fitting with 10 km s$^{-1}$ applied to COS spectra underestimate the true column densities. However, we deem the values derived from a 5 km s$^{-1}$ spacing implausible given the excellent agreement between the values derived from the AOD applied to STIS spectra for \niiis and their counterpart derived from COS spectra and the profile fitting method with 10 km s$^{-1}$ component spacing.\\
\indent Finally, for \ci, the values of $N$(\ci)$_{\mathrm{tot}}$, $f_1$ and $f_2$ are all within uncertainties whether a component spacing of 5 km s$^{-1}$ or 10 km s$^{-1}$ is assumed.

\subsection{Comparison of column densities derived from the AOD and profile fitting applied to STIS spectra}

\begin{figure}
\centering
\includegraphics[width=8cm]{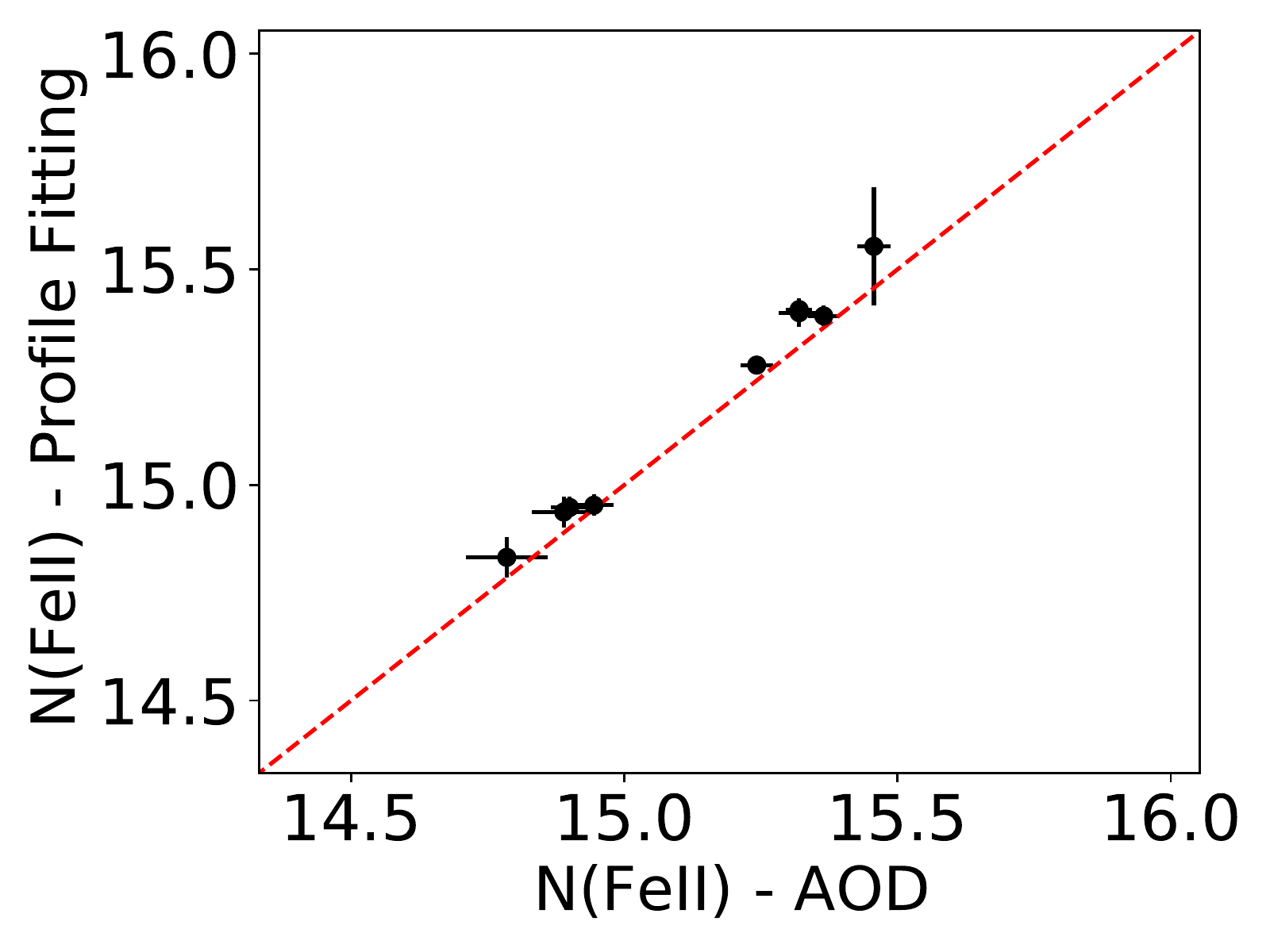}
\includegraphics[width=8cm]{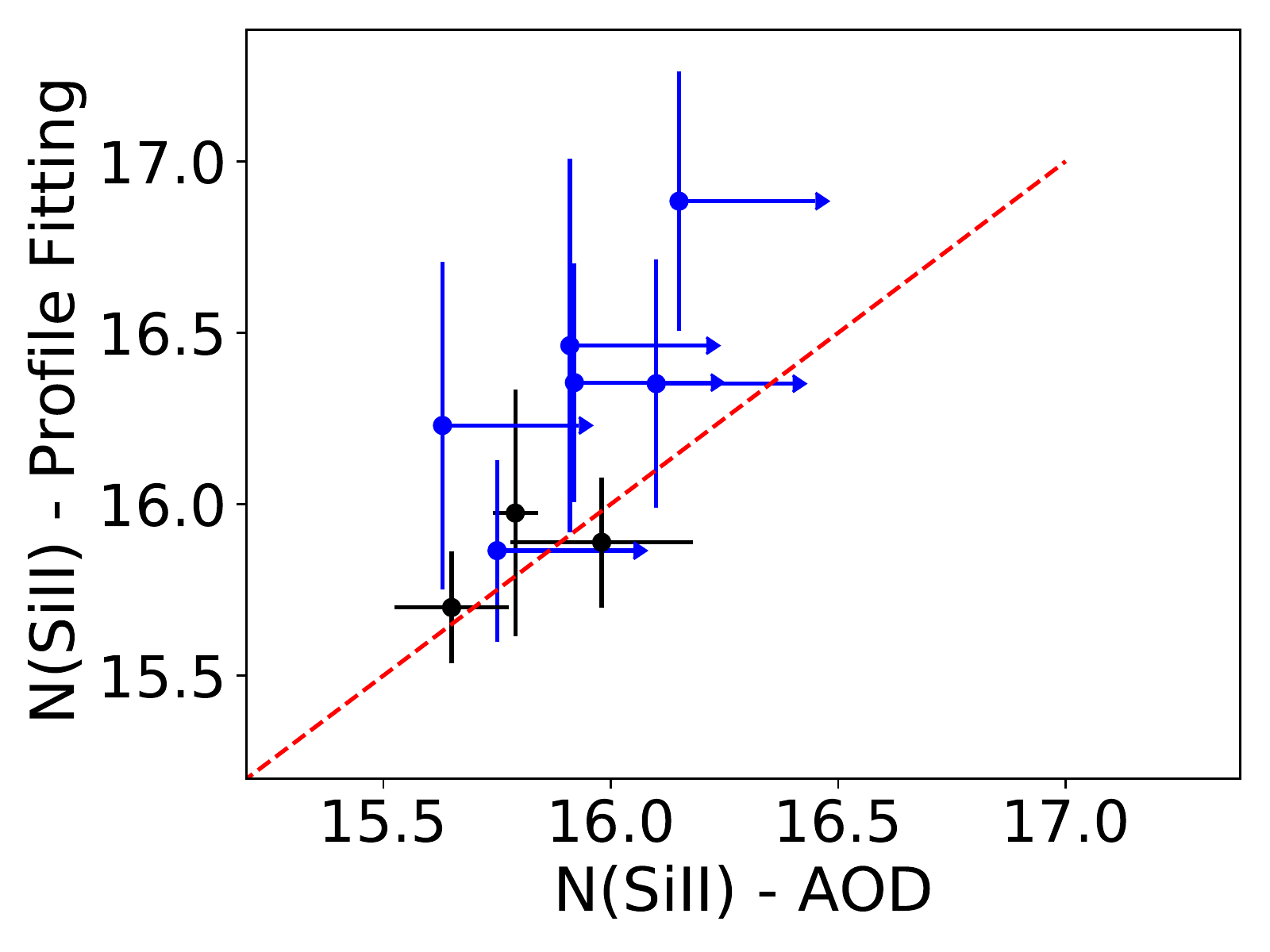}

\caption{Comparison of \feiis (left) and \siiis (right) column densities derived from the AOD (x-axis) and profile fitting (y-axis) methods applied to STIS spectra (with 10 km s$^{-1}$ component spacing for the profile fitting method). The red dashed line indicates a 1:1 comparison.}
\label{plot_aod_profile_fitting}
\end{figure}

\indent Owing to the different spectral resolutions of COS and STIS, different approaches were selected to determine column densities (AOD for the higher resolution STIS spectra, profile fitting for the lower resolution COS spectra). In order to evaluate systematic differences between the two methods, we applied the profile fitting method to the \feiis and \siiis lines for a few STIS spectra. These lines were selected as benchmarks because 1) \feiis column densities are very robustly determined from the AOD thanks to the many lines available and the lack of saturation deduced from the consistency between weak and strong lines, and 2) \siiis column density measurements with the AOD are difficult due to the availability of only one transition ($\lambda$ 1808 \AA), requiring further analysis involving a COG analysis to several elements (see Section \ref{section_siii}).\\
\indent The results of this comparison are shown in Figure \ref{plot_aod_profile_fitting}. For both lines, the column density derived from both methods are in very good agreement, as well as within uncertainties.

\subsection{Comparison between COS-, STIS-, and FUSE-based column densities}\label{compare_cos_stis}

\begin{figure}
\centering
\includegraphics[width=10cm]{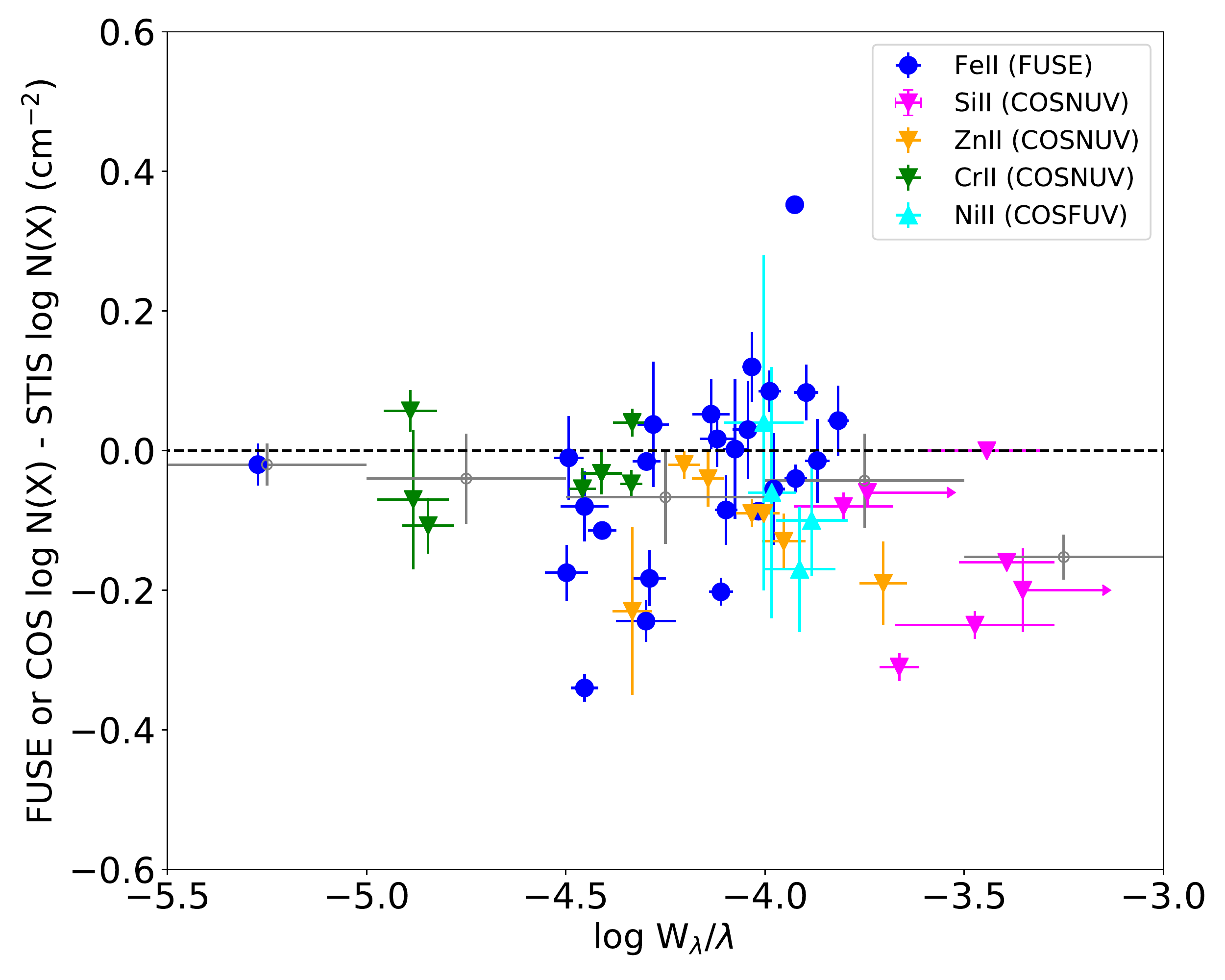}
\caption{Difference between column densities of various low ions $X$ measured in COS or FUSE spectra ($R$ $\sim$ 15,000-20,000) with respect to column densities measured in higher resolution STIS spectra (E140M, E230M, E140H, or E230H, $R$ $>$ 30,000), as a function of the equivalent width $W_{\lambda}/\lambda$ measured with the higher resolution STIS instrument. The FUSE and COS/NUV-based column densities are from \citet{tchernyshyov2015}, while the COS/FUV-based column densities are measured in METAL spectra (\niiis lines only). All STIS-based measurements are from METAL (Table 3). The empty gray circles represent binned averages, with the error in x being the width of the bin and the error in y the error on the mean in each bin. As the central optical depth (and hence equivalent width) increases, the lower resolution data (FUSE or COS) increasingly underestimates the true column density (as estimated from the higher resolution STIS data), due to unresolved saturation.  }
\label{plot_comparison_cos_fuse_stis}
\end{figure}

\indent A few sight-lines in the METAL sample were observed with COS in the FUV (G130M and G160M), and STIS/E230M/1978 in the NUV \citep[see Table 7 in][]{RD2019}. For those sight-lines, the \niiis column densities estimated from the $\lambda$1317 \mAA and $\lambda$1370 \mAA lines in the COS FUV spectra (R $\sim$ 18,000), and from the $\lambda$1709 \AA, $\lambda$1741 \AA, and $\lambda$1751 \mAA in the STIS E230M spectra (R $\sim$ 30,000), can be compared in order to evaluate the effects of unresolved saturation on column density measurements in lower resolution spectra. Additionally, \citet{tchernyshyov2015} measured column densities of \feii, \siii, \znii, and \criis in FUSE and COS NUV spectra, a few of which were also observed with STIS as part of the METAL survey. We can therefore also compare the column density measurements of these ions obtained at two different resolutions. The central optical depth of a line, $\tau_0$, is determined by the column density, oscillator strength $f_{\lambda}$, and velocity dispersion $b$ through $\tau_0 = 1.497\times10^{-15} N f_{\lambda}\lambda/b$. The velocity dispersions for our sight-lines are not known, but assuming that they do not exhibit significant variations from one case to the next, we can state that $W_\lambda/\lambda = 2 b F(\tau_0)/c$, where the function F increases monotonically \citep[see Eq. 4 of][]{jenkins1996}. We thus compare the column densities derived from higher resolution (STIS) observations to those from lower resolution (COS, FUSE) as a function of equivalent width. The results of these comparisons are shown in Figure \ref{plot_comparison_cos_fuse_stis}.\\
\indent In Figure \ref{plot_comparison_cos_fuse_stis}, it is clear that, as the strength of a line increases, measurements made in lower resolution spectra (FUSE, COS) increasingly underestimate the column density obtained from the higher resolution (STIS) spectra. While the vast majority of column density measurements presented here were performed using STIS spectra, the effects of unresolved saturation constitute a limitation of our measurements for a few sight-lines for which the \mgii, \siis and/or \niiis column densities were measured with COS only (see Table 3).  \\
\indent For \mgii, column densities derived from COS range between $\log$ N(\mgii) $=$ 15.77 and 16.06 cm$^{-2}$. With $\log f_{\lambda} \lambda$ $=$ $-$0.106 \mAA for the $\lambda$1239 \mAA line, \mgiis measurements occupy the range $\log W_{\lambda}/\lambda$ $<$ $-$4, which empirically appears to be less susceptible to unresolved saturation according to the differences in column densities derived at different resolutions in Figure \ref{plot_comparison_cos_fuse_stis}. All the \siis column density measurements (except one) derived from COS only yielded lower limits and thus account for the significant effects of saturation. Lastly, the \niiis measurements derived from COS spectra only ($\lambda$ 1370 \mAA line with $\log f_{\lambda} \lambda$ $=$ 2.02 \AA) have a range of COS-based $W_{\lambda}/\lambda$ $=$ $-$4 to $-$3.4 , which is similar to the range of equivalent widths observed in the sight-lines probed with both COS and STIS, and which exhibit significant unresolved saturation (with underestimation of the COS column density by up to 0.5 dex with respect to STIS, albeit with large error bars). Thus, the effects of saturation in our COS-based \niiis column density estimates are probably relatively large, but our large error bars should capture this uncertainty.

\begin{figure}
\centering
\includegraphics[width=10cm]{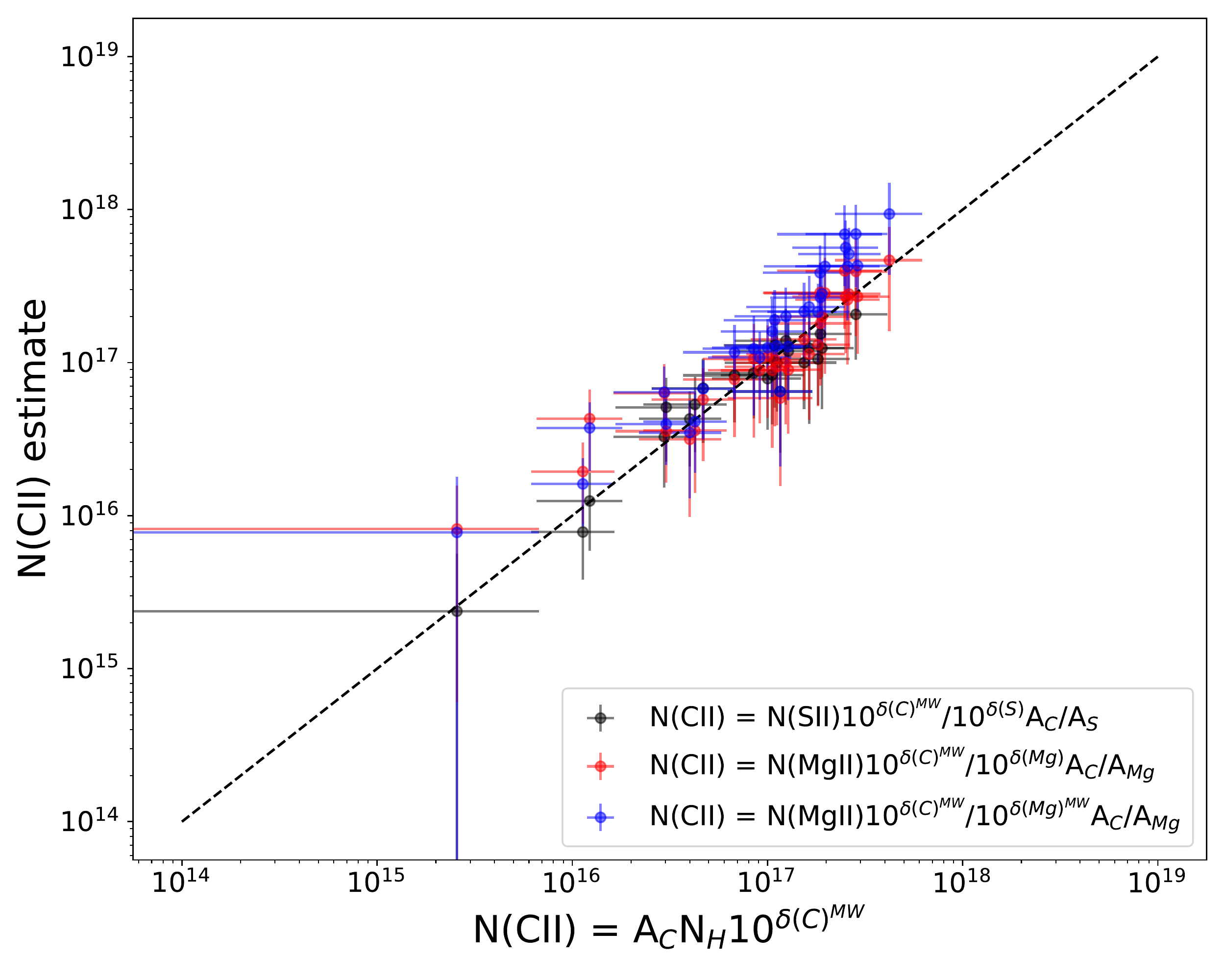}
\caption{Comparison of the \ciis column density estimates obtained from different methods: 1) scaling the hydrogen column density by the stellar abundance of carbon and carbon depletion, obtained from the relation between iron and carbon depletions in the Milky Way and the iron depletion measured in the LMC (x-axis), 2) scaling the column densities of magnesium (red)  and sulfur (black) measured in the LMC with the ratio of the LMC carbon/magnesium or carbon/sulfur stellar abundances and the ratio of the carbon and magnesium or sulfur depletions (y-axis). In both cases, the carbon depletions are estimated from the iron depletions measured in the LMC and the relation between iron and carbon depletions established in the Milky Way \citep{jenkins2009}. The blue points correspond to the second approach, but in this case, the magnesium depletions are estimated from the iron depletions measured in the LMC and the relation between iron and magnesium depletions established in the Milky Way by \citep{jenkins2009}.}
\label{plot_cii_estimates}
\end{figure}

\subsection{Comparison of different estimations of the column density of \cii}

\indent in calculating the radiation field intensity and volume density in Section \ref{ci_section}, we estimate the column density of \ciis by scaling the hydrogen column density with the total (stellar) carbon abundance in the LMC and the depletion of carbon derived from the relation between iron depletion and carbon depletion established in the Milky Way \citep{jenkins2009}. To validate this estimation, we have compared this approach with scaling the \mgiis and \siis column densities measured in the LMC (Table 5) with the ratio of the LMC carbon/magnesium or carbon/sulfur stellar abundances and the ratio of the carbon and magnesium or sulfur depletions. In this case also, the carbon depletions are estimated from the iron depletions measured in the LMC and the relation between iron and carbon depletions established in the Milky Way \citep{jenkins2009}. The magnesium and sulfur depletions are estimated from the relation between iron and magnesium (or sulfur) depletions established in the LMC from this study (Table 7).\\ 
\indent Owing to the weak carbon lines not being observable in the LMC and the strong carbon lines always being badly saturated, we have no other option than to use the relation between iron and carbon depletions established in the Milky Way applied to the iron depletions measured in the LMC in order to estimate the carbon depletions in the LMC.  However, it is possible that the relation between carbon and iron depletion in the LMC could differ from the one in the Milky Way. While we cannot compare the relations between carbon and iron depletions in the Milky Way and LMC, we can perform this exercise for magnesium, which was measured in the Milky Way by \citet{jenkins2009} and in the LMC by this study. Therefore, in order to gauge the impact of possible differences in depletion patterns between the LMC and Milky Way on the estimate of \ciis column densities, we also scaled the measured LMC \mgiis column densities (Table 5) by the ratio of carbon/magnesium stellar abundances and the ratio of carbon/magnesium depletions as described above, but this time using the iron depletion measured in the LMC applied to the relation between iron and magnesium depletions established in the Milky Way by \citep{jenkins2009}. \\
\indent The comparison of these different approaches is shown in Figure \ref{plot_cii_estimates}. We found no differences in the \ciis column density outcomes between these different approaches. This is a direct consequence of the result that will be published in an upcoming METAL paper that the relation between iron depletion and the depletions of other elements vary only very weakly between the MW, LMC, and SMC (even though the relation between depletions and N(H) does vary significantly between these galaxies as shown in \citet{RD2019} for silicon). Therefore the most simple approach of scaling the column density of hydrogen by the carbon abundance and depletions is preferred since it can be applied to all sight-lines (not all sight-lines have measured sulfur and magnesium column densities).

\end{document}